\documentclass[fleqn,usenatbib]{mnras}

\usepackage{newtxtext,newtxmath}
\usepackage[T1]{fontenc}
\usepackage{graphicx}
\usepackage{txfonts}
\usepackage{hyperref}
\usepackage{adjustbox}
\usepackage{url}
\usepackage{rotating}

\DeclareRobustCommand{\VAN}[3]{#2}
\let\VANthebibliography\thebibliography
\def\thebibliography{\DeclareRobustCommand{\VAN}[3]{##3}\VANthebibliography}

\usepackage{graphicx}	
\usepackage{amsmath}	
\usepackage{amssymb}	

\title[LAMOST Fe(42) Strong Stars]{Searching for shell stars in LAMOST DR4 by probing the Fe 42 multiplet lines}

\author[S.~H{\"u}mmerich et al.]{
Stefan H{\"u}mmerich,$^{1,2}$\thanks{E-mail: ernham@rz-online.de (SH)}
Ernst Paunzen,$^{3}$ and 
Klaus Bernhard$^{1,2}$
\\
$^{1}$Bundesdeutsche Arbeitsgemeinschaft f{\"u}r Ver{\"a}nderliche Sterne e.V. (BAV), Munsterdamm 90, D-12169 Berlin, Germany\\
$^{2}$American Association of Variable Star Observers (AAVSO), 49 Bay State Rd, Cambridge, MA 02138, USA\\
$^{3}$Department of Theoretical Physics and Astrophysics, Masaryk University, Kotl\'a\v{r}sk\'a 2, 611\,37 Brno, Czech Republic}

\date{Accepted 2022 October 6; Revised: 2022 September 29; Received: 2022 September 8}

\pubyear{2022}

\begin{document}
\label{firstpage}
\pagerange{\pageref{firstpage}--\pageref{lastpage}}
\maketitle

\begin{abstract}
Shell stars, in particular the cooler ones, often do not show conspicuous Balmer-line emission and may consequently be missed in surveys that specifically search for emission signatures in the H$\alpha$ line. The present work is aimed at identifying stars with shell-signatures via a search for strong \ion{Fe}{II} multiplet 42 lines at $\lambda\lambda$4924, 5018, 5169\,\AA\ in archival LAMOST spectra. Candidates were selected by probing the \ion{Fe}{II}\,(42) lines in the spectra of a sample of colour-preselected early-type stars using a modified version of the MKCLASS code and then categorised by visual inspection of their spectra. We identified 75 stars showing conspicuous shell features, 43 Am/CP1 stars, 12 Ap/CP2 stars, and three objects with composite spectra. Spectral types and equivalent width measurements of the \ion{Fe}{II}\,(42) lines are presented for the sample of shell stars. Except for three objects, all shell stars appear significantly removed from the ZAMS in the colour-magnitude diagram, which is likely due to extinction by circumstellar material. 
We find a correlation between the equivalent width of the $\lambda$5169\,\AA\ line and the distance to the locus of the main-sequence stars (the larger the IR-excess, the stronger the $\lambda$5169\,\AA\ line) and studied the variability of the shell star sample using TESS data, identifying a very high proportion of double stars. All but 14 shell stars are new discoveries, which highlights the efficiency of the here presented novel approach to identify stars with subtle shell features. This study may be used as a blueprint for discovering these objects in massive spectral databases.
\end{abstract}




\section{Introduction}

A certain set of B, A, and, much more rarely, F stars shows spectroscopic evidence for the presence of circumstellar shells, which manifests itself in enhanced lines of \ion{Fe}{II} and \ion{Ti}{II}, emission, or deep and narrow absorption cores in the hydrogen Balmer lines, and, quite often, strong and peculiar \ion{Ca}{II} K and weak \ion{Mg}{II} $\lambda$4481\,\AA\ lines \citep{gray09}. A quite heterogeneous group of objects may exhibit some or all of these features, such as B-type and A-type shell stars, and Herbig Ae/Be stars.

B-type shell stars are generally regarded as classical Be stars seen edge-on \citep[e.g.][]{porter03,rivinius06}, with the latter being defined as non-supergiant B stars that show (or have shown at some time) emission in one or more of the Balmer lines \citep{jaschek81}. Several O and A emission-line stars are also included in this group of objects \citep{jaschek86,negueruela04,li18}.

Classical Be and B-type shell stars are quite numerous and allow the investigation of the interplay of such diverse phenomena as, for example, mass loss, the development and dispersion of circumstellar disks or shells, and pulsation. They have therefore been in the focus of many studies in the past \citep{underhill82,rivinius13,baade16,labadiebartz17,labadiebartz18,bernhard18,labadiebartz22}. Recently, the study of these objects benefited greatly from the advent of large-scale spectroscopic surveys \citep[e.g.][]{chojnowski15,anusha21,zhang22,shridharan21}.

As has been hinted at above, the shell phenomenon also extends to later spectral types, that is, into the realm of the A- and even F-type stars \citep{slettebak82,gray09}. These cooler objects, which are in the focus of the present investigation, are generally thought to represent the less massive counterparts of the classical Be and B-type shell stars \citep{abt73,gray09,bohlender16}. Other physical mechanisms, however, can be responsible for the development of a circumstellar shell, such as binarity or evaporating, cometary-like bodies \citep{ferlet87}. Evidence for the presence of the latter is found in the distinct class of the $\beta$ Pictoris shell stars, which possess protoplanetary disks and show variable and narrow absorption features in, for example, \ion{Fe}{II}, \ion{Ti}{II}, and the \ion{Ca}{II} H\&K lines.

Generally, the strength of the shell features encompasses a broad range. In extreme cases, they can completely overwhelm the stellar spectrum, rendering the classification of the central star impossible \citep{gray09}. At the other extremity there are objects that show no other peculiarities than slightly enhanced lines of \ion{Fe}{II} and \ion{Ti}{II}, as compared to the luminosity class derived from the hydrogen lines, mostly in combination with a weak $\lambda$4481\,\AA\ line. These stars seem to only possess weak or tenuous shells and are sometimes referred to as ``proto-shell'' stars \citep{gray87}.

The \ion{Fe}{II} and \ion{Ti}{II} lines (or, rather, the non-photospheric contributions to these lines) arise from metastable states in the extended circumstellar shells and are commonly termed ``shell lines''. Of particular interest is the \ion{Fe}{II} multiplet 42, which consists of three lines at $\lambda\lambda$4924, 5018, 5169\,\AA. Shell stars can be readily recognized by their strongly enhanced \ion{Fe}{II}\,(42) lines \citep{gray09}. This is particulary helpful for the identification of (cool) objects that do not show conspicuous Balmer-line profiles, let alone any appreciable Balmer-line emission. However, most recent surveys that target emission-line and shell stars specifically search for emission signatures in the H$\alpha$ line \citep{anusha21,zhang22,shridharan21}, and these objects will consequently be missed.

Here we present our efforts at identifying shell-signatures via a search for strong \ion{Fe}{II}\,(42) lines using archival spectra from the fourth data release (DR4) of the Large Sky Area Multi-Object Fiber Spectroscopic Telescope
(LAMOST) of the Chinese Academy of Science \citep{lamost1,lamost2}.

\section{Spectroscopic data and target selection} \label{dataanalysis}

This section describes the LAMOST spectral archive, the MKCLASS code, the sample selection process, and the spectral classification workflow.

\subsection{The Large Sky Area Multi-Object Fiber Spectroscopic Telescope (LAMOST)} \label{LAMOST}

The LAMOST telescope, also referred to as the Guo Shou Jing\footnote{Guo Shou Jing was a Chinese astronomer, mathematician and hydraulic engineer of the Yuan Dynasty.} telescope, is based at Xinglong Observatory in Beijing, China \citep{lamost1,lamost2}. It is a Schmidt telescope design with an effective aperture of 3.6$-$4.9\,m and its optical axis fixed along the north-south meridian. With its ability to take 4000 spectra in a single exposure, the LAMOST telescope is specially suited to carry out large-scale spectral surveys and is currently involved in a survey of the entire available northern sky. The LAMOST low-resolution spectra, which form the basis of our study, have a spectral resolution of R\,$\sim$\,1800, cover the wavelength range from about 3700 to 9000\,\AA\, and are available to a limiting magnitude of $r$\,$\sim$\,19\,mag. The spectra are released to the public in consecutive data releases, which can be accessed via the LAMOST spectral archive.\footnote{\url{http://www.lamost.org}} The LAMOST archive is a prime resource for researchers, whose exploitation has been gathering pace during the recent years.

\subsection{The MKCLASS code} \label{MKCLASS}

\subsubsection{General overview} \label{MKCLASS_1}

The MKCLASS code was conceived and written by Richard O. Gray to classify stellar spectra on the MK system. It imitates the methodology of a human classifier and directly compares the input spectrum to a set of standard star spectra \citep{gray14}. Provided the input spectra are of sufficient signal-to-noise (S/N), the classifications derived by MKCLASS agree very well with the classifications derived by expert human classifiers \citep[e.g.][]{gray14,gray16,huemmerich18,huemmerich20,paunzen21}. Typical uncertainties for the derived temperature and luminosity classes amount to 0.6 and 0.5 spectral subclasses, respectively \citep{gray14}.

MKCLASS is inherently able to identify certain spectral peculiarities, such as those found, for example, in Am/CP1 stars, Ap/CP2 stars, barium stars, and carbon-rich giants. More information on the MKCLASS code is provided in \citet{gray14} and on the corresponding website.\footnote{\url{http://www.appstate.edu/~grayro/mkclass/}}

\subsubsection{Targeting the \ion{Fe}{II} (42) triplet lines with the MKCLASS code} \label{MKCLASS_2}

We successfully employed modified versions of the MKCLASS code to search for several groups of chemically peculiar (CP) stars in spectra from LAMOST Data Release (DR) 4 (magnetic CP stars, i.e. Ap/CP2 and He peculiar stars, \citealt{huemmerich20}; HgMn/CP3 stars, \citealt{paunzen21}). To this end, the code was altered to probe additional spectral features relevant to the identification of these objects, such as certain \ion{Si}{II}, \ion{Cr}{II}, \ion{Sr}{II}, and \ion{Eu}{II} lines (Ap/CP2 stars), and \ion{Hg}{II} and \ion{Mn}{II} lines (HgMn/CP3 stars).

\begin{figure}
        \includegraphics[width=\columnwidth]{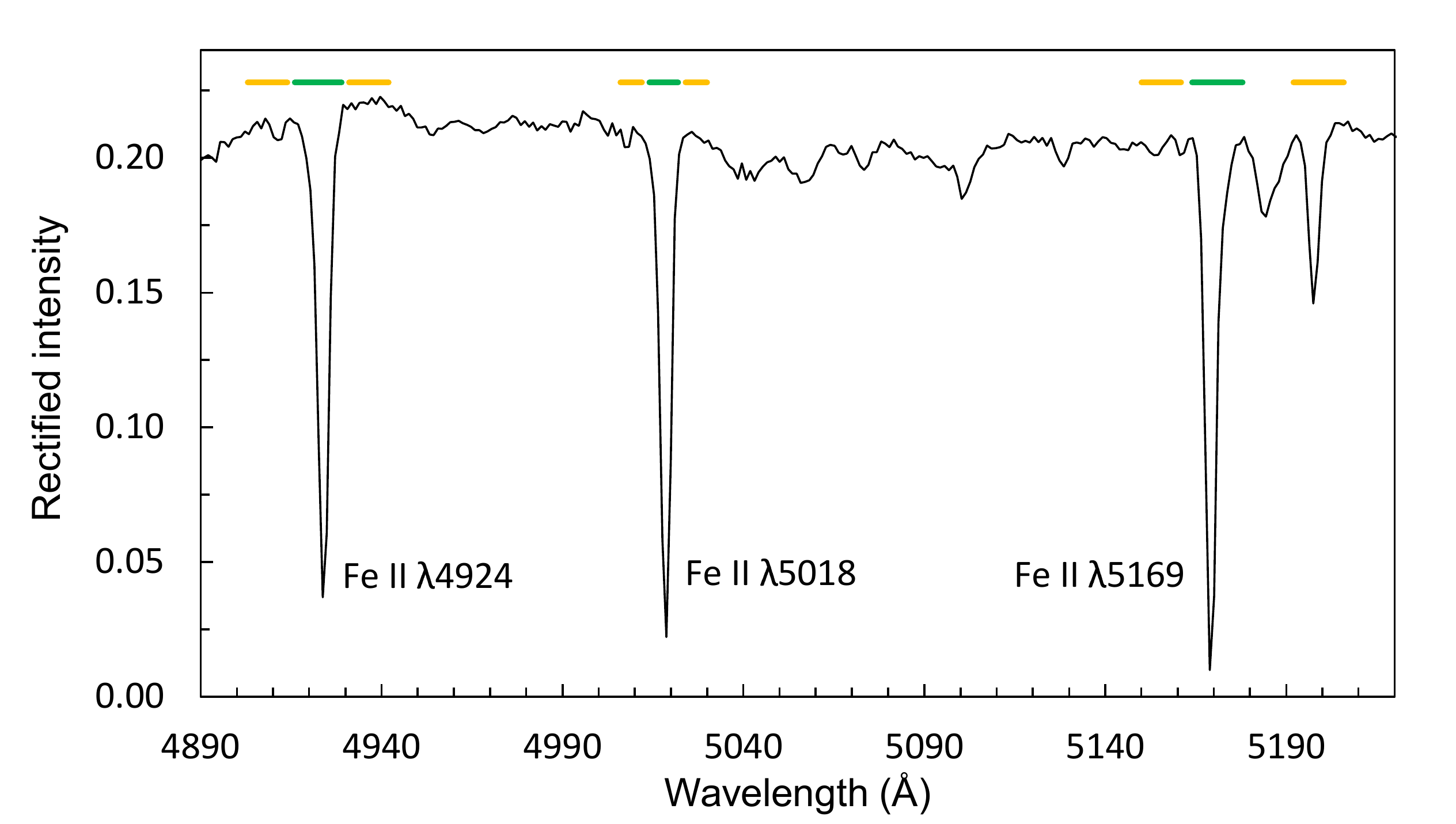}
    \caption{Graphic representation of the setup adopted to probe the \ion{Fe}{II} (42) triplet lines at $\lambda\lambda$4924, 5018, 5169\,\AA\ in the low-resolution LAMOST spectra. Indicated are the spectral regions chosen to probe the corresponding absorption lines (green) and the neighbouring continuum flux (orange). The results were corrected for the rather prominent absorption line in the 5192 $-$ 5206\,\AA\ region.}
    \label{fig_MKCLASS}
\end{figure}

To suit the needs of the present study, the MKCLASS code was altered in a similar manner to probe the \ion{Fe}{II} (42) triplet lines at $\lambda\lambda$4924, 5018, 5169\,\AA\ as they appear in the low-resolution LAMOST spectra (R\,$\sim$\,1800). The corresponding flux was obtained by direct integration and then substracted from the continuum measured in suitable adjacent spectral regions without, or with a minimum of, lines. Some experimentation was required to come up with a solution that is robust against contamination by other lines and blends in the corresponding spectral regions. As no suitable line-free neighbouring continuum region is available redwards of the $\lambda$5169\,\AA\ line, we resorted to the spectral region from $\lambda\lambda$5192 $-$ 5206\,\AA\ and corrected for the rather prominent blend at around $\lambda$5198\,\AA. The finally adopted solution is given in Table \ref{table_MKCLASS} and graphically represented in Fig. \ref{fig_MKCLASS}.

\begin{table}
\caption{Spectral regions chosen to probe the \ion{Fe}{II} (42) triplet lines at $\lambda\lambda$4924, 5018, 5169\,\AA\ and the neighbouring continuum flux.}
\label{table_MKCLASS}
\begin{center}
\begin{tabular}{cccc}
\hline
\hline
(1) & (2) & (3) & (4) \\
line & continuum bluewards & line region & continuum redwards \\
\hline
$\lambda$4924 & $\lambda$$\lambda$4903 $-$ 4916 & $\lambda$$\lambda$4916 $-$ 4929 & $\lambda$$\lambda$4929 $-$ 4942 \\
\hline
$\lambda$5018 & $\lambda$$\lambda$5006 $-$ 5014 & $\lambda$$\lambda$5014 $-$ 5022 & $\lambda$$\lambda$5022 $-$ 5030 \\
\hline
$\lambda$5169 & $\lambda$$\lambda$5150 $-$ 5164 & $\lambda$$\lambda$5164 $-$ 5178 & $\lambda$$\lambda$5192 $-$ 5206 \\
\hline
\hline
\end{tabular}
\end{center}
\end{table}

\subsection{Sample selection} \label{sample_selection}

We initially cross-matched the complete LAMOST DR4 catalogue with the $Gaia$ DR2 catalogue \citep{gaia1,gaia2,gaia3} and then employed a $G$ versus $(BP-RP)$ diagram to select objects with $(BP-RP)$\,$<$\,0.45\,mag, with the aim of restricting the initial sample to stars hotter than a spectral type of about mid F \citep{huemmerich20,paunzen21}. As this approach is bound to miss highly reddened early-type objects or objects with bad $Gaia$ photometry, we selected additional B, A and F stars via the spectral types listed in the DR4 VizieR online catalogue \citep{DR4}.\footnote{\url{http://cdsarc.u-strasbg.fr/viz-bin/cat/V/153}} 

As next step, all corresponding spectra were downloaded from the LAMOST DR4 archive, requiring a S/N of at least 50 in the Sloan $g$ band. In the case of objects with more than one spectrum available, only the spectrum with the highest S/N was considered in the analysis.

Finally, the spectra were classified with the modified version of the MKCLASS code using the four different standard star libraries \textit{libr18}, \textit{libnor36}, \textit{libsynth}, and \textit{liblamost}, as described in \citet{huemmerich20}. To identify spurious detections, the number of detections of the \ion{Fe}{II} (42) triplet lines with the different standard star libraries was counted, which yielded $N$\textsubscript{det}($\lambda$) for each line. As in our previous studies (\citealt{huemmerich20} and \citealt{paunzen21}), the $N$\textsubscript{det}($\lambda$) values were interpreted as an estimation of significance. Only objects were chosen that satisfied the criterion $N$\textsubscript{det}($\lambda$4924)\,>\,1 OR $N$\textsubscript{det}($\lambda$5018)\,>\,1 OR $N$\textsubscript{det}($\lambda$5169)\,>\,1, that is, we required that at least one of the \ion{Fe}{II} (42) triplet lines was detected with at least two different libraries. The stars satisfying this criterion were assigned candidate status.

The spectra of all candidates ($N$\,=\,158) were visually inspected to sort out artefacts and contamination by other groups of stars with prominent lines in the investigated spectral region. Spectra containing strong artefacts or other issues, such as null flux at the wavelengths of one of the \ion{Fe}{II} (42) triplet lines, and spectra with very low S/N that entered the sample despite the selection criterion S/N $g$ > 50 were sorted out ($N$\,=\,25). The remaining sample consists of three distinct groups of stars: (i) a diverse set of stars with clearly defined \ion{Fe}{II} (42) triplet lines of various strength ($N$\,=\,75), (ii) a homogeneous set of stars that shows rather prominent \ion{Fe}{II} (42) triplet lines together with a multitude of other lines and blends in the corresponding spectral region ($N$\,=\,55), and (iii) several stars showing composite spectra that combine characteristic lines of a hot and a cool object ($N$\,=\,3). Group (i) will be referred to in the following as the sample of shell stars, even though it contains two Herbig Ae/Be stars (cf. Section \ref{sample_shell_stars}), which are not shell stars in the narrower sense of the term. Group (ii) consists of Am/CP1 ($N$\,=\,43) and Ap/CP2 ($N$\,=\,12) stars.

\subsection{Spectral classification} \label{spectral_classification}

Spectral types were derived using using the same methodology as in our previous MKCLASS-based studies \citep{huemmerich20,paunzen21}. In a nutshell, spectral types were preferred in the order \textit{liblamost} > \textit{libsynth} > \textit{libnor36} > \textit{libr18}, except when a spectral type was derived more than once across the four spectral libraries. In that case, the most common spectral type was selected as the final classification (hereafter ``MKCLASS final type''). This procedure is is exemplarily illustrated in Table \ref{MKCLASS_output}. In the rare case of striking discrepancies between the spectral types derived with the different libraries, the best-fitting spectral type was adopted after a visual inspection of the spectra. To comply with MK standards, for the final notation, the term ``shell'' was appended to the spectral type instead of the MKCLASS output \citep{gray87,gray09}. We did not use parentheses to indicate the strength of the shell features \citep[cf. e.g.][]{gray87}. 

\begin{table*}
\caption{Spectral classification procedure based on the raw output of the modified MKCLASS code. Spectral types were preferred in the order \textit{liblamost} > \textit{libsynth} > \textit{libnor36} > \textit{libr18}. If common classifications exist, the most common spectral type was selected as the final classification. The columns denote: (1) LAMOST identifier. (2) MKCLASS output using the standard star libraries \textit{lib18}, \textit{libnor36}, \textit{libsynth}, and \textit{liblamost}. (3) finally adopted spectral type. The spectral types on which the final classification was based are highlighted with bold font in column (2).}
\label{MKCLASS_output}
\begin{center}
\begin{tabular}{|l|l|l|}
\hline
(1) & (2) & (3) \\
\hline
LAMOST ID	&	Output using \textit{libr18/\textit{libnor36}/\textit{libsynth}/\textit{liblamost}} &	SpT\_final \\
\hline
J002559.10+555631.3	 &	A0 II-III  Fe4923 Fe5018 Fe5171  & A0 II-III  shell \\	
& A0 II-III  Fe4923 Fe5018 Fe5171 & \\
& A0 II-III  Fe4923 Fe5018 Fe5171 & \\
& A0 III  Fe4923 Fe5018 Fe5171	& \\
\hline
J003933.02+273029.5 & kA4hA9mA8  Fe4923 Fe5018 Fe5171 &  A5 III-IV  shell \\
& kA4hA8mF0  Fe5018 Fe5171 & \\
& A5 III-IV  Fe4923 Fe5018 Fe5171 & \\
& A5 III-IV  Fe4923 Fe5018 Fe5171 & \\
\hline
J004127.44+220716.9 &  kA2hA3mA6  Fe4923 Fe5018 Fe5171 &  A2 IV-V  shell \\
&  A2 IV-V  Fe5018 Fe5171 & \\
&  kA2hA4mA6  Fe4923 Fe5018 Fe5171 & \\
&  A2 IV-V  Fe4923 Fe5018 Fe5171 & \\
\hline
J020138.33+561624.4 &  A2 IV-V  Fe4923 Fe5018 Fe5171  &  A2 IV-V  shell \\
&  A2 IV-V  Fe4923 Fe5018 Fe5171  & \\
&  A3 IV-V  Fe4923 Fe5018 Fe5171  & \\
&  A2 IV-V  Fe4923 Fe5018 Fe5171  & \\
\hline
J020207.64+543927.9 &  kA1hA9mA8  Fe4923 Fe5018 Fe5171 &  A1 III-IV  shell  \\
&  A1 III  Fe4923 Fe5018 Fe5171  & \\
&  kA1hA3mA6  Fe4923 Fe5018 Fe5171 & \\
&  A1 III-IV  Fe4923 Fe5018 Fe5171 & \\
\hline
\end{tabular}        
\end{center}
\end{table*}


\subsection{Equivalent widths} \label{equivalent_widths}

The equivalent widths of the \ion{Fe}{II}\,(42) lines at $\lambda\lambda$4924, 5018, 5169\,\AA\ were measured using the program package ROBOSPEC\footnote{\url{http://www.ifa.hawaii.edu/users/watersc1/robospect/}} \citep{2013PASP..125.1164W}, which is based on the assumption that all spectra are comprised of three components: the continuum
level, the line solution relative to that continuum, and an error component that contains the deviation between
the true spectrum and the current model. By iterating the fit of these components, it is ensured that the line
and continuum solutions are not biased by each other. Prior to this process, an initial estimate of the fit parameters is constructed for each line under the assumption that the line profile is Gaussian -- a good approximation for the investigated \ion{Fe}{II}\,(42) lines.

As a prerequisite for using ROBOSPEC, the spectra were normalised in the corresponding spectral range using the Astropy package SPECUTILS\footnote{\url{https://specutils.readthedocs.io/}}, which yields very robust results. Nevertheless, all normalised spectra were visually inspected before the calculation of the equivalent widths.

As a final step, we compared the results from ROBOSPEC for ten stars using the basic IRAF routine SPLOT. The agreement was excellent. The equivalent width measurements are included in the presentation of results in Table \ref{table_results}.

\section{Results} \label{results}

In this section, we present the list of shell stars identified with the specified approach, a list containing the ``bycatch'' of other astrophysically interesting objects, an investigation of the shell star sample in colour-magnitude and near-infrared colour-colour diagrams, and a photometric variability analysis.

\subsection{The shell star sample} \label{sample_shell_stars}

In total, we identified 75 objects with clearly defined \ion{Fe}{II} (42) triplet lines of various strength whose spectra show shell-features of various strength (cf. Section \ref{sample_selection}).

\begin{figure*}
        \includegraphics[width=\textwidth]{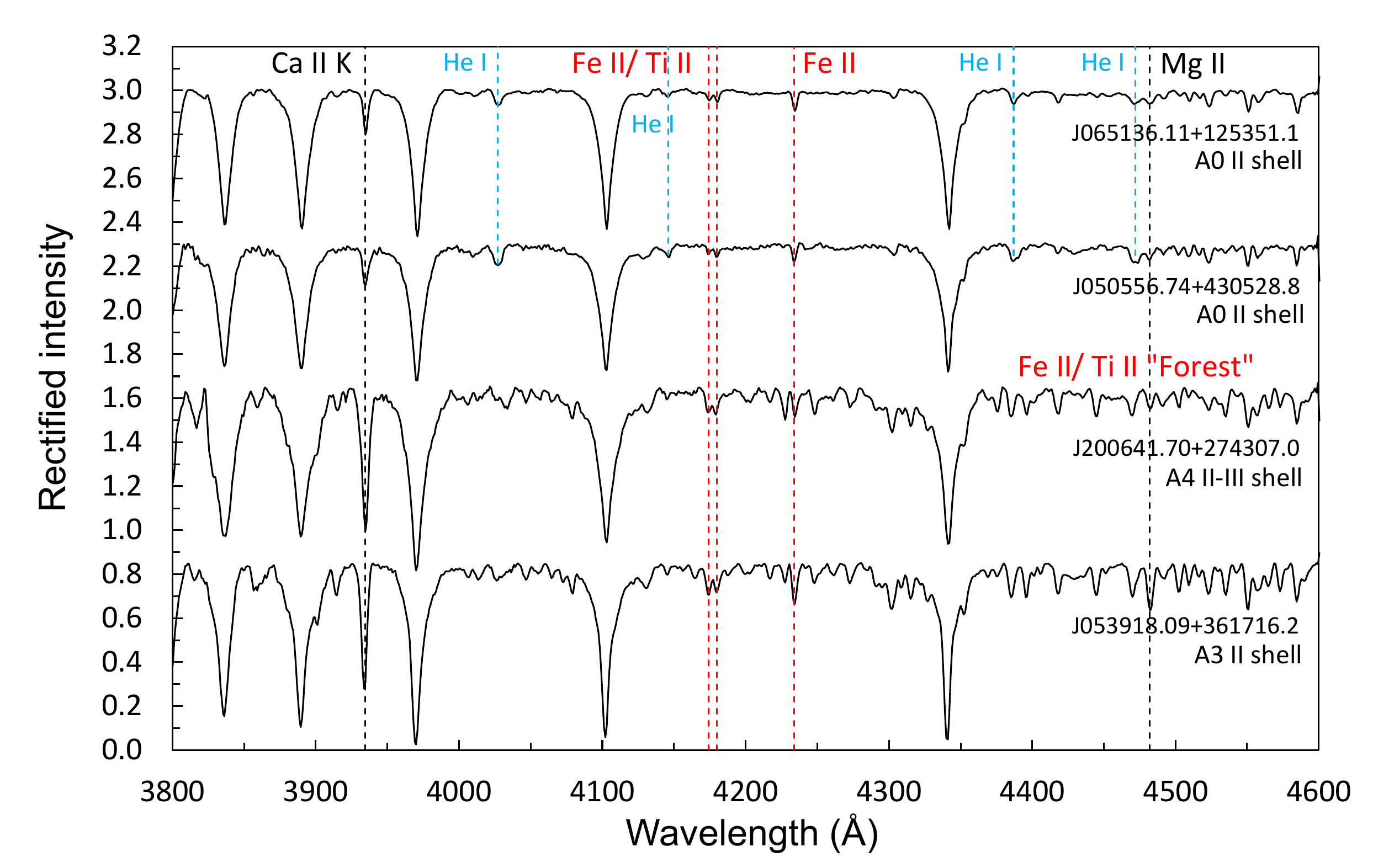}
    \caption{Blue-violet spectral region of four example spectra with shell features, based on spectra from LAMOST DR4. Some prominent lines of interest are identified.}
    \label{fig_showcase1}
\end{figure*}

In the blue-violet ($\lambda\lambda$3800$-$4600\,\AA) spectral region, all stars show typical shell lines, that is, enhanced \ion{Fe}{II} and \ion{Ti}{II} lines, which is most readily seen in the $\lambda\lambda$4172-9 blend due to \ion{Fe}{II} and \ion{Ti}{II}, the \ion{Fe}{II} $\lambda$4233 line, and the ``forest'' of \ion{Fe}{II} and \ion{Ti}{II} lines from about $\lambda$4400 to $\lambda$4600\,\AA\ \citep{gray09}. Figure \ref{fig_showcase1} illustrates the blue-violet spectral region of four example stars. Other characteristics commonly encountered in these objects -- and readily visible in this plot -- are weak \ion{Mg}{II} $\lambda$4481 lines; peculiarly strong \ion{Ca}{II} K lines; and rather broad lines of \ion{He}{I} $\lambda\lambda$4026, 4144, 4387, 4471 in the hotter stars (the upper two objects).

\begin{figure*}
        \includegraphics[width=\textwidth]{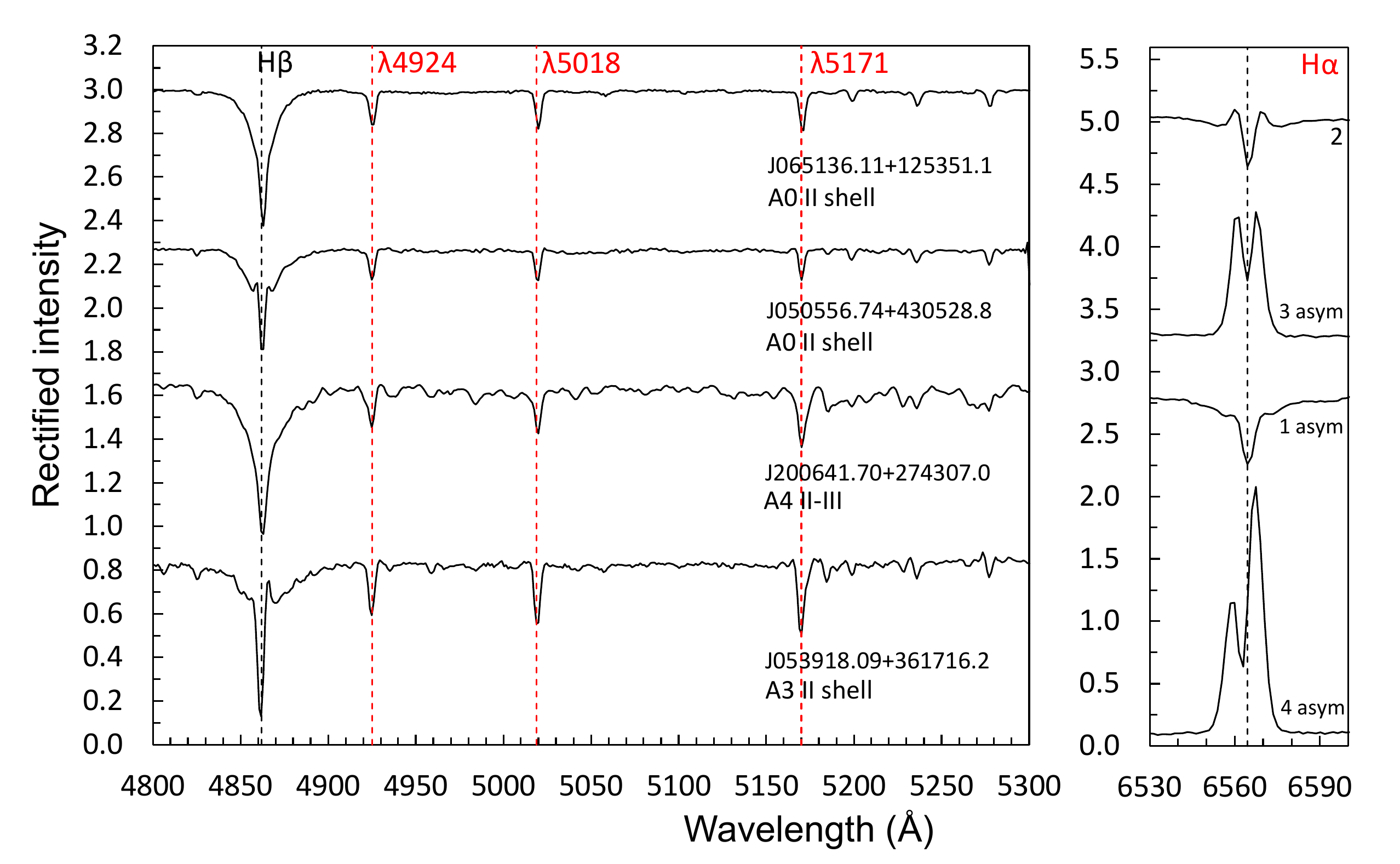}
    \caption{Spectral region of the \ion{Fe}{II}\,(42) triplet lines including the H$\beta$ line (left panel) and the corresponding region of the H$\alpha$ line (right panel), based on spectra from LAMOST DR4. Some prominent lines of interest are identified. The objects are the same as shown in Fig. \ref{fig_showcase1}. The numbers next to the H$\alpha$ line profiles correspond to the notation used to describe the H$\alpha$ line profile characteristics (cf. Table \ref{table_Halpha_characteristics}).}
    \label{fig_showcase23}
\end{figure*}

Figure \ref{fig_showcase23} shows the corresponding region of the \ion{Fe}{II}\,(42) triplet lines including the H$\beta$ line ($\lambda\lambda$4800$-$5300\,\AA; left panel) and the region of the H$\alpha$ line ($\lambda\lambda$6530$-$6600\,\AA; right panel). All objects show strongly enhanced \ion{Fe}{II}\,(42) triplet lines as well as peculiar H$\beta$ profiles characterised by shell absorption cores of varying strength. The star J053918.09+361716.2\footnote{Unless indicated otherwise, all coordinate-based identifiers in this paper refer to LAMOST identifiers.} is a Herbig Ae/Be star \citep{zhang22}.

The H$\alpha$ line profiles shown in the right panel of Figure \ref{fig_showcase23} are equally characteristic. In particular the hotter shell stars show conspicuous emission features with a central absorption core, as seen here in all stars but the cool J200641.70+274307.0, which only shows very weak emission wings. The Herbig Ae/Be star J053918.09+361716.2 exhibits strongly asymmetric profiles in both the H$\beta$ and H$\alpha$ lines.

\begin{figure}
    \includegraphics[width=0.45\textwidth]{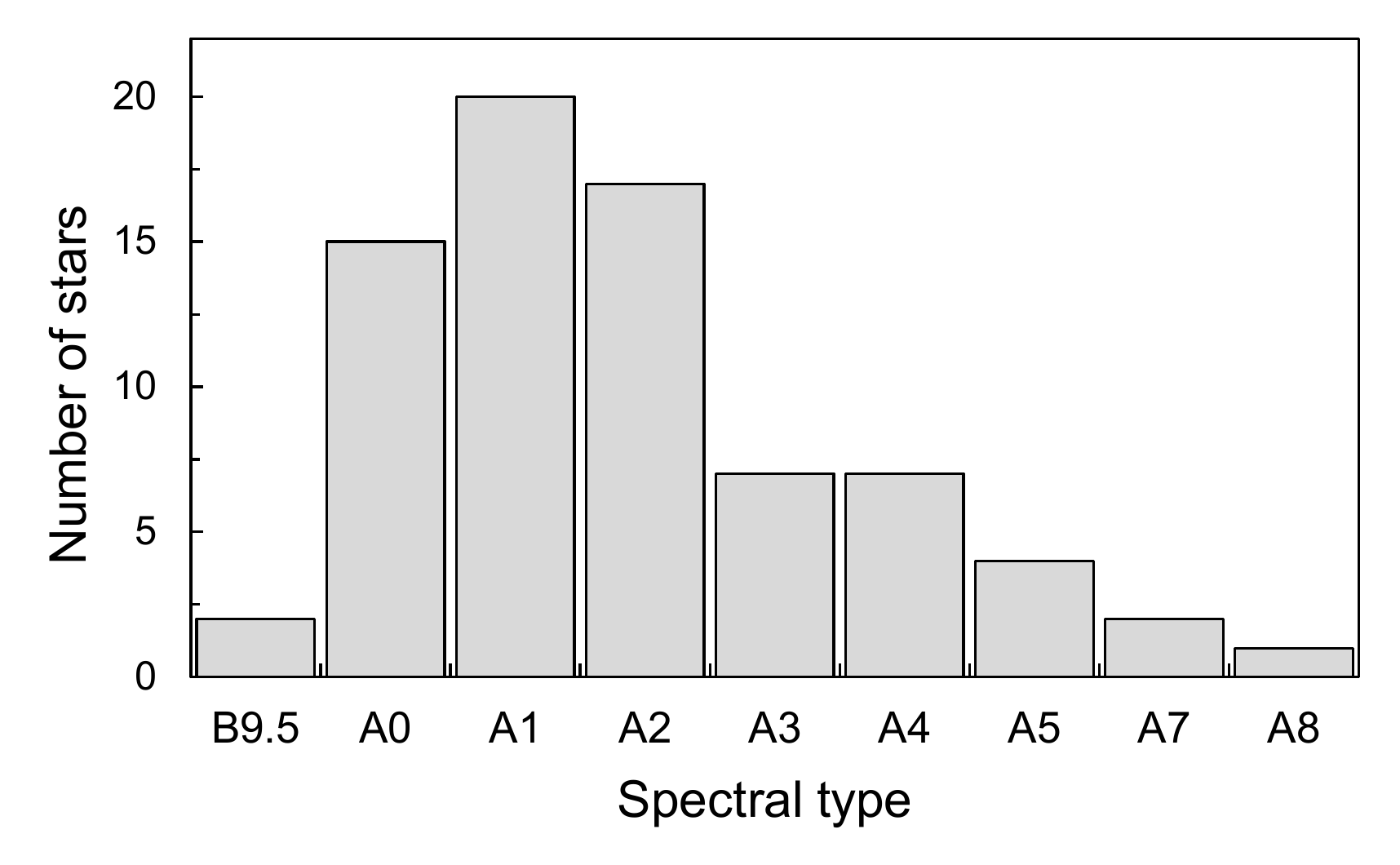}
    \caption{Distribution of spectral types among the sample of shell stars.}
    \label{fig_histogramSpT}
\end{figure}

The final sample of shell stars is presented in Table \ref{table_results},  which contains LAMOST identifiers, GAIA EDR3 coordinates, brightness measurements in the GAIA $G$ band, MK spectral types, equivalent width measurements of the \ion{Fe}{II}\,(42) triplet lines, and a description of the H$\alpha$ line profile characteristics. The notation used to describe the latter is presented in Table \ref{table_Halpha_characteristics}. The distribution of spectral types is shown in Fig. \ref{fig_histogramSpT}.

\begin{table}
\caption{Notation used to describe the H$\alpha$ line profile characteristics of the final sample of shell stars.}
\label{table_Halpha_characteristics}
\begin{center}
\begin{tabular}{ll}
\hline
\hline
Notation & Description \\
\hline
0 & arrow line, no emission, no absorption core, \\              & often filled-in/flat-bottomed \\
0(+) & conspicuously filled-in \\
\hline
1 & narrow absorption core, emission wings absent \\
    or very weak \\
1(-) & weak absorption core \\
1(+) & conspicuous absorption core \\
\hline
2 & emission wings and central absorption core \\
2(-) & weak emission wings \\
2(+) & strong emission wings \\
\hline
3 & emission with central absorption core \\
\hline
4 & strong emission with central absorption core \\
\hline
asym & asymmetric profile \\
\hline
\end{tabular}
\end{center}                               
\end{table}

All but 14 stars are new discoveries. 13 objects are contained in the list of \citet{zhang22}, five of which are also included in the sample of \citet{shridharan21}. One additional star (J200641.70+274307.0) is contained in the GAIA DR2-based catalogue of \citet{vioque20}. These objects are identified by the footnotes provided after the LAMOST identifiers in Column 1 of Table \ref{table_results}. As LAMOST spectra have been thoroughly searched for emission-line stars \citep{anusha21,zhang22,shridharan21}, this is a remarkable but not unexpected result because our approach is based on a search for strong \ion{Fe}{II}\,(42) lines and does not specifically target emission-line features, which may be inconspicuous or even absent in shell stars. This highlights the efficiency of the chosen approach to find shell stars in massive spectral databases.

\begin{figure}
    \includegraphics[width=0.45\textwidth]{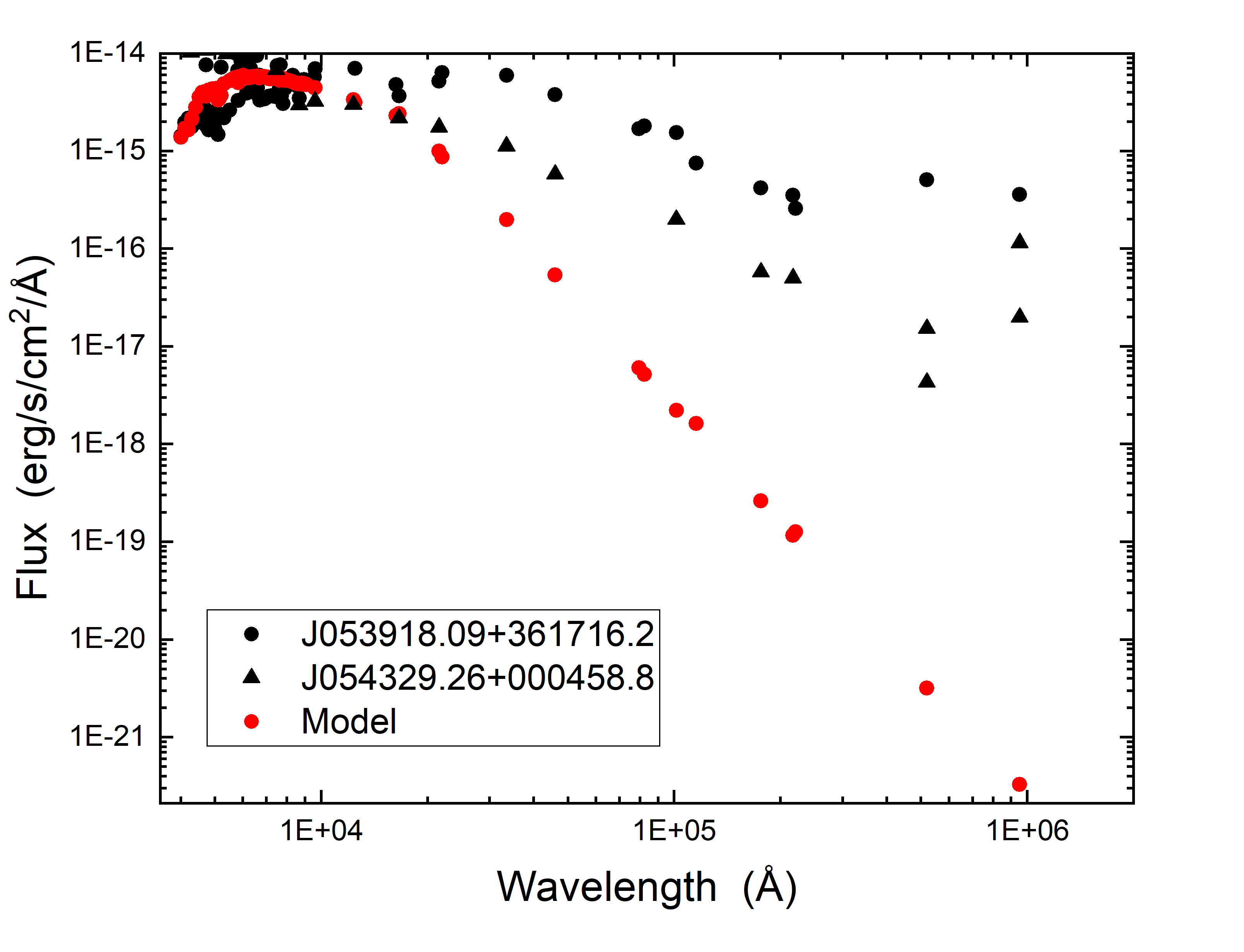}
    \caption{IR excesses for the Herbig Ae/Be stars J053918.09+361716.2 and J054329.26+000458.8, which are typical for a circumstellar disk.}
    \label{fig_IR_excess}
\end{figure}

Interestingly, two of the stars contained in the sample of \citet{zhang22} (J053918.09+361716.2 and J054329.26+000458.8) have been identified as Herbig Ae/Be stars. To further investigate the spectral energy distribution (SED) of these objects, we have employed
the VO Sed Analyzer tool, VOSA\footnote{http://svo2.cab.inta-csic.es/theory/vosa/} 
v7.5 \citep{2008A&A...492..277B} for fitting the available photometry.
In Fig. \ref{fig_IR_excess} we show the SED of the two Herbig Ae/Be stars, which clearly exhibit an IR-excess redwards of 1\,$\mu m$, which is typical for a
circumstellar disk \citep{1998A&A...329..131M}.

\subsection{Other astrophysically interesting objects} \label{sample_other_stars}

Apart from the shell stars, our sample also contains 55 stars with prominent \ion{Fe}{II} (42) triplet features and a multitude of other lines and blends in the corresponding spectral region. A careful study of the corresponding spectra revealed that this group mostly consists of CP1 and CP2 stars. The CP1 stars exhibit significant underabundances of Ca and Sc and a general enhancement of the ironpeak
and heavier elements in the stellar photosphere. The CP2 stars, on the other hand, are characterised by significant overabundances of selected elements such as Si, Sr, Eu, or the rare-earth elements. Both types of star show characteristic ``forests'' of lines in the region of the \ion{Fe}{II} (42) triplet. Their detection by the modified MKCLASS code, therefore, does not come as a surprise.

\begin{figure*}
        \includegraphics[width=\textwidth]{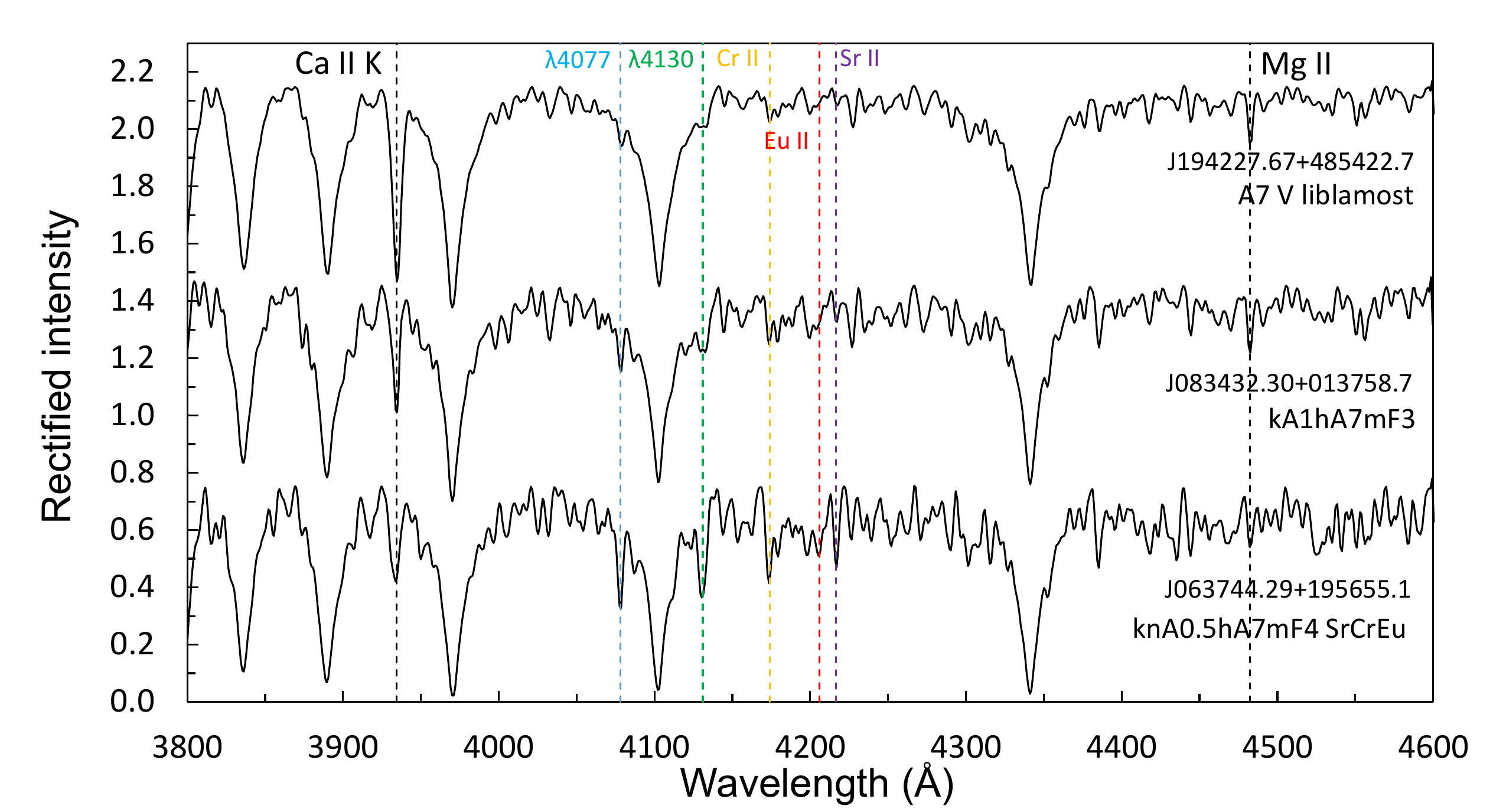}
    \caption{Blue-violet spectral region of, from top to bottom, the A7 V standard star from the $liblamost$ library, an example CP1/Am star, and an example CP2/Ap star, based on spectra from LAMOST DR4. Some prominent lines of interest are identified.}
    \label{fig_showcase3}
\end{figure*}

Example spectra of a late A-type CP1 and a late A-type CP2 star in comparison to the A7 V standard star from the \textit{liblamost} library \citep{huemmerich20} are shown in Fig. \ref{fig_showcase3}. Both CP stars have characteristically weak \ion{Ca}{II} K lines. While the CP1 star shows a general overabundace of the heavy elements, there is a strong selective enhancement of Sr, Cr, and Eu features in the CP2 star. This is particularly visible in the strong $\lambda$4077\,\AA\ blend, which contains contributions from \ion{Si}{II} $\lambda$4076, \ion{Cr}{II} $\lambda$4077, and \ion{Sr}{II} $\lambda$4077; the $\lambda$4130\,\AA\ blend, which contains contributions from \ion{Si}{II} $\lambda\lambda$4128-30 and \ion{Eu}{II} $\lambda\lambda$4130; the \ion{Cr}{II} $\lambda$4172\,\AA\ line; the \ion{Eu}{II} $\lambda$4205\,\AA\ line; and the \ion{Sr}{II} $\lambda$4216\,\AA\ \citep{gray09}.\footnote{ Cf. also the Atlas of LAMOST Low-Resolution Spectra of Chemically Peculiar Stars by one of the authors (SH), which is available at \url{http://www.appstate.edu/~grayro/mkclass/LAMOST_CP_Atlas_v1.pdf}}

Table \ref{table_stars_other} contains elementary data for the CP1 and CP2 stars as well as the stars showing composite spectra. We also cross-matched this subsample with LAMOST-based catalogues of CP1 and CP2 stars \citep{RM09,scholz2019,qin19,huemmerich20,shang22}, the results of which are included in Column 9 of Table \ref{table_stars_other}. It is noteworthy that almost all the CP stars have been identified correctly in former studies.

\begin{figure}
    \includegraphics[width=0.45\textwidth]{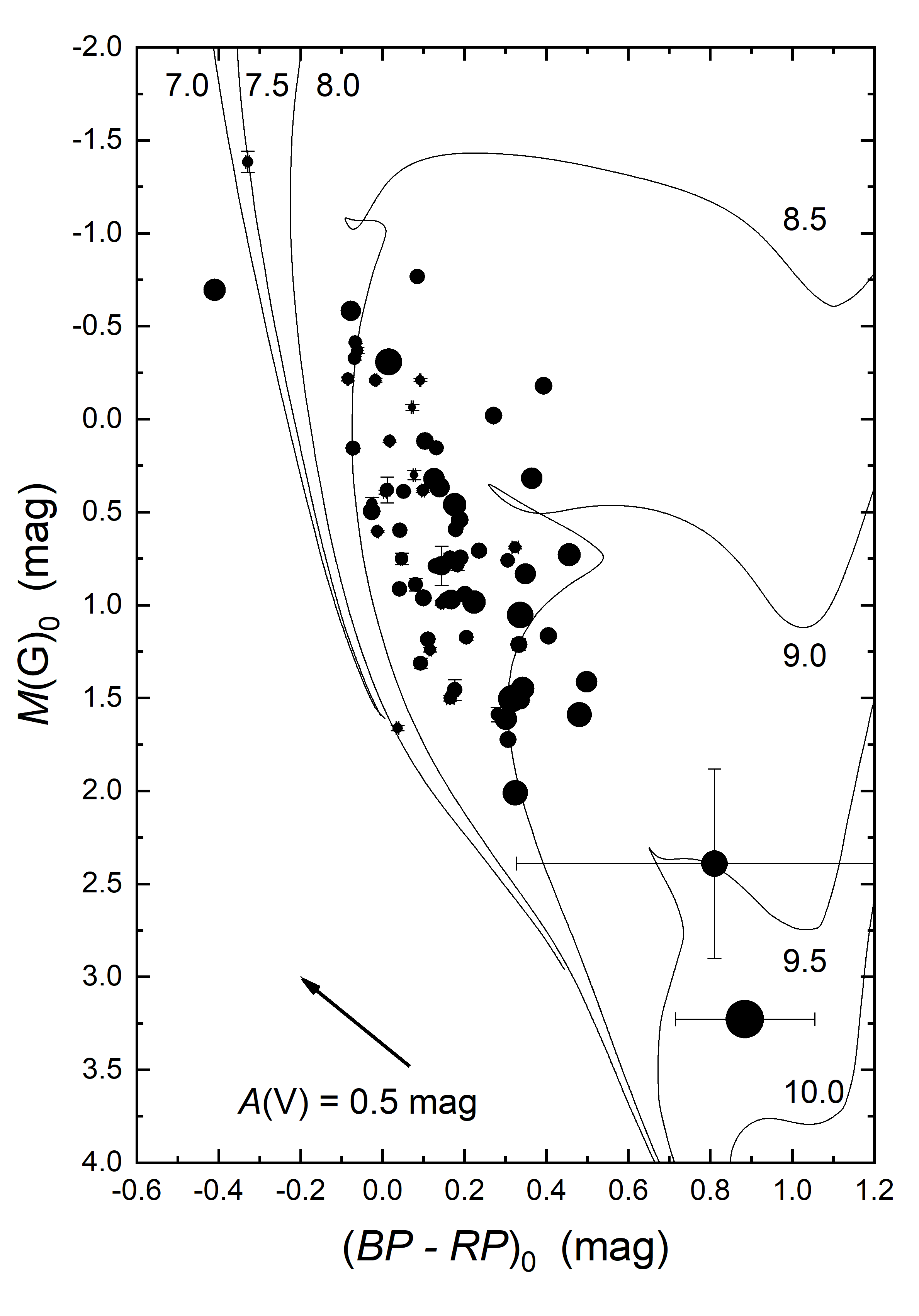}
    \caption{The CMD for our target star sample together with main-sequence PARSEC isochrones \citep{2012MNRAS.427..127B} for a solar metallicity of [Z]\,=\,0.0152. The symbol sizes are 
    proportional to equivalent widths of the  \ion{Fe}{II}\,(42) line at $\lambda\lambda$5169\,\AA\,
    (Table \ref{table_results}). Also shown is the reddening vector for $A$(V)\,=\,0.5\,mag.}
    \label{fig_CMD}
\end{figure}

\subsection{Astrophysical parameters, equivalent widths, binarity, and IR-excess of the shell star sample} \label{astrophysical_parameters}

We constructed the colour-magnitude diagram (CMD) for the final sample of shell stars using the homogeneous $Gaia$ DR2 photometry from \citet{gaia3}. The interstellar reddening (absorption) needs to be taken into account because
most of our sample stars are members of the Galactic disk and situated beyond 500\,pc from the Sun.
Unfortunately, Str{\"o}mgren-Crawford indices, which allow a reliable reddening estimation, 
are not available \citep{2015A&A...580A..23P}. We therefore relied on the three-dimensional reddening map of
\citet{2019ApJ...887...93G}. For the distances, we used the values published 
by \citet{2018AJ....156...58B} who applied a weak distance prior that varies smoothly as a 
function of Galactic longitude and latitude according to a Galaxy model. The estimated extinctions
were then transformed to the $Gaia$ photometric systems using the coefficients listed in 
\citet{2018A&A...616A..10G}. The distances were directly converted into the absolute
magnitudes $M$(G).

In Fig. \ref{fig_CMD}, we present the $M$(G)$_\mathrm{0}$ versus ($BP-RP$)$_\mathrm{0}$ diagram 
together with the main-sequence PARSEC isochrones \citep{2012MNRAS.427..127B} for a solar metallicity of [Z]\,=\,0.0152. 
The Herbig Ae/Be stars J053918.09+361716.2 and J054329.26+000458.8 
(see Section \ref{sample_shell_stars}) are clearly separated from the other objects.
Besides three stars, all objects are located quite far away from the zero-age main sequence (ZAMS). This is not 
expected because they should cover the whole area up to the ZAMS \citep{1988A&AS...72..505J}.
This discrepancy can be explained by ``the missing'' extinction due to the circumstellar
material. In our analysis, we have only corrected for the interstellar absorption. An additional
0.5\,mag in $V$ would bring most of the stars very close to the ZAMS (see Fig. \ref{fig_CMD}).
Without knowledge about the strength of this additional contribution, it does not make sense to further
calibrate any astrophysical parameters. Only fitting the hydrogen lines and, to some extent, the
SED could further narrow down the effective temperature range for the individual stars.

\begin{figure}
    \includegraphics[width=0.45\textwidth]{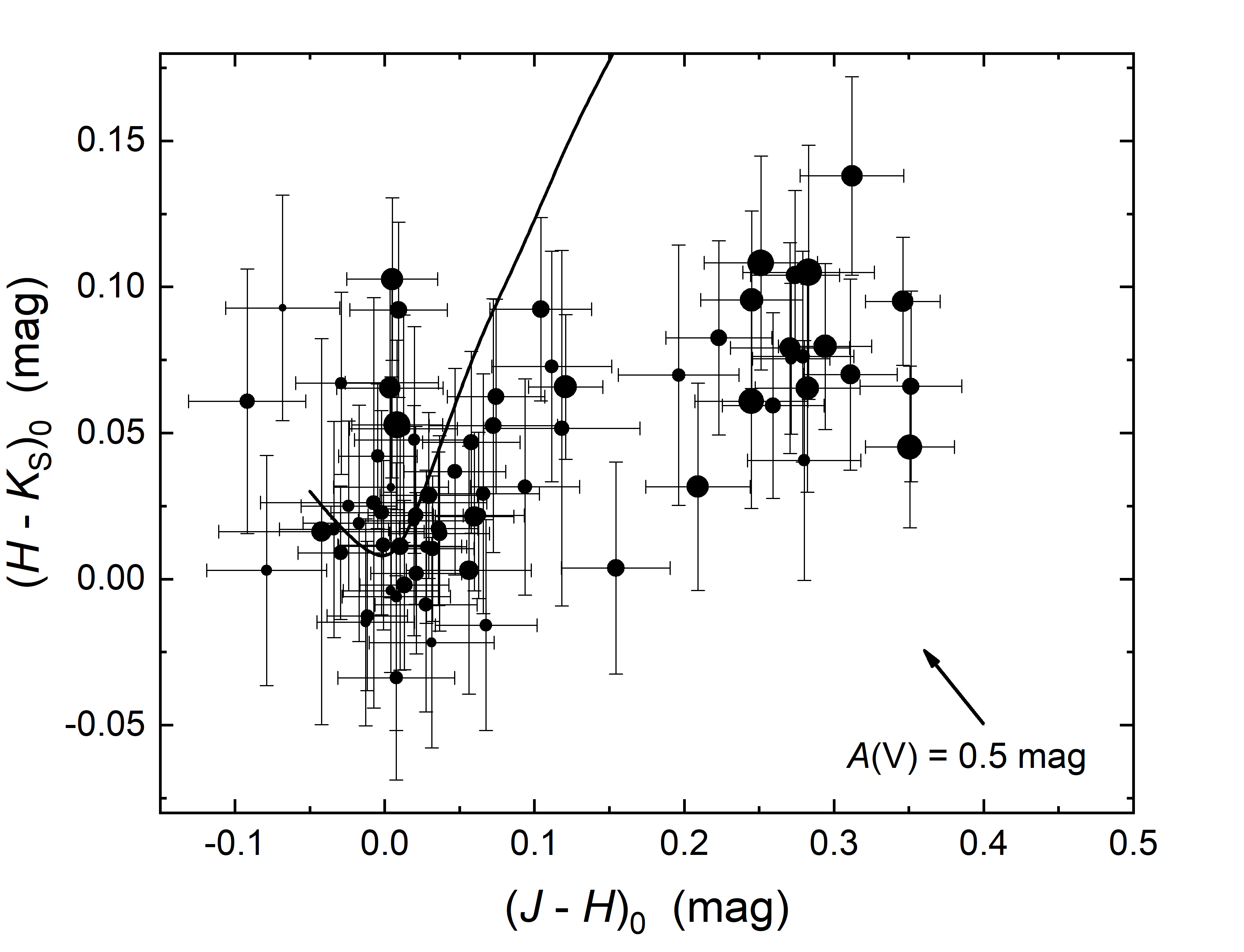}
    \caption{The ($J - H$)$_\mathrm{0}$ versus ($H - K_\mathrm{S}$)$_\mathrm{0}$ diagram showing
    two groups of objects. The line represents the locus of the normal type main sequence stars 
    \citep{1988PASP..100.1134B}. The symbol sizes are 
    proportional to equivalent widths of the  \ion{Fe}{II}\,(42) line at $\lambda\lambda$5169\,\AA\,
    (Table \ref{table_results}). 
    The two outliers J053918.09+361716.2 and J054329.26+000458.8 were omitted.}
    \label{fig_2MASS}
\end{figure}

To further investigate the IR-excess of the shell star sample, we performed a cross-match with the 
$JHK_\mathrm{S}$
2MASS survey \citep{2006AJ....131.1163S}. We then constructed a ($J - H$)$_\mathrm{0}$ versus ($H - K_\mathrm{S}$)$_\mathrm{0}$ diagram using the corresponding
absorption coefficients from \citet{2013MNRAS.430.2188Y}. We note that the reddening is smaller than in the optical region, therefore the above mentioned effect is not so severe. The result
is plotted in Fig. \ref{fig_2MASS}, in which two distinct groups separated at 
($J - H$)$_\mathrm{0}$\,=\,0.2\,mag are visible. The reddening vector \citep{1988PASP..100.1134B}
points toward the 
line of the normal type main-sequence stars. To shift the stars of the second group close to
the main sequence, correcting for an absorption of two to three magnitudes in $V$ is necessary. Such high values are very unlikely taking
into account the derived spectral types and the corresponding shift in the CMD (Figure \ref{fig_CMD}).

There is a correlation of the ($J - H$)$_\mathrm{0}$ colour with EQW$_{\lambda5169}$, whereas the other two lines do not show a correlation (Table \ref{table_correlation}). 
Figure \ref{fig_2MASS} illustrates that stars which are more distant to the standard line tend
to have larger equivalent widths, that means the larger the IR-excess, the stronger the 
$\lambda$5169\,\AA\ line. A similar effect can be seen in the CMD (Fig. \ref{fig_CMD}). Objects closer to the ZAMS tend to have weaker $\lambda$5169\,\AA\ lines.

\begin{figure}
    \includegraphics[width=0.45\textwidth]{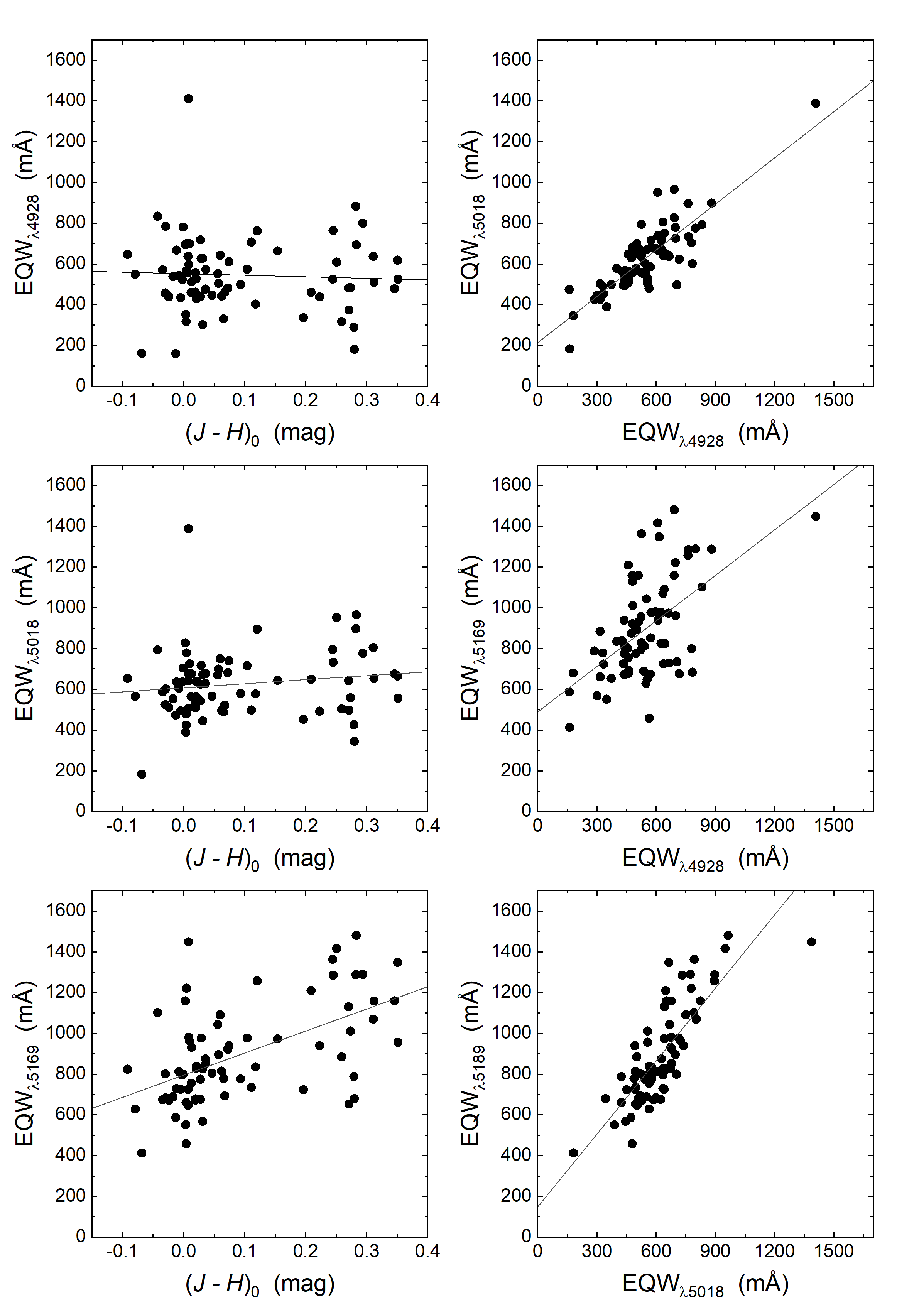}
    \caption{The correlations of the ($J - H$)$_\mathrm{0}$ magnitudes with the 
    different equivalent widths (left columns) and the equivalent widths among themselves (right columns).
    The $R^2$ values of the individual fits are listed in Table \ref{table_correlation}.}
    \label{fig_Correlations}
\end{figure}

As described in Section \ref{equivalent_widths}, we measured the equivalent widths of the \ion{Fe}{II}\,(42) triplet lines of the shell star sample (Table \ref{table_results}). For the calculation of the correlations between the corresponding equivalent widths, we use the $R^2$ value (Table \ref{table_correlation}) which is a measure of the goodness of fit. As expected, the three equivalent widths are correlated, with EQW$_{\lambda4924}$ versus 
EQW$_{\lambda5169}$ showing the weakest, but still statistically significant, correlation (Figure \ref{fig_Correlations}).
This can be interpreted as an indication that these two lines are formed in different stellar environments/regions
\citep{2021A&A...653A.115E}. Although the here employed LAMOST spectra have only classification resolution (R\,$\sim$\,1800), our results prove that they can be used in future efforts to model the environments of shell 
stars in more detail \citep{1996ARep...40..509M}. 

Finally, we checked the $Gaia$ DR2 and DR3 for indications of binarity among the shell star sample. To this end, we used the ``Dup'' flag and the RUWE parameter, which are good indicators \citep{2022AJ....163...33Z}. In Table \ref{table_binaries}, we list the corresponding indicators for the stars that show conspicuous values (RUWE\,$>$\,1.6 was adopted as threshold). Six of these stars have been identified as eclipsing binaries (J032704.99+434028.5, J040713.90+291832.1, J074455.91+042805.1, J075535.94-033455.4, J092110.64+283147.9) or ellipsoidal variables (J232258.57+541829.2; cf. also Table \ref{table_VAR}).

\begin{table}
\caption{$R^2$ values as a measure of the goodness of fit for the linear regression between the 
tabulated parameters. The two outlying datapoints of the Herbig Ae/Be stars J053918.09+361716.2 and J054329.26+000458.8 were excluded from the statistical analysis.}
\label{table_correlation}
\begin{center}
\begin{tabular}{cccc}
\hline
\hline
& EQW$_{\lambda4924}$ & EQW$_{\lambda5018}$ & EQW$_{\lambda5169}$ \\
\hline
EQW$_{\lambda4924}$ & & 0.7058	& 0.3149 \\
EQW$_{\lambda5018}$ & & & 0.6574 \\
($J - H$)$_\mathrm{0}$ & 0.0025 &	0.0216	& 0.3012 \\
\hline
\end{tabular}
\end{center}                               
\end{table}

\begin{table*}
\caption{Binarity indicators from $Gaia$ DR2 and DR3.}
\label{table_binaries}
\begin{center}
\begin{tabular}{cccccccc}
\hline
\hline
ID\_LAMOST & RUWE & Dup(DR3) & Dup(DR2) & ID\_LAMOST & RUWE & 
Dup(DR3) & Dup(DR2) \\
\hline
J002559.10+555631.3	&	0.951	&	0	&	1	&	J054358.96+491845.6	&	0.954	&	1	&	1	\\
J020138.33+561624.4	&	0.998	&	0	&	1	&	J054834.63+234141.4	&	1.120	&	1	&	1	\\
J020207.64+543927.9	&	1.919	&	0	&	0	&	J064230.24+092626.7	&	0.845	&	1	&	1	\\
J020820.76+562433.0	&	0.952	&	0	&	1	&	J074455.91+042805.1	&	1.078	&	1	&	1	\\
J032704.99+434028.5	&	0.965	&	1	&	1	&	J075535.94$-$033455.4	&	1.405	&	0	&	1	\\
J033909.14+523714.5	&	0.931	&	0	&	1	&	J092110.64+283147.9	&	3.920	&	0	&	0	\\
J035033.65+525943.2	&	1.906	&	0	&	1	&	J180605.37+020543.8	&	1.843	&	0	&	0	\\
J040713.90+291832.1	&	1.592	&	1	&	1	&	J195131.48+484558.9	&	4.257	&	0	&	0	\\
J041308.41+534045.9	&	0.922	&	1	&	1	&	J200641.70+274307.0	&	1.157	&	0	&	1	\\
J053235.07+495218.6	&	1.838	&	0	&	0	&	J231357.71+545812.0	&	0.903	&	1	&	1	\\
J054329.26+000458.8	&	2.059	&	0	&	0	&	J232258.57+541829.2	&	1.033	&	0	&	1	\\
\hline
\end{tabular}
\end{center}                               
\end{table*}

\subsection{Variability analysis} \label{variability_analysis}

The Transiting Exoplanet Survey Satellite (TESS) aims at the discovery of transiting exoplanets \citep{TESS3,TESS1,TESS2}.\footnote{\url{https://heasarc.gsfc.nasa.gov/docs/tess/}} To this end, four cameras (effective aperture size 10\,cm) equipped with f/1.4 lenses and MIT/Lincoln Lab CCD detectors (4096x4096 pixels; imaging area of 2048x2048 pixels; remaining pixels used as frame-store for rapid shutterless readout) are used, which produce single-passband (6000$-$10000\,\AA) light curves. TESS data have a cadence of 30 minutes and 2 minutes for the summed full-frame images and the ``postage stamp'' observations centered on 20000 preselected stars. With their wealth of high-cadence high-precision time-series data, the TESS archives are an excellent starting point for stellar variability studies.

The full-frame TESS images have been processed by \citet{huang2020_1,huang2020_2} and \citet{kunimoto21} to provide quick-look pipeline light curves. Data from all available sectors (up to sector 35) for the final shell star sample were downloaded via MAST\footnote{\url{https://archive.stsci.edu/hlsp/qlp}} and included in the analysis. In total, data from 72 sectors could be retrieved for 48 stars.

After a visual inspection of all light curves and the removal of apparent outliers, a period analysis of the uncorrected SAP flux data was performed using the Lomb-Scargle method as implemented in the software package \textsc{PERANSO} \citep{PERANSO}.

Significant periods as well as preliminary types are given in Table \ref{table_VAR}, which also lists periods and types from the International Variable Star Index (VSX; \citealt{VSX}) of the American Association of Variable Star Observers (AAVSO). Due to the limited observation time in a TESS sector ($\sim$27d on average), only periods of a maximum of 14\,d could be reliably recorded.

In addition to a few objects with irregular variability ($N$\,=\,3) and presumed pulsators ($N$\,=\,3), there is a high number of eclipsers and presumably ellipsoidal variables ($N$\,=\,18) among the stars of our final sample. This indicates a very high proportion of double stars among the shell star sample, which is not surprising, as binarity can be responsible for the development of a circumstellar shell \citep{ferlet87}.

The JD light curves of all variables among the final sample of shell stars are provided in the Appendix (Section \ref{appendix}).

\section{Conclusions}

We carried out a search for shell stars by targeting the \ion{Fe}{II}\,(42) lines at $\lambda\lambda$4924, 5018, 5169\,\AA\ in LAMOST DR4 spectra with a modified version of the MKCLASS code. This led to the identification of three distinct groups of stars. 75 stars show shell features, that is, enhanced \ion{Fe}{II} and \ion{Ti}{II} lines as well as typical H$\alpha$ profiles such as emission features, a central absorption core, and/or filled-in line cores, in agreement with a shell star classification. Two of these stars (J053918.09+361716.2 and J054329.26+000458.8) are Herbig Ae/Be stars \citep{zhang22}. Apart from that, we identified 43 Am/CP1 stars, 12 Ap/CP2 stars, as well as three stars showing composite spectra with lines indicative of a hot and a cool object.

We present MK class spectral types and equivalent width measurements of the \ion{Fe}{II}\,(42) lines for the sample of shell stars, which was also investigated in the $M$(G)$_\mathrm{0}$ vs. ($BP-RP$)$_\mathrm{0}$ and ($J - H$)$_\mathrm{0}$ vs. ($H - K_\mathrm{S}$)$_\mathrm{0}$ parameter spaces. Except for three objects, all stars appear significantly ($\sim$0.5 mag in $V$) removed from the ZAMS in the CMD, which is likely due to extinction caused by circumstellar material. This assumption is further corroborated by the results from the near-infrared colour-colour diagram, which shows two distinct groups of stars separated at ($J - H$)$_\mathrm{0}$\,=\,0.2\,mag. As expected, the Herbig Ae/Be stars are clearly separated from the other objects in both parameter spaces.

We find a correlation between the position of a star in both diagrams and the equivalent width of the $\lambda$5169\,\AA\ line, in the sense that the further a star is removed from the locus of the main-sequence stars, the larger is EQW$_{\lambda5169}$. Put differently, the larger the IR-excess, the stronger is the $\lambda$5169\,\AA\ line. The strongest correlation was found between ($J - H$)$_\mathrm{0}$ and EQW$_{\lambda5169}$. We find evidence that the $\lambda$4924\,\AA\ and $\lambda$5169\,\AA\ lines are formed in different stellar environments/regions and show that the low-resolution LAMOST spectra are generally suited for modeling the environments of shell stars in more detail.

Using TESS data, we investigated the photometric variability of the shell stars. With 18 eclipsers and ellipsoidal variables, we find a very high proportion of double stars. Likely, binarity is responsible for the development of the circumstellar shell in these objects.

All but 14 shell stars are new discoveries, which highlights the efficiency of the chosen approach to identify objects with subtle shell features that, for the most part, do not show strong Balmer-line emission and will consequently be missed in surveys that specifically search for emission signatures in the H$\alpha$ line. Therefore, by opening up a new way of identification, our study adds a further piece to the puzzle of understanding the various manifestations of the shell star phenomenon and may be used as a blueprint for discovering these objects in massive spectral databases.

\section{DATA AVAILABILITY}

The data underlying this article are available in the article and in its online supplementary material.

\begin{table*}
\caption{Essential data for our sample stars, sorted by increasing right ascension. The columns denote: (1) LAMOST identifier. (2) LAMOST observation ID. (3) Alternativ identifier. (4) Right ascension (J2000; GAIA EDR3). (5) Declination (J2000; GAIA EDR3). (6) $G$\,mag (GAIA EDR3). (7) Spectral type. (8) Sloan $g$ band S/N ratio of the analysed spectrum. (9) Equivalent width (m\AA) of the $\lambda$4924\,\AA\ line. (10) $\sigma$EQW (m\AA). (11) Equivalent width (m\AA) of the $\lambda$5018\,\AA\ line. (12) $\sigma$EQW (m\AA). (13) Equivalent width (m\AA) of the $\lambda$5169\,\AA\ line. (14) $\sigma$EQW (m\AA). (15) H$\alpha$ line profile characteristics.}
\label{table_results}
\begin{adjustbox}{max width=\textwidth}
\begin{tabular}{lllllllllllllll}
\hline
\hline
(1) & (2) & (3) & (4) & (5) & (6) & (7) & (8) & (9) & (10) & (11) & (12) & (13) & (14) & (15) \\
ID\_LAMOST	&	ObsID	&	ID\_alt	&	RA(J2000)	&	Dec(J2000)	&	$G$\,mag	&	SpT\_final	&	S/N\,$g$	&	EQW$_{\lambda4924}$	&	$\sigma$EQW	&	EQW$_{\lambda5018}$	&	$\sigma$EQW	&	EQW$_{\lambda5169}$	&	$\sigma$EQW	&	H$\alpha$	\\
\hline
J002559.10+555631.3	&	171103150	&	GSC 03657-00187	&	00 25 59.097	&	+55 56 31.114	&	11.999	&	 A0 II-III  shell  	&	178.2	&	570.4	&	2.9	&	586.0	&	0.0	&	673.1	&	0.1	&	2	\\
J003933.02+273029.5	&	157503180	&	GSC 01744-02329	&	00 39 33.025	&	+27 30 29.417	&	12.503	&	 A5 III-IV  shell  	&	154.8	&	616.8	&	0.6	&	663.9	&	0.3	&	1347.0	&	1.3	&	2(-)	\\
J004127.44+220716.9$^{3a}$	&	194501119	&	GSC 01193-00234	&	00 41 27.442	&	+22 07 16.944	&	11.712	&	 A2 IV-V  shell  	&	230.1	&	483.3	&	0.9	&	558.1	&	0.5	&	1011.0	&	0.8	&	2(+)	\\
J020138.33+561624.4	&	380711095	&	UCAC4 732-019419	&	02 01 38.341	&	+56 16 24.493	&	14.307	&	 A2 IV-V  shell  	&	119.5	&	481.4	&	0.5	&	681.1	&	0.5	&	922.4	&	0.4	&	1(-)	\\
J020207.64+543927.9	&	380804167	&	GSC 03689-00919	&	02 02 07.642	&	+54 39 28.010	&	11.422	&	 A1 III-IV  shell  	&	419.1	&	573.1	&	1.2	&	678.2	&	0.3	&	851.9	&	0.1	&	1 asym	\\
J020820.76+562433.0	&	380812126	&	HD 12921	&	02 08 20.633	&	+56 24 35.290	&	10.119	&	 A0 III-IV  shell  	&	445.7	&	538.6	&	0.9	&	551.6	&	0.0	&	689.3	&	0.1	&	1 asym	\\
J030240.87+522334.6$^{2,3a}$	&	253614211	&	GSC 03322-00936	&	03 02 40.875	&	+52 23 34.692	&	12.112	&	 A1 II  shell  	&	103.1	&	1410.0	&	0.1	&	1388.0	&	0.0	&	1448.0	&	0.1	&	4 asym	\\
J031003.35+551804.4	&	244009232	&	HD 232754	&	03 10 03.359	&	+55 18 01.367	&	9.564	&	 A2 III-IV  shell	&	570.6	&	541.6	&	0.9	&	605.1	&	0.0	&	812.4	&	0.1	&	1(-) asym	\\
J031638.78+494256.2	&	393114190	&	GSC 03319-00602	&	03 16 38.777	&	+49 42 55.989	&	11.106	&	 A2 IV-V  shell  	&	279.9	&	512.1	&	0.3	&	674.7	&	0.0	&	930.6	&	0.3	&	1(-)	\\
J031841.75+462842.1	&	408714242	&	GSC 03311-01885	&	03 18 41.776	&	+46 28 40.974	&	10.913	&	 A2 IV  shell  	&	134.8	&	476.2	&	0.2	&	628.6	&	0.1	&	873.3	&	0.2	&	0	\\
J031909.14+192811.7	&	387704144	&	BD+18 457	&	03 19 08.931	&	+19 28 11.753	&	9.537	&	 A5 III  shell  	&	503.0	&	642.0	&	0.2	&	750.0	&	0.1	&	1090.0	&	0.4	&	0	\\
J032704.99+434028.5	&	96515069	&	GSC 02873-01933	&	03 27 04.997	&	+43 40 28.556	&	10.181	&	 A4 IV-V  shell  	&	304.2	&	480.6	&	0.4	&	641.9	&	1.2	&	1129.0	&	0.5	&	2	\\
J033909.14+523714.5	&	374616043	&	GSC 03716-00037	&	03 39 09.142	&	+52 37 14.550	&	12.146	&	 A0 II-III  shell  	&	152.7	&	557.9	&	1.3	&	528.2	&	0.5	&	672.0	&	0.1	&	2(-)	\\
J034636.05+441159.3	&	185201110	&	GSC 02875-01237	&	03 46 36.054	&	+44 11 59.403	&	11.520	&	 A1 IV  shell  	&	237.3	&	523.6	&	1.7	&	635.7	&	0.0	&	795.1	&	0.1	&	0(+)	\\
J035000.80+580406.0	&	157904161	&	GSC 03725-01226	&	03 50 00.807	&	+58 04 06.118	&	11.470	&	 A1 II-III  shell  	&	230.0	&	350.8	&	0.6	&	389.4	&	0.4	&	550.9	&	0.1	&	1(-) asym	\\
J035033.65+525943.2	&	374611112	&	GSC 03717-00421	&	03 50 33.658	&	+52 59 43.279	&	13.000	&	 A7 III-IV  shell  	&	110.2	&	551.6	&	1.4	&	669.5	&	0.2	&	1043.0	&	1.1	&	0	\\
J035205.73+351059.1	&	181504205	&	GSC 02364-02043	&	03 52 05.733	&	+35 10 59.126	&	14.084	&	 A3 III-IV  shell  	&	152.2	&	330.7	&	0.4	&	488.6	&	0.1	&	777.6	&	0.4	&	0(+)	\\
J040713.90+291832.1	&	370010068	&	GSC 01826-00710	&	04 07 13.878	&	+29 18 32.426	&	11.397	&	 A2 IV  shell  	&	369.8	&	524.1	&	0.3	&	556.7	&	0.2	&	955.2	&	0.7	&	1	\\
J041308.41+534045.9	&	354108184	&	GSC 03718-00545	&	04 13 08.415	&	+53 40 45.956	&	12.785	&	 A2 III  shell  	&	115.4	&	497.8	&	0.3	&	578.8	&	0.1	&	775.9	&	0.3	&	1(-)	\\
J041638.47+530600.9$^{3a}$	&	169203247	&	GSC 03719-00690	&	04 16 38.473	&	+53 06 00.846	&	12.139	&	 A0 II  shell	&	138.5	&	718.3	&	1.2	&	624.1	&	0.5	&	675.2	&	0.1	&	3	\\
J043336.04+451232.6$^{2,3a}$	&	275204156	&	GSC 03342-00138	&	04 33 36.042	&	+45 12 32.709	&	11.803	&	 A0 Ib-II  shell  	&	133.3	&	783.5	&	0.2	&	601.3	&	0.1	&	683.5	&	0.2	&	3	\\
J043415.66+405139.5	&	295303158	&	UCAC4 655-022462	&	04 34 15.666	&	+40 51 39.552	&	14.088	&	 A1 III  shell	&	121.9	&	439.5	&	1.0	&	543.3	&	0.1	&	773.1	&	0.2	&	1	\\
J043948.40+474848.5	&	217903222	&	GSC 03346-00854	&	04 39 48.401	&	+47 48 48.587	&	11.051	&	 A4 III  shell  	&	247.8	&	662.9	&	1.3	&	642.3	&	0.1	&	972.4	&	0.6	&	1	\\
J044307.81+571907.8	&	409716058	&	GSC 03741-00200	&	04 43 07.813	&	+57 19 07.806	&	12.550	&	 A1 III  shell  	&	119.3	&	373.1	&	0.7	&	497.5	&	0.2	&	652.9	&	0.1	&	1(-)	\\
J050435.48+402559.7$^{3a}$	&	315109045	&	UCAC4 653-028138	&	05 04 35.489	&	+40 25 59.682	&	14.133	&	 A1 II  shell  	&	129.3	&	556.1	&	1.2	&	505.3	&	0.5	&	646.4	&	0.2	&	2(+)	\\
J050556.74+430528.8$^{2,3a}$	&	364906120	&	GSC 02903-00983	&	05 05 56.760	&	+43 05 28.871	&	13.310	&	 A0 II  shell  	&	124.4	&	564.8	&	1.4	&	479.2	&	0.6	&	457.1	&	0.1	&	3 asym	\\
J051428.45+274324.3	&	15502153	&	GSC 01854-00033	&	05 14 28.465	&	+27 43 24.555	&	11.595	&	 A2 III-IV  shell  	&	141.4	&	798.9	&	1.3	&	775.1	&	1.1	&	1289.0	&	0.9	&	2(-)	\\
J053235.07+495218.6	&	190911156	&	GSC 03367-01174	&	05 32 35.073	&	+49 52 18.684	&	11.361	&	 A1 III-IV  shell  	&	162.1	&	434.2	&	0.9	&	493.7	&	0.0	&	724.6	&	0.2	&	1(-)	\\
J053414.62+215213.6	&	195010034	&	GSC 01309-01353	&	05 34 14.626	&	+21 52 13.641	&	11.387	&	 A4 III  shell  	&	231.0	&	442.2	&	0.3	&	495.1	&	0.2	&	814.4	&	0.3	&	0 asym	\\
J053918.09+361716.2$^{2,3b}$	&	380515103	&	GSC 02416-00657	&	05 39 18.088	&	+36 17 16.214	&	14.710	&	 A3 II  shell  	&	137.2	&	936.5	&	1.4	&	1080.0	&	0.1	&	1426.0	&	0.3	&	4 asym	\\
J054113.82+351910.7$^{3a}$	&	297004149	&	GSC 02412-01090	&	05 41 13.827	&	+35 19 10.725	&	11.020	&	 B9.5 III  shell  	&	268.0	&	780.0	&	1.0	&	703.6	&	0.0	&	799.2	&	0.1	&	3 asym	\\
J054239.89+174652.7	&	393313172	&	HD 246608	&	05 42 39.841	&	+17 46 53.528	&	10.920	&	 A1 III  shell	&	395.1	&	526.6	&	1.3	&	639.4	&	0.4	&	828.5	&	0.3	&	1	\\
J054329.26+000458.8$^{2,3b}$	&	326912175	&	HD 290828	&	05 43 29.252	&	+00 04 58.970	&	11.676	&	 A7 Ib-II  shell  	&	115.7	&	1164.0	&	1.0	&	1448.0	&	0.4	&	2064.0	&	1.4	&	4 asym	\\
J054358.96+491845.6	&	191012008	&	GSC 03368-01318	&	05 43 58.958	&	+49 18 45.706	&	11.424	&	 A8 II  shell  	&	112.9	&	762.4	&	1.8	&	895.4	&	0.4	&	1257.0	&	0.9	&	0(+)	\\
J054834.63+234141.4	&	116210055	&	BD+23 1058	&	05 48 34.637	&	+23 41 41.489	&	10.921	&	 A1 III-IV  shell  	&	171.8	&	300.8	&	0.3	&	445.2	&	0.1	&	567.7	&	0.2	&	0(+)	\\
J055140.83+414108.5	&	124303201	&	GSC 02920-02782	&	05 51 40.835	&	+41 41 08.500	&	12.615	&	 A0 III  shell  	&	65.8	&	162.3	&	6.3	&	183.1	&	5.9	&	413.0	&	3.5	&	0	\\
J055340.45+332003.9	&	269313053	&	UCAC4 617-029482	&	05 53 40.456	&	+33 20 03.995	&	14.269	&	 A3 III-IV  shell  	&	105.6	&	706.8	&	0.8	&	496.5	&	0.8	&	734.3	&	0.3	&	0 asym	\\
J055441.84+223317.7	&	434407220	&	GSC 01863-02412	&	05 54 41.850	&	+22 33 17.763	&	11.185	&	 A3 III  shell	&	130.1	&	596.3	&	1.3	&	676.3	&	0.1	&	979.7	&	0.4	&	0 asym	\\
J055907.94-060737.6$^{3a}$	&	211510003	&	GSC 04781-00953	&	05 59 07.949	&	-06 07 37.675	&	12.108	&	 A3 IV  shell  	&	127.9	&	509.9	&	3.8	&	653.5	&	1.3	&	1158.0	&	1.3	&	2(+)	\\
J060706.46+245547.1	&	165716036	&	UCAC4 575-024208	&	06 07 06.470	&	+24 55 47.138	&	13.847	&	 A0 II  shell  	&	61.4	&	459.0	&	1.4	&	564.3	&	0.1	&	755.4	&	0.2	&	2	\\
J061810.01+411332.0	&	378007213	&	GSC 02930-00867	&	06 18 10.012	&	+41 13 32.113	&	11.615	&	 A3 III  shell  	&	268.1	&	693.1	&	0.1	&	825.9	&	0.1	&	1158.0	&	0.4	&	1	\\
J062416.04+422125.4	&	378006164	&	GSC 02935-01774	&	06 24 16.047	&	+42 21 25.477	&	11.073	&	 A4 IV  shell  	&	304.3	&	478.3	&	1.1	&	675.9	&	0.7	&	1158.0	&	0.5	&	1	\\
J062618.73+174526.9	&	435515074	&	HD 257117	&	06 26 18.745	&	+17 45 27.254	&	11.345	&	 A2 IV  shell  	&	427.0	&	625.1	&	2.9	&	716.2	&	0.0	&	976.0	&	0.2	&	0	\\
J062724.55+234931.0	&	201604230	&	UCAC4 570-028755	&	06 27 24.551	&	+23 49 31.048	&	15.252	&	 A2 IV  shell  	&	105.8	&	401.0	&	0.3	&	577.0	&	0.2	&	833.9	&	0.2	&	1(-)	\\
J064142.55+031421.9	&	385412092	&	GSC 00151-01200	&	06 41 42.552	&	+03 14 21.954	&	13.017	&	 A1 III  shell  	&	106.6	&	610.1	&	0.1	&	739.4	&	0.0	&	938.6	&	0.2	&	1	\\
J064230.24+092626.7	&	37103019	&	HD 262371	&	06 42 30.247	&	+09 26 26.726	&	9.922	&	 A0 III  shell  	&	338.4	&	438.6	&	0.5	&	509.8	&	0.0	&	671.3	&	0.1	&	0	\\
J064751.97+522321.9	&	264709144	&	GSC 03402-01356	&	06 47 51.974	&	+52 23 21.876	&	12.838	&	 A5 IV-V  shell  	&	108.6	&	525.6	&	2.0	&	794.4	&	0.4	&	1363.0	&	0.6	&	0(+)	\\
J065136.11+125351.1$^{3a}$	&	38714083	&	HD 265136	&	06 51 36.153	&	+12 53 51.562	&	10.277	&	 A0 II  shell  	&	312.1	&	667.3	&	1.0	&	636.3	&	0.4	&	729.2	&	0.0	&	2	\\
J065141.20+070357.7	&	34803143	&	HD 265284	&	06 51 41.202	&	+07 03 57.750	&	10.883	&	 A2 III-IV  shell	&	215.9	&	444.5	&	0.2	&	566.5	&	0.1	&	804.8	&	0.3	&	0	\\
J065300.06+010730.6	&	368901030	&	GSC 00149-01072	&	06 53 00.062	&	+01 07 30.643	&	12.472	&	 A3 II-III  shell  	&	105.5	&	574.7	&	0.1	&	715.6	&	0.1	&	975.9	&	0.5	&	0	\\
J065606.26+033757.5	&	368904202	&	GSC 00153-01723	&	06 56 06.264	&	+03 37 57.524	&	13.402	&	 A1 III  shell  	&	100.6	&	461.0	&	1.8	&	521.7	&	0.1	&	692.8	&	0.3	&	1	\\
J070219.92+162845.4	&	84409116	&	GSC 01344-00359	&	07 02 19.921	&	+16 28 45.453	&	13.394	&	 A2 IV shell 	&	50.3	&	181.2	&	1.6	&	344.6	&	0.6	&	678.7	&	2.0	&	0(+)	\\
J071200.59+241137.2	&	177916248	&	GSC 01896-00475	&	07 12 00.593	&	+24 11 37.376	&	12.633	&	 A1 IV  shell  	&	150.8	&	316.4	&	1.6	&	424.2	&	0.1	&	659.3	&	0.3	&	0(+)	\\
J071340.29+380623.9$^{3a}$	&	184912129	&	HD 55200	&	07 13 40.582	&	+38 06 30.260	&	8.347	&	 A0 II-III  shell  	&	443.6	&	832.6	&	0.7	&	792.7	&	0.0	&	1102.0	&	0.5	&	2(+)	\\
J071630.57+053405.5	&	369602069	&	GSC 00172-00143	&	07 16 30.576	&	+05 34 05.521	&	11.043	&	 A1 III  shell  	&	441.4	&	460.4	&	0.6	&	508.1	&	0.0	&	677.9	&	0.1	&	1	\\
J072719.01+015251.6	&	88611215	&	GSC 00169-00831	&	07 27 19.013	&	+01 52 51.702	&	11.286	&	 A2 III  shell  	&	149.4	&	427.6	&	0.3	&	564.5	&	0.1	&	837.8	&	0.5	&	0 asym	\\
J074455.91+042805.1	&	263916013	&	GSC 00187-01505	&	07 44 55.913	&	+04 28 05.207	&	12.467	&	 A2 III-IV  shell	&	173.4	&	437.9	&	0.6	&	492.2	&	0.8	&	938.7	&	0.9	&	2 asym	\\
J075535.94-033455.4	&	170212127	&	GSC 04837-02525	&	07 55 35.947	&	-03 34 55.562	&	10.675	&	 A0 IV-V  shell  	&	141.0	&	316.9	&	0.5	&	502.3	&	0.2	&	883.6	&	0.2	&	1(-)	\\
J080809.68-052950.7	&	93715174	&	GSC 04855-02705	&	08 08 09.691	&	-05 29 50.777	&	11.292	&	 A4 III-IV  shell  	&	119.7	&	503.1	&	0.3	&	699.0	&	0.2	&	895.4	&	0.6	&	0	\\
J081646.65-040220.3	&	196814212	&	GSC 04856-00003	&	08 16 46.658	&	-04 02 20.468	&	13.763	&	 A1 IV-V  shell  	&	88.3	&	335.4	&	1.4	&	452.7	&	0.9	&	722.4	&	0.4	&	0	\\
J085216.65+090518.7	&	46315140	&	BD+9 2072 	&	08 52 16.652	&	+09 05 18.779	&	10.305	&	 A4 IV-V  shell  	&	113.5	&	608.8	&	1.2	&	951.1	&	2.7	&	1415.0	&	0.6	&	1	\\
J090754.79+212319.3	&	187405093	&	GSC 01407-00362	&	09 07 54.796	&	+21 23 19.322	&	11.285	&	 A5 IV  shell 	&	491.9	&	459.7	&	1.3	&	648.4	&	0.3	&	1210.0	&	0.7	&	1(-) asym	\\
J092110.64+283147.9	&	231808034	&	HD 80535	&	09 21 10.656	&	+28 31 48.053	&	9.481	&	 A2 III-IV  shell  	&	660.7	&	693.1	&	1.0	&	965.1	&	0.5	&	1480.0	&	1.2	&	0(+)	\\
J093958.52-005332.0	&	41910246	&	GSC 04894-02078	&	09 39 58.526	&	-00 53 31.465	&	13.026	&	 A0 V  shell	&	20.7	&	160.2	&	0.5	&	473.7	&	1.4	&	587.2	&	2.3	&	0	\\
J180605.37+020543.8	&	458609157	&	GSC 00434-03484	&	18 06 05.379	&	+02 05 43.834	&	13.018	&	 A2 IV  shell  	&	125.9	&	764.2	&	0.7	&	732.5	&	0.3	&	1285.0	&	1.3	&	1	\\
J190401.96+415350.8	&	458112119	&	GSC 03128-02038	&	19 04 01.952	&	+41 53 50.944	&	13.618	&	 A1 IV-V  shell  	&	117.2	&	636.0	&	0.4	&	804.3	&	0.1	&	1070.0	&	0.2	&	1(-) asym	\\
J195131.48+484558.9	&	462802091	&	GSC 03565-01258	&	19 51 31.489	&	+48 45 58.977	&	11.435	&	 A1 IV-V  shell  	&	246.8	&	456.7	&	0.2	&	523.1	&	0.0	&	799.6	&	0.1	&	1	\\
J200641.70+274307.0$^{1}$	&	461511071	&	GSC 02162-01397	&	20 06 41.705	&	+27 43 07.005	&	12.463	&	 A4 II-III  shell  	&	103.0	&	699.0	&	0.8	&	777.9	&	0.3	&	1220.0	&	1.0	&	1 asym	\\
J213345.79+423744.6	&	255304062	&	GSC 03191-01168	&	21 33 45.789	&	+42 37 44.569	&	12.490	&	 A1 II-III  shell  	&	171.4	&	699.8	&	1.0	&	724.7	&	0.0	&	962.0	&	0.3	&	1	\\
J213720.80+132827.0	&	259901129	&	BD+12 4653	&	21 37 20.905	&	+13 28 28.499	&	10.307	&	 A2 III-IV  shell  	&	265.0	&	881.8	&	0.3	&	897.3	&	0.2	&	1287.0	&	0.9	&	1	\\
J225304.50+544503.0	&	260909237	&	GSC 03988-01889	&	22 53 04.500	&	+54 45 03.047	&	12.922	&	 A0 III  shell  	&	124.0	&	645.8	&	1.0	&	651.7	&	0.6	&	822.1	&	0.2	&	2(-)	\\
J225542.92+555816.9	&	260912220	&	GSC 03989-00149	&	22 55 42.921	&	+55 58 16.904	&	12.080	&	 B9.5 II-III  shell	&	165.5	&	550.3	&	1.4	&	565.8	&	0.5	&	627.2	&	0.1	&	2(-)	\\
J225954.10+421753.8	&	180214011	&	GSC 03223-03570	&	22 59 54.112	&	+42 17 53.841	&	13.801	&	 A1 IV-V shell	&	59.8	&	288.7	&	0.6	&	425.3	&	0.4	&	787.4	&	1.6	&	0	\\
J231357.71+545812.0$^{3a}$	&	392502193	&	GSC 04002-01262	&	23 13 57.525	&	+54 58 12.204	&	10.628	&	 A0 II  shell  	&	229.9	&	636.6	&	0.7	&	641.2	&	0.0	&	723.7	&	0.1	&	2	\\
J232258.57+541829.2	&	392501164	&	GSC 03999-01276	&	23 22 58.579	&	+54 18 29.297	&	10.962	&	 A1 II-III  shell	&	204.1	&	626.2	&	1.3	&	672.5	&	0.6	&	825.5	&	0.1	&	1 asym	\\
\hline
\multicolumn{15}{l}{Remarks:} \\
\multicolumn{15}{l}{1 \citet{vioque20} 2 \citet{shridharan21} 3a Emission-line star according to \citet{zhang22}. 3b Herbig Ae/Be star according to \citet{zhang22}.} \\
\end{tabular}
\end{adjustbox}
\end{table*}

\begin{table*}
\caption{Essential data for the CP1 and CP2 stars and the stars with composite spectra, sorted by increasing right ascension. The columns denote: (1) LAMOST identifier. (2) LAMOST observation ID. (3) Alternativ identifier. (4) Right ascension (J2000; GAIA EDR3). (5) Declination (J2000; GAIA EDR3). (6) $G$\,mag (GAIA EDR3). (7) Sloan $g$ band S/N ratio of the analyzed spectrum. (8) Type. (9) References.}
\label{table_stars_other}
\begin{adjustbox}{max width=0.9\textwidth}
\begin{tabular}{lllllllll}
\hline
\hline
(1) & (2) & (3) & (4) & (5) & (6) & (7) & (8) & (9) \\
ID\_LAMOST	&	ObsID	&	ID\_alt	&	RA(J2000)	&	Dec(J2000)	&	$G$\,mag	&	S/N\,$g$	&	Type	&	Ref	\\
\hline
J010651.35+154426.9	&	409510074	&	HD 6590	&	01 06 51.168	&	+15 44 26.921	&	9.998	&	416.0	&	CP2	&	1,2,4	\\
J031043.71+480727.8	&	263604028	&	GSC 03314-01599	&	03 10 43.717	&	+48 07 27.866	&	12.714	&	160.4	&	CP2	&	2,4	\\
J045926.29+535030.4	&	377907100	&	GSC 03734-01043	&	04 59 26.172	&	+53 50 30.516	&	10.918	&	291.8	&	CP2	&	2,4	\\
J052739.41+533935.4	&	425715081	&	GSC 03748-01827	&	05 27 39.457	&	+53 39 35.696	&	16.891	&	385.1	&	CP2	&	2,4	\\
J054547.46+400703.6	&	286206027	&	GSC 02915-01913	&	05 45 47.464	&	+40 07 03.687	&	11.925	&	191.9	&	CP2	&	2,4	\\
J060259.74+290339.2	&	3109241	&	UCAC4 596-027865	&	06 02 59.747	&	+29 03 39.287	&	14.446	&	24.0	&	CP2	&		\\
J063114.53+184730.7	&	435511229	&	HD 258682	&	06 31 14.464	&	+18 47 31.872	&	10.752	&	187.4	&	CP2	&	2,4	\\
J063744.29+195655.1	&	431606168	&	HD 47103	&	06 37 44.072	&	+19 56 55.155	&	9.176	&	811.4	&	CP2	&	1,2,3,4,5	\\
J065216.71+205412.6	&	410003112	&	GSC 01343-00132	&	06 52 16.711	&	+20 54 12.632	&	13.225	&	125.3	&	CP2	&	2,4	\\
J072040.31-000827.5	&	88814064	&	GSC 04816-00856	&	07 20 40.308	&	-00 08 27.521	&	13.135	&	62.0	&	CP2	&	2,4	\\
J214528.67+423410.2	&	172212022	&	GSC 03192-01238	&	21 45 28.674	&	+42 34 10.256	&	11.888	&	75.4	&	CP2	&		\\
J231412.30+232336.9	&	57503106	&	GSC 02236-01035	&	23 14 12.305	&	+23 23 36.978	&	11.207	&	161.5	&	CP2	&	2,4	\\
\hline																	
J012246.86+474343.9	&	82713145	&	GSC 03269-00936	&	01 22 46.858	&	+47 43 43.925	&	11.719	&	91.6	&	CP1	&	3,4	\\
J022830.23+351753.0	&	294714195	&	HD 15282	&	02 28 30.440	&	+35 17 52.382	&	19.845	&	545.5	&	CP1	&	3,4	\\
J023538.77+504829.1	&	206310079	&	GSC 03307-00765	&	02 35 38.776	&	+50 48 29.158	&	11.857	&	167.0	&	CP1	&	3,4	\\
J024916.34+474840.8	&	172014239	&	GSC 03301-00058	&	02 49 16.344	&	+47 48 40.875	&	12.288	&	98.0	&	CP1	&	3,4	\\
J033121.96+453110.7	&	408704129	&	GSC 03312-00096	&	03 31 21.960	&	+45 31 10.795	&	11.504	&	136.4	&	CP1	&	3,4	\\
J035817.16+353234.0	&	181913164	&	GSC 02365-00334	&	03 58 17.157	&	+35 32 34.115	&	12.885	&	100.7	&	CP1	&	3,4	\\
J041423.04+510959.8	&	379014063	&	GSC 03340-01232	&	04 14 23.039	&	+51 09 59.815	&	12.021	&	146.4	&	CP1	&	3,4	\\
J042256.78+441458.8	&	190312238	&	GSC 02891-02087	&	04 22 56.780	&	+44 14 58.944	&	11.982	&	157.8	&	CP1	&	3,4	\\
J051609.42+503544.8	&	168508158	&	HD 233093	&	05 16 09.425	&	+50 35 44.749	&	9.570	&	325.3	&	CP1	&	3,4	\\
J051716.27+320140.4	&	89405073	&	GSC 02394-01286	&	05 17 16.275	&	+32 01 40.532	&	12.720	&	108.7	&	CP1	&	3,4	\\
J052654.71-051454.6	&	181204075	&	GSC 04761-00694	&	05 26 54.707	&	-05 14 54.688	&	11.790	&	334.8	&	CP1	&	4	\\
J053442.48+050744.1	&	307004089	&	GSC 00122-00503	&	05 34 42.490	&	+05 07 44.114	&	11.704	&	125.6	&	CP1	&	3,4	\\
J053852.99+500732.1	&	191011033	&	GSC 03368-00681	&	05 38 53.025	&	+50 07 32.150	&	11.265	&	82.0	&	CP1	&	3,4	\\
J055621.82+215854.1	&	116301217	&	GSC 01324-02312	&	05 56 21.833	&	+21 58 54.152	&	11.796	&	142.0	&	CP1	&	3,4	\\
J055658.80+305431.7	&	127711190	&	GSC 02406-01503	&	05 56 58.805	&	+30 54 31.692	&	11.420	&	116.4	&	CP1	&	1,3,4	\\
J061118.40+513036.5	&	380005061	&	GSC 03387-00370	&	06 11 18.406	&	+51 30 36.523	&	11.905	&	289.7	&	CP1	&	3,4	\\
J061743.05+595315.3	&	169614033	&	GSC 03776-00815	&	06 17 43.066	&	+59 53 15.312	&	11.316	&	148.5	&	CP1	&	3	\\
J063358.86+162851.1	&	43802152	&	GSC 01329-00314	&	06 33 58.866	&	+16 28 51.167	&	11.133	&	106.7	&	CP1	&	3,4	\\
J063844.04+273220.6	&	189513009	&	UCAC4 588-033330	&	06 38 44.056	&	+27 32 20.719	&	14.855	&	60.8	&	CP1	&	3,4	\\
J064705.49+263321.7	&	100205069	&	GSC 01901-00721	&	06 47 05.498	&	+26 33 21.676	&	11.683	&	77.0	&	CP1	&	3,4	\\
J065620.88+234100.2	&	108810223	&	GSC 01894-00419	&	06 56 20.880	&	+23 41 00.159	&	12.613	&	122.6	&	CP1	&	3,4	\\
J065626.12+053853.9	&	368911033	&	GSC 00161-00337	&	06 56 26.129	&	+05 38 53.960	&	12.573	&	104.5	&	CP1	&	3,4	\\
J070252.43+211100.1	&	174915183	&	UCAC4 556-037142	&	07 02 52.425	&	+21 11 00.135	&	15.520	&	56.4	&	CP1	&	3	\\
J072913.48+242111.9	&	140416192	&	GSC 01910-00845	&	07 29 13.479	&	+24 21 11.988	&	12.082	&	112.7	&	CP1	&	3,4	\\
J082647.44+161418.4	&	432714119	&	GSC 01379-00095	&	08 26 47.367	&	+16 14 18.445	&	10.884	&	254.7	&	CP1	&	3,4	\\
J083432.30+013758.7	&	137215069	&	GSC 00210-00002	&	08 34 32.308	&	+01 37 58.699	&	9.404	&	386.8	&	CP1	&	3,4	\\
J083947.05-072231.4	&	111104095	&	GSC 04875-00091	&	08 39 47.053	&	-07 22 31.590	&	10.513	&	108.9	&	CP1	&	3,4	\\
J084251.20-015717.7	&	125411209	&	GSC 04867-01502	&	08 42 51.205	&	-01 57 17.861	&	10.952	&	68.8	&	CP1	&	3,4	\\
J084436.87+320618.2	&	130208121	&	GSC 02484-00400	&	08 44 36.869	&	+32 06 18.265	&	14.909	&	37.5	&	CP1	&	3,4	\\
J093730.37+321651.6	&	22615081	&	GSC 02501-00330	&	09 37 30.287	&	+32 16 51.008	&	11.196	&	432.2	&	CP1	&	3,4	\\
J093920.09-055123.1	& 83503117	&	GSC 04900-00542	&	09 39 20.089	&	-05 51 23.115	&	13.117	&	51.5	&	CP1?	&		\\
J115927.56+022614.0	&	237705132	&	BD+03 2582	&	11 59 27.564	&	+02 26 14.098	&	11.045	&	124.9	&	CP1	&	1,3,4	\\
J131759.16+170150.3	&	425015052	&	GSC 01451-00443	&	13 17 59.162	&	+17 01 50.379	&	12.247	&	175.4	&	CP1	&	3,4	\\
J191403.99+511612.5	&	250014100	&	GSC 03554-00217	&	19 14 04.024	&	+51 16 12.338	&	11.664	&	183.3	&	CP1	&	3,4	\\
J191519.54+434913.1	&	362008084	&	GSC 03133-02227	&	19 15 19.543	&	+43 49 13.107	&	11.478	&	447.6	&	CP1	&	3	\\
J192615.25+422836.6	&	154008048	&	GSC 03142-00964	&	19 26 15.257	&	+42 28 36.717	&	12.267	&	140.5	&	CP1	&	3,4	\\
J193201.84+392304.9	&	373105196	&	GSC 03139-00006	&	19 32 01.739	&	+39 23 05.786	&	9.904	&	292.6	&	CP1	&	3,4	\\
J193724.78+454300.2	&	247715139	&	GSC 03556-01543	&	19 37 24.820	&	+45 43 00.865	&	12.754	&	147.3	&	CP1	&	3,4	\\
J195842.74+403448.3	&	367806224	&	UCAC4 653-084816	&	19 58 42.740	&	+40 34 48.231	&	12.735	&	135.5	&	CP1	&	3,4	\\
J211930.96+244342.2	&	258415189	&	GSC 02190-02112	&	21 19 30.969	&	+24 43 42.321	&	11.864	&	150.0	&	CP1	&	3,4	\\
J213435.04+410543.0	&	172204169	&	GSC 03187-00774	&	21 34 35.005	&	+41 05 43.022	&	10.933	&	330.8	&	CP1	&	3,4	\\
J224509.70+081116.1	&	78213142	&	GSC 01152-01029	&	22 45 09.701	&	+08 11 16.000	&	14.255	&	92.2	&	CP1	&	3,4	\\
J225451.47+514110.6	&	361915072	&	GSC 03634-01072	&	22 54 51.473	&	+51 41 10.686	&	11.895	&	294.0	&	CP1	&	3,4	\\
\hline																	
J004534.75+452026.3	&	186715166	&	GSC 03262-00282	&	00 45 34.766	&	+45 20 25.352	&	12.471	&	172.6	&	comp?	&	4	\\
J050321.41+512856.5	&	168514062	&	GSC 03356-00169	&	05 03 21.421	&	+51 28 56.567	&	10.754	&	229.6	&	comp?	&		\\
J125004.42+550602.1	&	152807229	&	GSC 03845-00640	&	12 50 04.514	&	+55 06 01.635	&	12.665	&	116.3	&	comp?	&		\\

\hline
\multicolumn{9}{l}{References:} \\
\multicolumn{9}{l}{1 \citet{RM09} 2 \citet{huemmerich20} 3 \citet{qin19} 4 \citet{shang22} 5 \citet{scholz2019}} \\
\end{tabular}
\end{adjustbox}
\end{table*}

\begin{table*}
\caption{Variability analysis results. Stars are sorted by increasing right ascension. The columns denote: (1) LAMOST identifier. (2) Alternativ identifier. (3) Right ascension (J2000; GAIA EDR3). (4) Declination (J2000; GAIA EDR3). (5) VSX identifier. (6) VSX variability type. (7) VSX period (d). (8) TESS identifier. (9) Variability type deduced from analysis of TESS data. Types are explained in the remarks below the table. (10) Period (d) derived from TESS data.}
\label{table_VAR}
\begin{adjustbox}{max width=\textwidth}
\begin{tabular}{llllllllll}
\hline
\hline
(1) & (2) & (3) & (4) & (5) & (6) & (7) & (8) & (9) & (10) \\
ID\_LAMOST	&	ID\_alt	&	RA(J2000)	&	Dec(J2000)	&	ID\_VSX	&	Type\_VSX	&	$P$\_VSX	&	ID\_TIC	&	Type\_TESS	&	$P$\_TESS	\\
\hline
J002559.10+555631.3	&	GSC 03657-00187	&	00 25 59.097	&	+55 56 31.114	&		&		&		&	449983408	&	const?	&		\\
J003933.02+273029.5	&	GSC 01744-02329	&	00 39 33.025	&	+27 30 29.417	&	V0496 And	&	EA	&	4.40262	&	25756249	&	EA	&	4.40869	\\
J004127.44+220716.9	&	GSC 01193-00234	&	00 41 27.442	&	+22 07 16.944	&		&		&		&	434216252	&	eclipser	&	5.7012	\\
J020138.33+561624.4	&	UCAC4 732-019419	&	02 01 38.341	&	+56 16 24.493	&		&		&		&	445557996	&		&		\\
J020207.64+543927.9	&	GSC 03689-00919	&	02 02 07.642	&	+54 39 28.010	&		&		&		&	374568473	&	const?	&		\\
J020820.76+562433.0	&	HD 12921	&	02 08 20.633	&	+56 24 35.290	&		&		&		&		&		&		\\
J030240.87+522334.6	&	GSC 03322-00936	&	03 02 40.875	&	+52 23 34.692	&		&		&		&	116411454	&	ELL	&	1.04484	\\
J031003.35+551804.4	&	HD 232754	&	03 10 03.359	&	+55 18 01.367	&		&		&		&		&		&		\\
J031638.78+494256.2	&	GSC 03319-00602	&	03 16 38.777	&	+49 42 55.989	&		&		&		&	117621124	&	const	&		\\
J031841.75+462842.1	&	GSC 03311-01885	&	03 18 41.776	&	+46 28 40.974	&		&		&		&	192843390	&	irr	&		\\
J031909.14+192811.7	&	BD+18 457	&	03 19 08.931	&	+19 28 11.753	&		&		&		&		&		&		\\
J032704.99+434028.5	&	GSC 02873-01933	&	03 27 04.997	&	+43 40 28.556	&		&		&		&	456144376	&	eclipser	&	6.2902	\\
J033909.14+523714.5	&	GSC 03716-00037	&	03 39 09.142	&	+52 37 14.550	&		&		&		&	428250707	&	const	&		\\
J034636.05+441159.3	&	GSC 02875-01237	&	03 46 36.054	&	+44 11 59.403	&		&		&		&	431964351	&	ELL	&	0.43026	\\
J035000.80+580406.0	&	GSC 03725-01226	&	03 50 00.807	&	+58 04 06.118	&		&		&		&	86661062	&	const?	&		\\
J035033.65+525943.2	&	GSC 03717-00421	&	03 50 33.658	&	+52 59 43.279	&		&		&		&	449966196	&	const?	&		\\
J035205.73+351059.1	&	GSC 02364-02043	&	03 52 05.733	&	+35 10 59.126	&		&		&		&	94278532	&		&		\\
J040713.90+291832.1	&	GSC 01826-00710	&	04 07 13.878	&	+29 18 32.426	&	IL Tau	&	EA/SD	&	5.36062	&	348952490	&		&		\\
J041308.41+534045.9	&	GSC 03718-00545	&	04 13 08.415	&	+53 40 45.956	&		&		&		&	267695404	&	const?	&		\\
J041638.47+530600.9	&	GSC 03719-00690	&	04 16 38.473	&	+53 06 00.846	&		&		&		&	104848807	&	EW?	&	1.02151	\\
J043336.04+451232.6	&	GSC 03342-00138	&	04 33 36.042	&	+45 12 32.709	&		&		&		&	28752151	&	ELL	&	1.08241	\\
J043415.66+405139.5	&	UCAC4 655-022462	&	04 34 15.666	&	+40 51 39.552	&		&		&		&	155600657	&		&		\\
J043948.40+474848.5	&	GSC 03346-00854	&	04 39 48.401	&	+47 48 48.587	&		&		&		&	348102309	&	const?	&		\\
J044307.81+571907.8	&	GSC 03741-00200	&	04 43 07.813	&	+57 19 07.806	&	Mis V1381	&	EA	&	4.65	&	9953640	&	EA	&	4.64024	\\
J050435.48+402559.7	&	UCAC4 653-028138	&	05 04 35.489	&	+40 25 59.682	&		&		&		&	122098910	&		&		\\
J050556.74+430528.8	&	GSC 02903-00983	&	05 05 56.760	&	+43 05 28.871	&		&		&		&	261149041	&	const?	&		\\
J051428.45+274324.3	&	GSC 01854-00033	&	05 14 28.465	&	+27 43 24.555	&	AS Tau	&	EA/SD	&	3.483328	&	61770518	&		&		\\
J053235.07+495218.6	&	GSC 03367-01174	&	05 32 35.073	&	+49 52 18.684	&		&		&		&	310350759	&	const	&		\\
J053414.62+215213.6	&	GSC 01309-01353	&	05 34 14.626	&	+21 52 13.641	&		&		&		&	437804190	&		&		\\
J053918.09+361716.2	&	GSC 02416-00657	&	05 39 18.088	&	+36 17 16.214	&	ASASSN-V J053918.09+361716.3	&	VAR	&	0	&	239106869	&		&		\\
J054113.82+351910.7	&	GSC 02412-01090	&	05 41 13.827	&	+35 19 10.725	&		&		&		&	116394274	&	puls?	&	0.36919	\\
J054239.89+174652.7	&	HD 246608	&	05 42 39.841	&	+17 46 53.528	&		&		&		&	247473690	&		&		\\
J054329.26+000458.8	&	HD 290828	&	05 43 29.252	&	+00 04 58.970	&	GT Ori	&	UXOR	&	0	&	199955590	&	irr	&		\\
J054358.96+491845.6	&	GSC 03368-01318	&	05 43 58.958	&	+49 18 45.706	&		&		&		&	311231207	&	puls?	&	1.15559	\\
J054834.63+234141.4	&	BD+23 1058	&	05 48 34.637	&	+23 41 41.489	&		&		&		&	358224196	&		&		\\
J055140.83+414108.5	&	GSC 02920-02782	&	05 51 40.835	&	+41 41 08.500	&		&		&		&	266348210	&	DSCT	&	0.15266	\\
J055340.45+332003.9	&	UCAC4 617-029482	&	05 53 40.456	&	+33 20 03.995	&		&		&		&	312221983	&		&		\\
J055441.84+223317.7	&	GSC 01863-02412	&	05 54 41.850	&	+22 33 17.763	&	NSV 2719	&	0	&	0	&	118897886	&		&		\\
J055907.94-060737.6	&	GSC 04781-00953	&	05 59 07.949	&	-06 07 37.675	&		&		&		&	67279509	&	ELL	&	6.9642	\\
J060706.46+245547.1	&	UCAC4 575-024208	&	06 07 06.470	&	+24 55 47.138	&		&		&		&	81186298	&		&		\\
J061810.01+411332.0	&	GSC 02930-00867	&	06 18 10.012	&	+41 13 32.113	&		&		&		&	258356786	&		&		\\
J062416.04+422125.4	&	GSC 02935-01774	&	06 24 16.047	&	+42 21 25.477	&		&		&		&	189945062	&	EA	&	10.687	\\
J062618.73+174526.9	&	HD 257117	&	06 26 18.745	&	+17 45 27.254	&		&		&		&	438232591	&		&		\\
J062724.55+234931.0	&	UCAC4 570-028755	&	06 27 24.551	&	+23 49 31.048	&	HU Gem	&	EA	&	4.206669	&	426585626	&		&		\\
J064142.55+031421.9	&	GSC 00151-01200	&	06 41 42.552	&	+03 14 21.954	&		&		&		&	301494903	&	const?	&		\\
J064230.24+092626.7	&	HD 262371	&	06 42 30.247	&	+09 26 26.726	&		&		&		&	231162744	&	const	&		\\
J064751.97+522321.9	&	GSC 03402-01356	&	06 47 51.974	&	+52 23 21.876	&		&		&		&	453343612	&	eclipser	&	1.42251	\\
J065136.11+125351.1	&	HD 265136	&	06 51 36.153	&	+12 53 51.562	&		&		&		&	155306152	&	const	&		\\
J065141.20+070357.7	&	HD 265284	&	06 51 41.202	&	+07 03 57.750	&		&		&		&	235258580	&	const?	&		\\
J065300.06+010730.6	&	GSC 00149-01072	&	06 53 00.062	&	+01 07 30.643	&		&		&		&	237564914	&	const?	&		\\
J065606.26+033757.5	&	GSC 00153-01723	&	06 56 06.264	&	+03 37 57.524	&		&		&		&	237736667	&	irr	&		\\
J070219.92+162845.4	&	GSC 01344-00359	&	07 02 19.921	&	+16 28 45.453	&		&		&		&	457088125	&		&		\\
J071200.59+241137.2	&	GSC 01896-00475	&	07 12 00.593	&	+24 11 37.376	&		&		&		&	101573787	&		&		\\
J071340.29+380623.9	&	HD 55200	&	07 13 40.582	&	+38 06 30.260	&		&		&		&		&		&		\\
J071630.57+053405.5	&	GSC 00172-00143	&	07 16 30.576	&	+05 34 05.521	&		&		&		&	284555430	&	const	&		\\
J072719.01+015251.6	&	GSC 00169-00831	&	07 27 19.013	&	+01 52 51.702	&		&		&		&	318579975	&	const	&		\\
J074455.91+042805.1	&	GSC 00187-01505	&	07 44 55.913	&	+04 28 05.207	&		&		&		&	266811412	&	EA	&	5.9205	\\
J075535.94-033455.4	&	GSC 04837-02525	&	07 55 35.947	&	-03 34 55.562	&		&		&		&	123194618	&	EA	&	4.02277	\\
J080809.68-052950.7	&	GSC 04855-02705	&	08 08 09.691	&	-05 29 50.777	&		&		&		&	88684812	&	ELL	&	1.48389	\\
J081646.65-040220.3	&	GSC 04856-00003	&	08 16 46.658	&	-04 02 20.468	&		&		&		&	178544181	&		&		\\
J085216.65+090518.7	&	BD+9 2072 	&	08 52 16.652	&	+09 05 18.779	&		&		&		&	459056359	&		&		\\
J090754.79+212319.3	&	GSC 01407-00362	&	09 07 54.796	&	+21 23 19.322	&		&		&		&	386890261	&		&		\\
J092110.64+283147.9	&	HD 80535	&	09 21 10.656	&	+28 31 48.053	&		&		&		&	149418554	&	EA	&	3.18875	\\
J093958.52-005332.0	&	GSC 04894-02078	&	09 39 58.526	&	-00 53 31.465	&		&		&		&	62730967	&	ELL	&	1.35403	\\
J180605.37+020543.8	&	GSC 00434-03484	&	18 06 05.379	&	+02 05 43.834	&		&		&		&	390639438	&		&		\\
J190401.96+415350.8	&	GSC 03128-02038	&	19 04 01.952	&	+41 53 50.944	&		&		&		&	120687891	&		&		\\
J195131.48+484558.9	&	GSC 03565-01258	&	19 51 31.489	&	+48 45 58.977	&		&		&		&	28308522	&	const?	&		\\
J200641.70+274307.0	&	GSC 02162-01397	&	20 06 41.705	&	+27 43 07.005	&		&		&		&	244659647	&	const?	&		\\
J213345.79+423744.6	&	GSC 03191-01168	&	21 33 45.789	&	+42 37 44.569	&		&		&		&	240465526	&	const?	&		\\
J213720.80+132827.0	&	BD+12 4653	&	21 37 20.905	&	+13 28 28.499	&	AQ Peg	&	EA/SD	&	5.5485028	&		&		&		\\
J225304.50+544503.0	&	GSC 03988-01889	&	22 53 04.500	&	+54 45 03.047	&		&		&		&	388968595	&	EA	&	1.02991	\\
J225542.92+555816.9	&	GSC 03989-00149	&	22 55 42.921	&	+55 58 16.904	&		&		&		&	367689958	&	const	&		\\
J225954.10+421753.8	&	GSC 03223-03570	&	22 59 54.112	&	+42 17 53.841	&		&		&		&	154930789	&		&		\\
J231357.71+545812.0	&	GSC 04002-01262	&	23 13 57.525	&	+54 58 12.204	&		&		&		&	371823816	&	artefact?	&		\\
J232258.57+541829.2	&	GSC 03999-01276	&	23 22 58.579	&	+54 18 29.297	&		&		&		&	319506298	&	ELL	&	8.8806	\\
\hline
\multicolumn{10}{l}{Variability types in Column 9 are as follows:} \\
\multicolumn{10}{l}{const = constant; irr = irregular variability; puls = pulsational variability; eclipser = eclipsing binary; EA = eclipsing binary of Algol-type; EW = eclipsing binary of W UMa-type; ELL = ellipsoidal variable;} \\
\multicolumn{10}{l}{DSCT = delta Scuti star}\\
\end{tabular}
\end{adjustbox}
\end{table*}

\section*{Acknowledgements}

Guoshoujing Telescope (the Large Sky Area Multi-Object Fiber Spectroscopic Telescope LAMOST) is a National Major Scientific Project built by the Chinese Academy of Sciences. Funding for the project has been provided by the National Development and Reform Commission. LAMOST is operated and managed by the National Astronomical Observatories, Chinese Academy of Sciences. This work has made use of data from the European Space Agency (ESA) mission {\it Gaia} (\url{https://www.cosmos.esa.int/gaia}), processed by the {\it Gaia} Data Processing and Analysis Consortium (DPAC, \url{https://www.cosmos.esa.int/web/gaia/dpac/consortium}). Funding for the DPAC has been provided by national institutions, in particular, the institutions participating in the {\it Gaia} Multilateral Agreement. This research has made use of the SIMBAD database, operated at CDS, Strasbourg, France.



\bibliographystyle{mnras}
\bibliography{main} 




\appendix

\clearpage

\section{TESS light curves}
\label{appendix}

This section contains the TESS JD light curves of the final sample of shell stars (cf. Section \ref{variability_analysis}).

\begin{figure*}
\includegraphics[width=0.68\columnwidth]{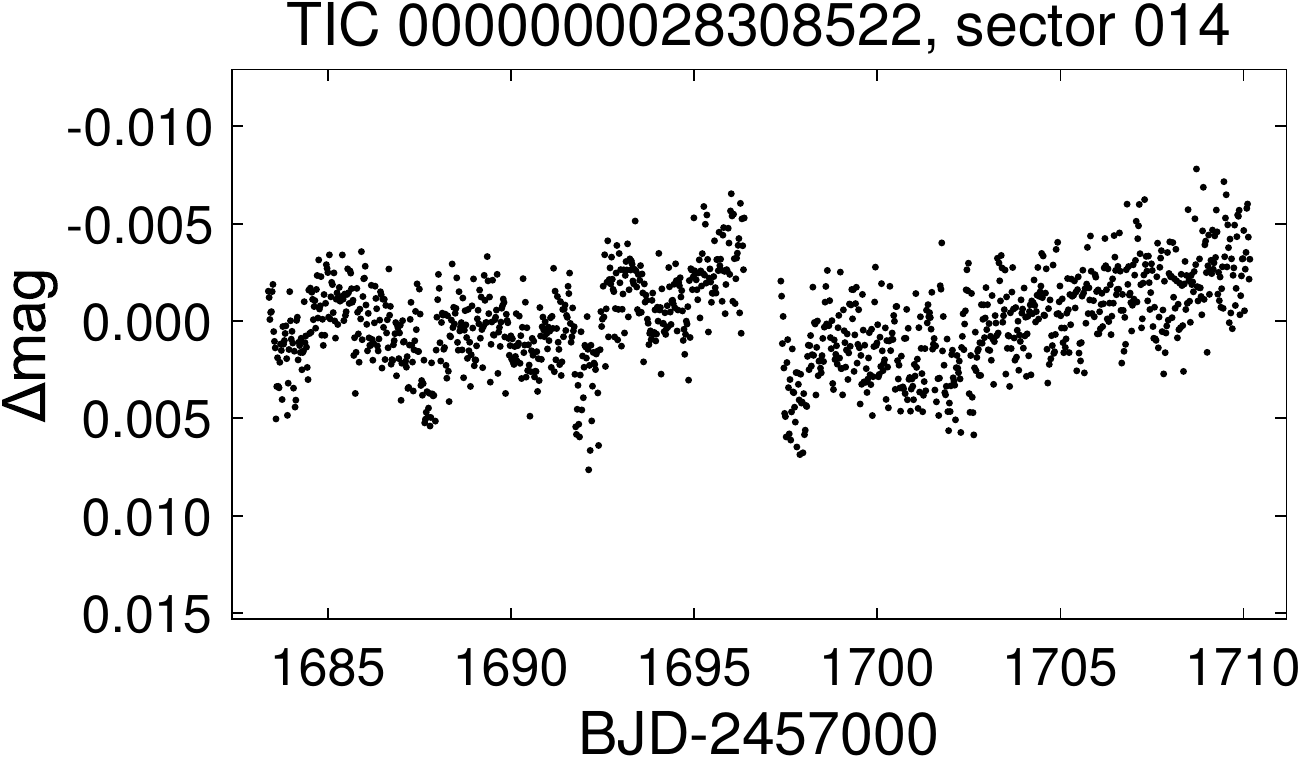}
\includegraphics[width=0.68\columnwidth]{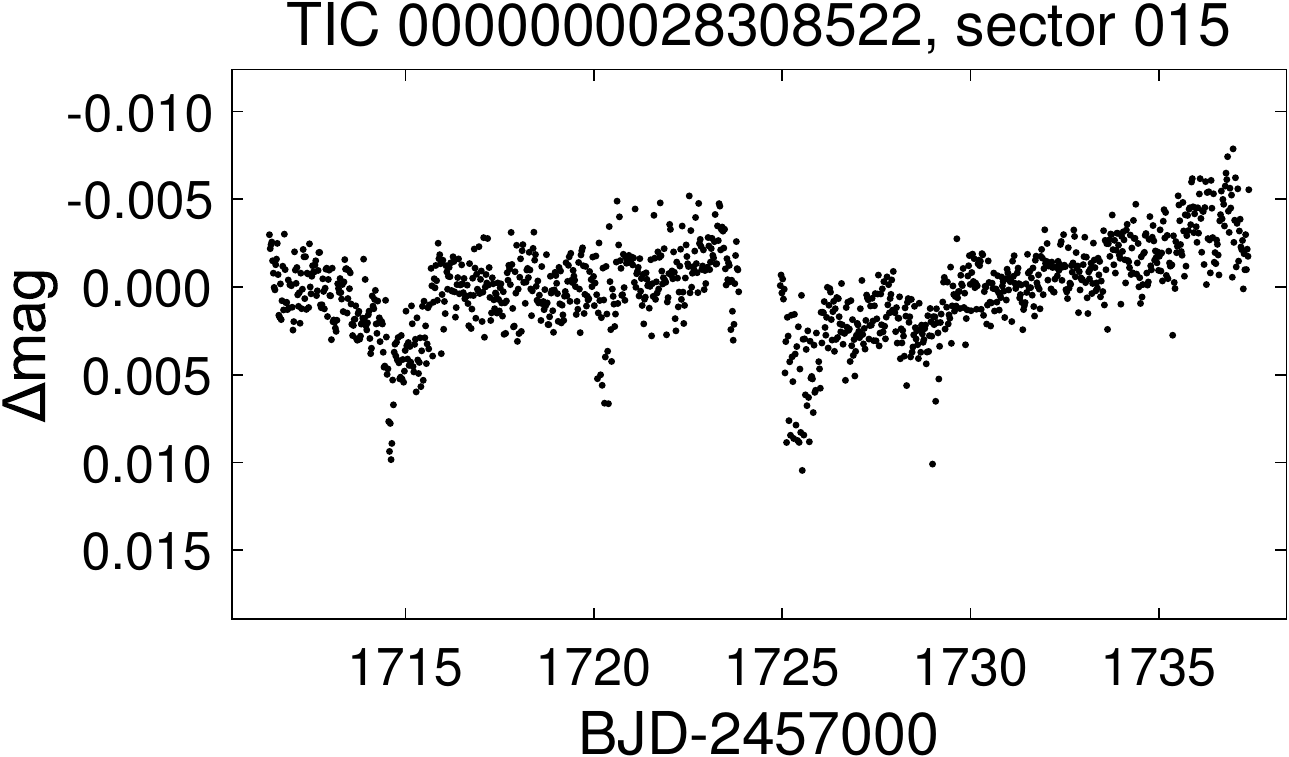}
\includegraphics[width=0.68\columnwidth]{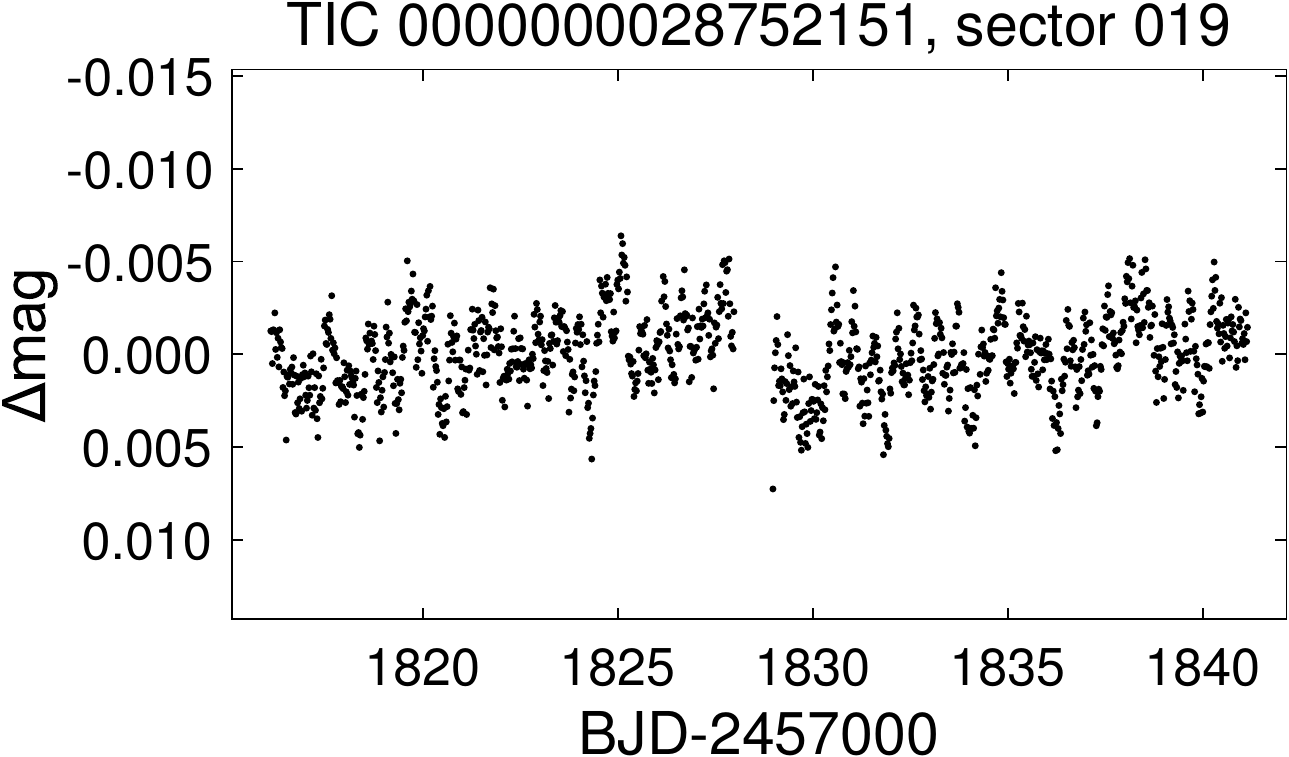}
\includegraphics[width=0.68\columnwidth]{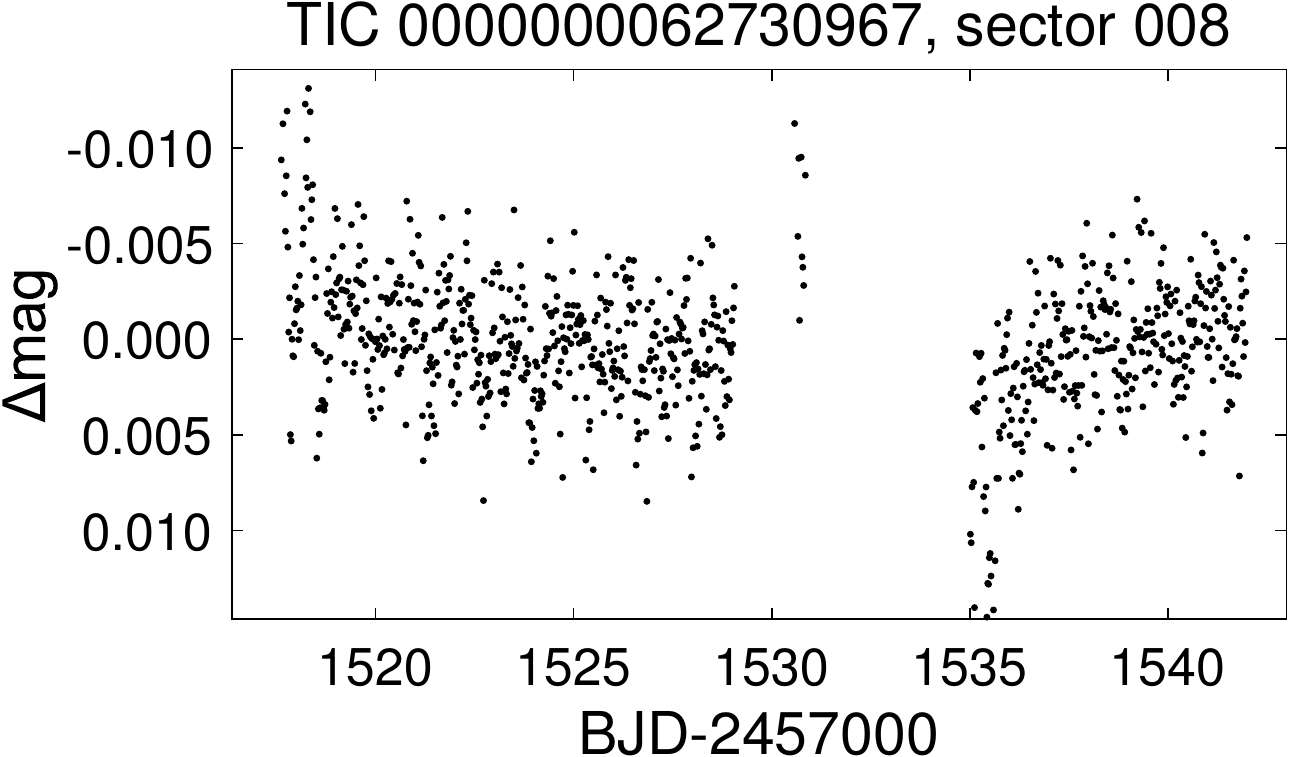}
\includegraphics[width=0.68\columnwidth]{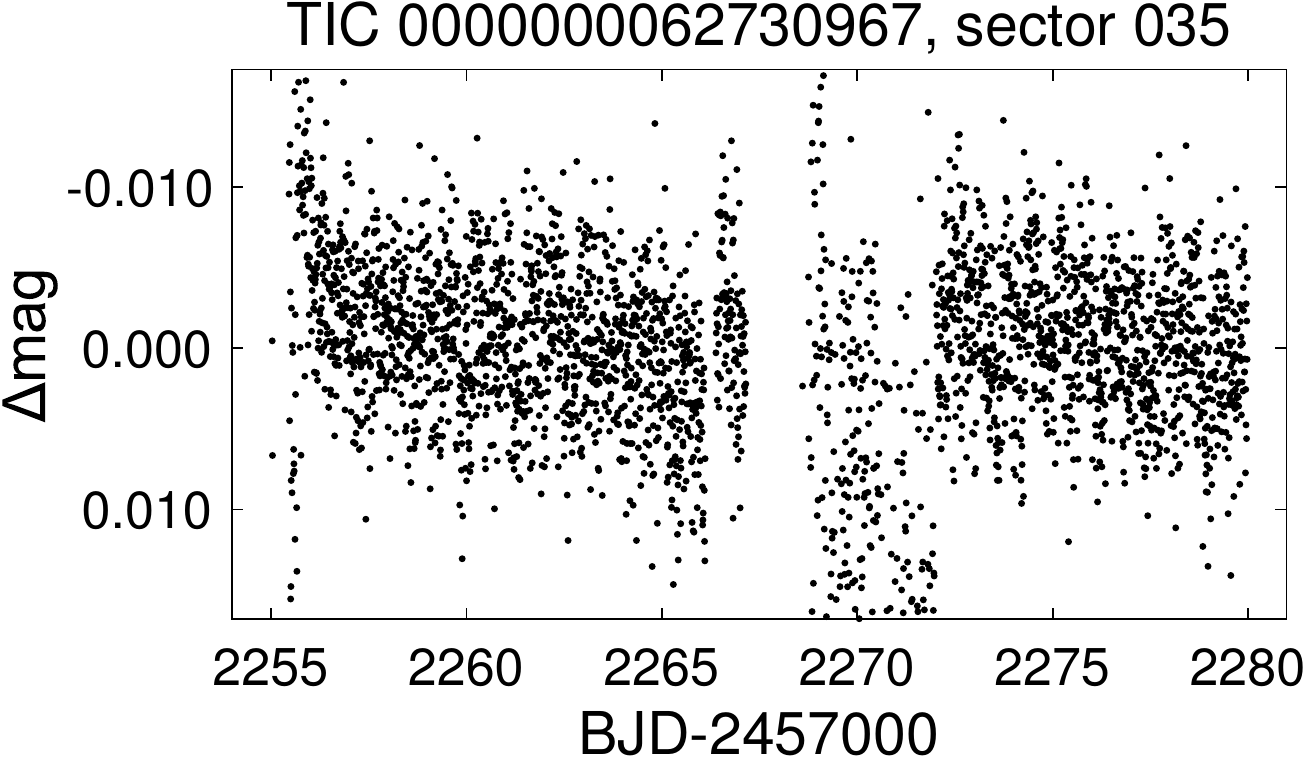}
\includegraphics[width=0.68\columnwidth]{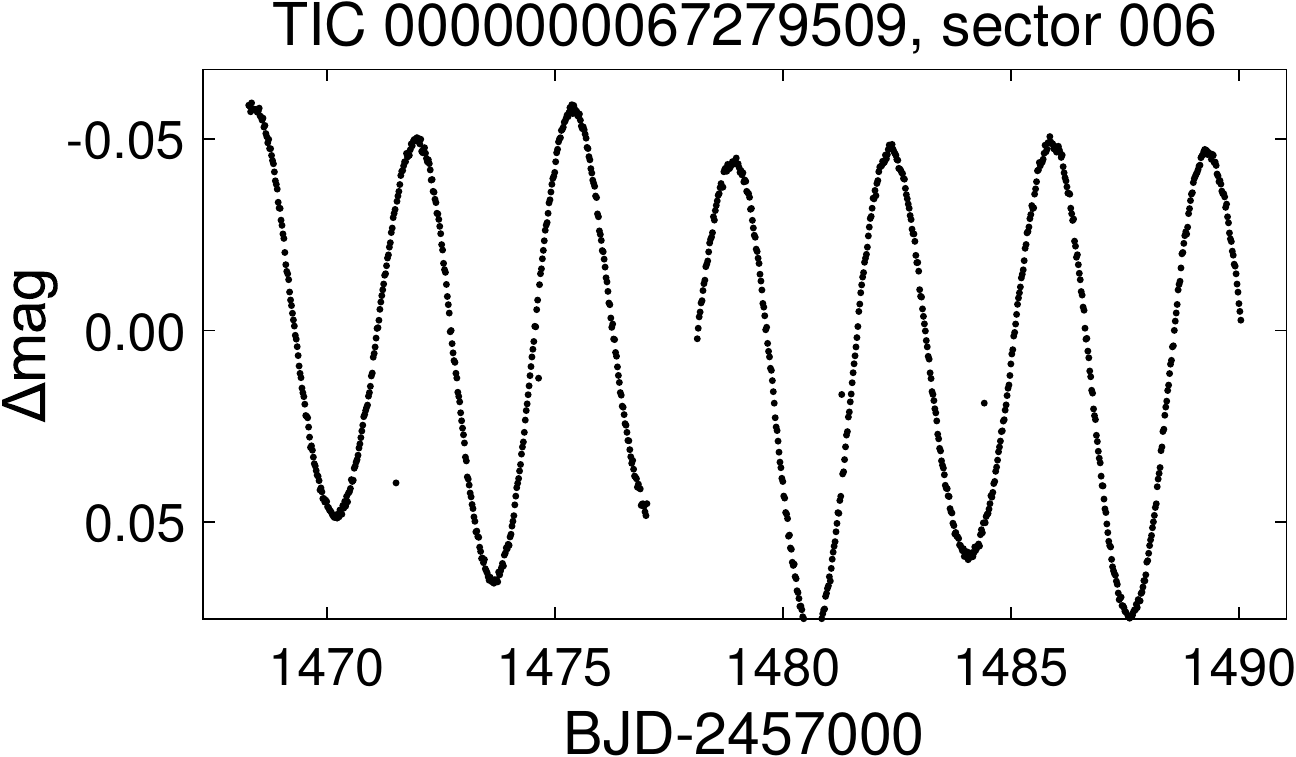}
\includegraphics[width=0.68\columnwidth]{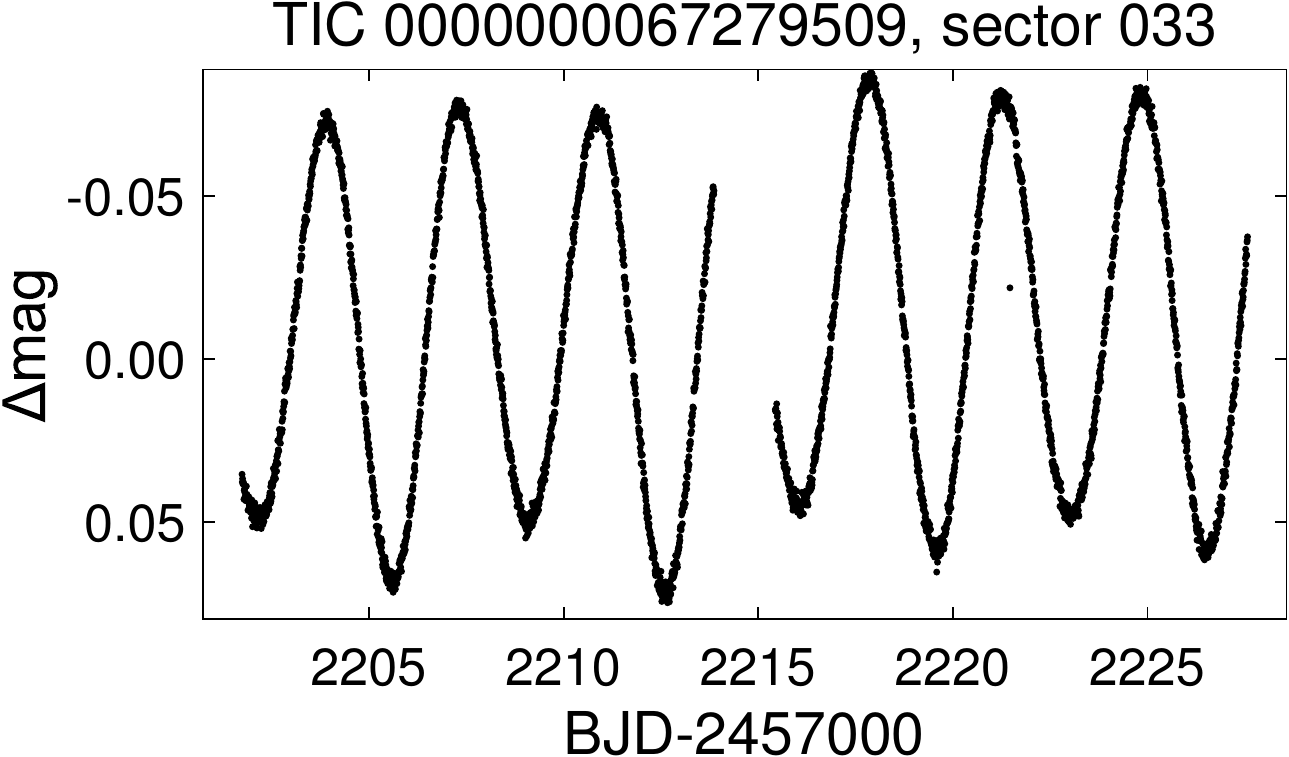}
\includegraphics[width=0.68\columnwidth]{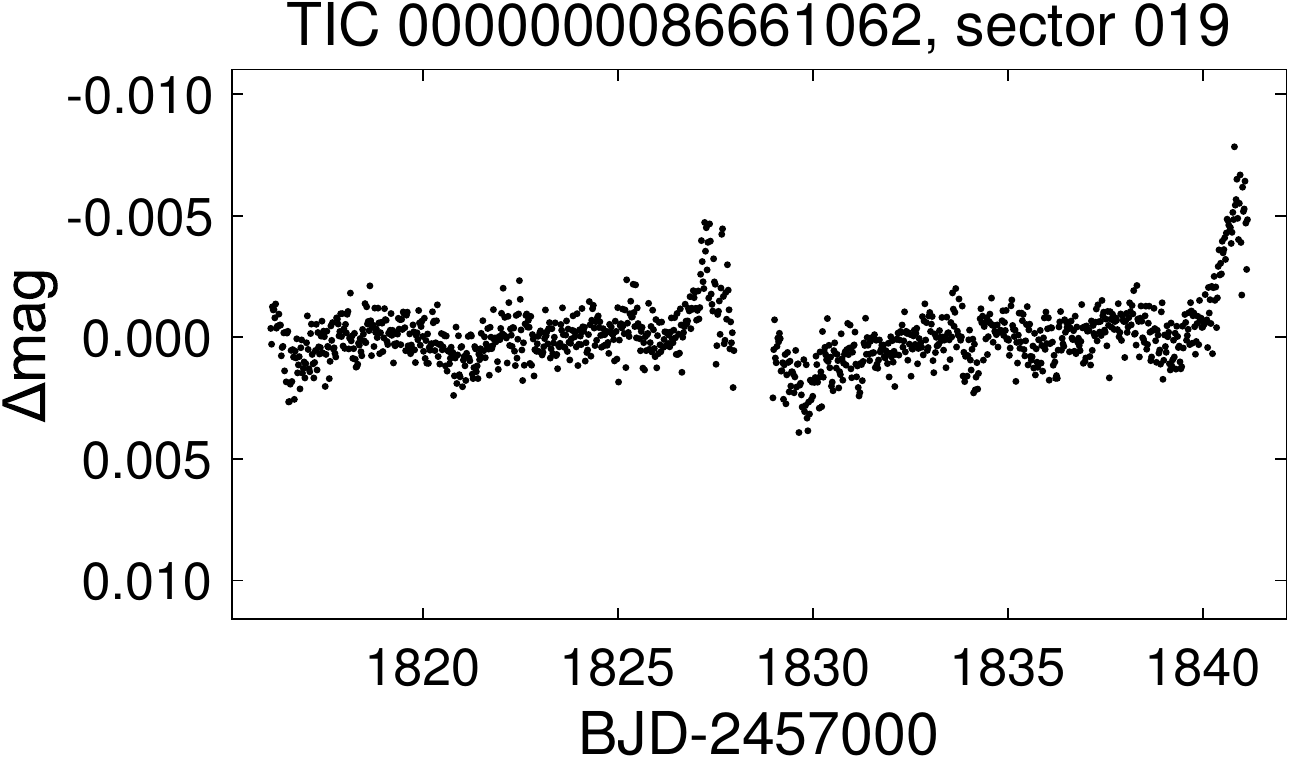}
\includegraphics[width=0.68\columnwidth]{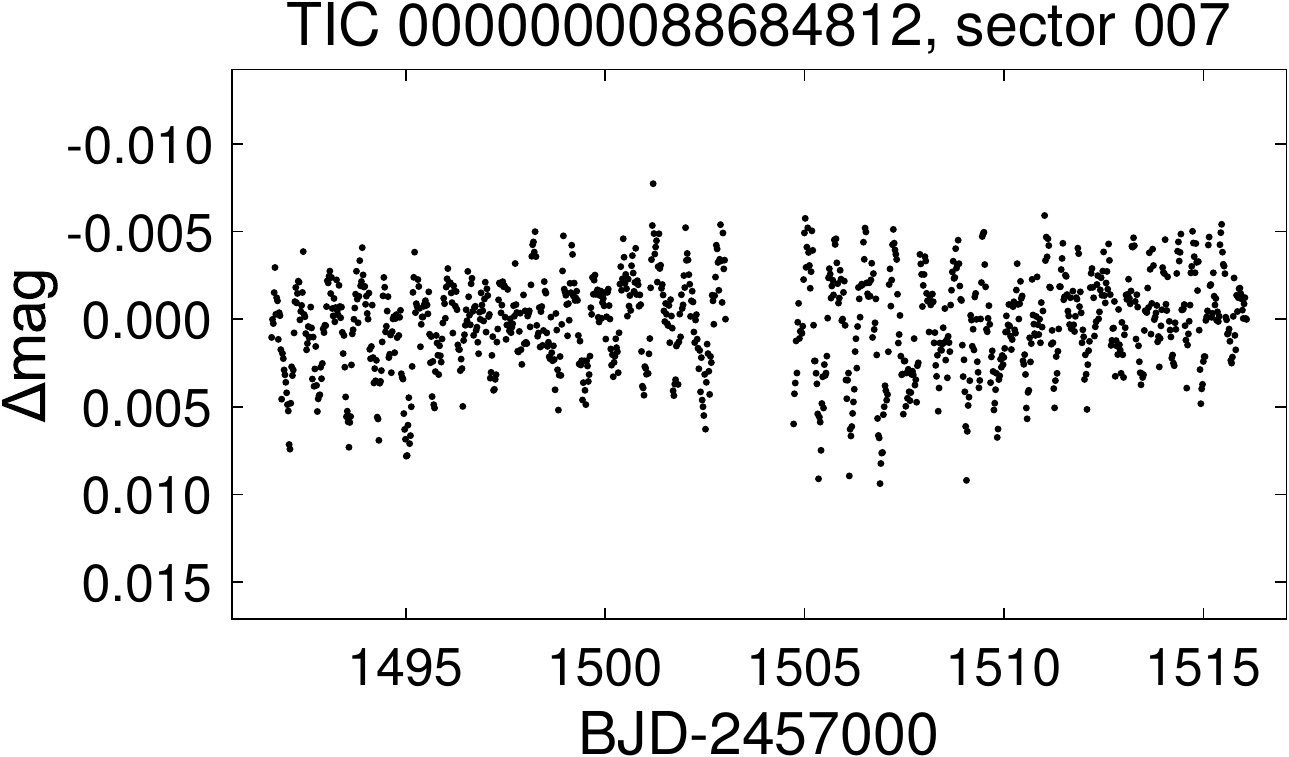}
\includegraphics[width=0.68\columnwidth]{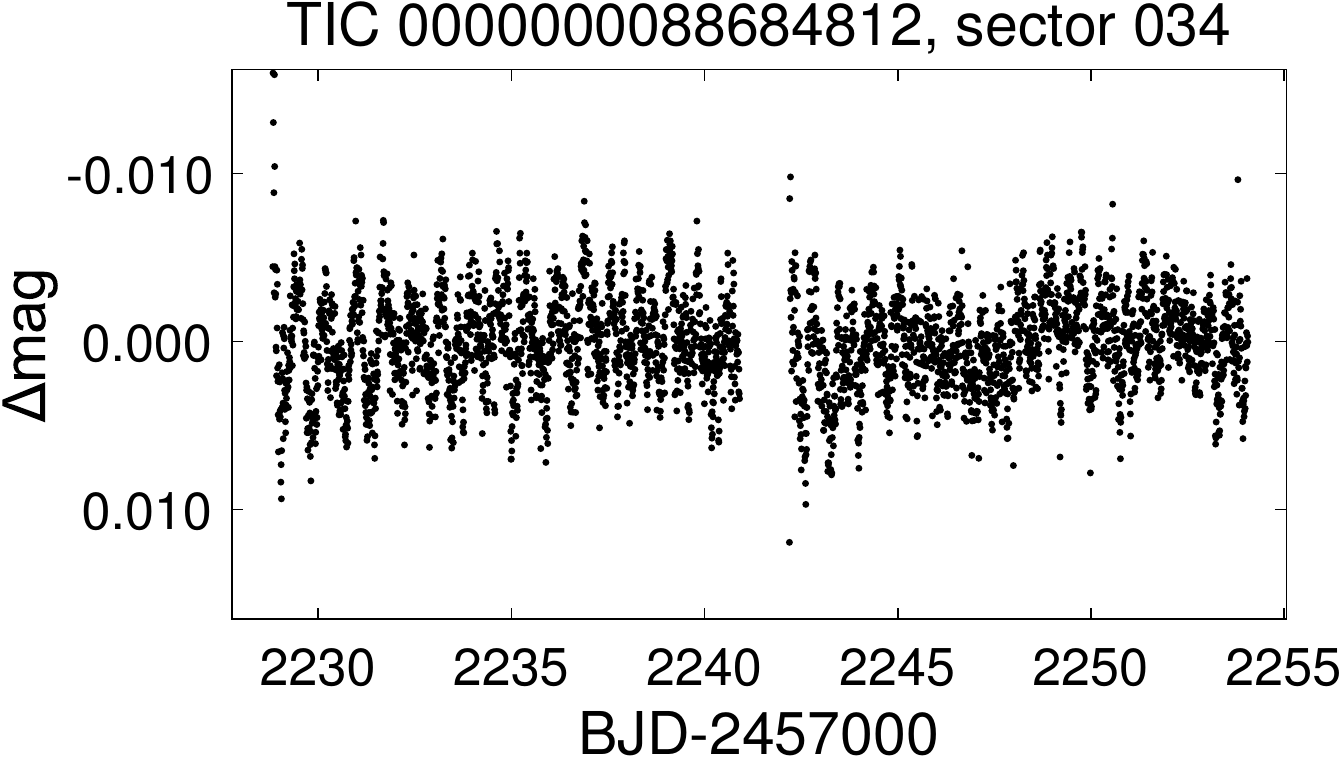}
\includegraphics[width=0.68\columnwidth]{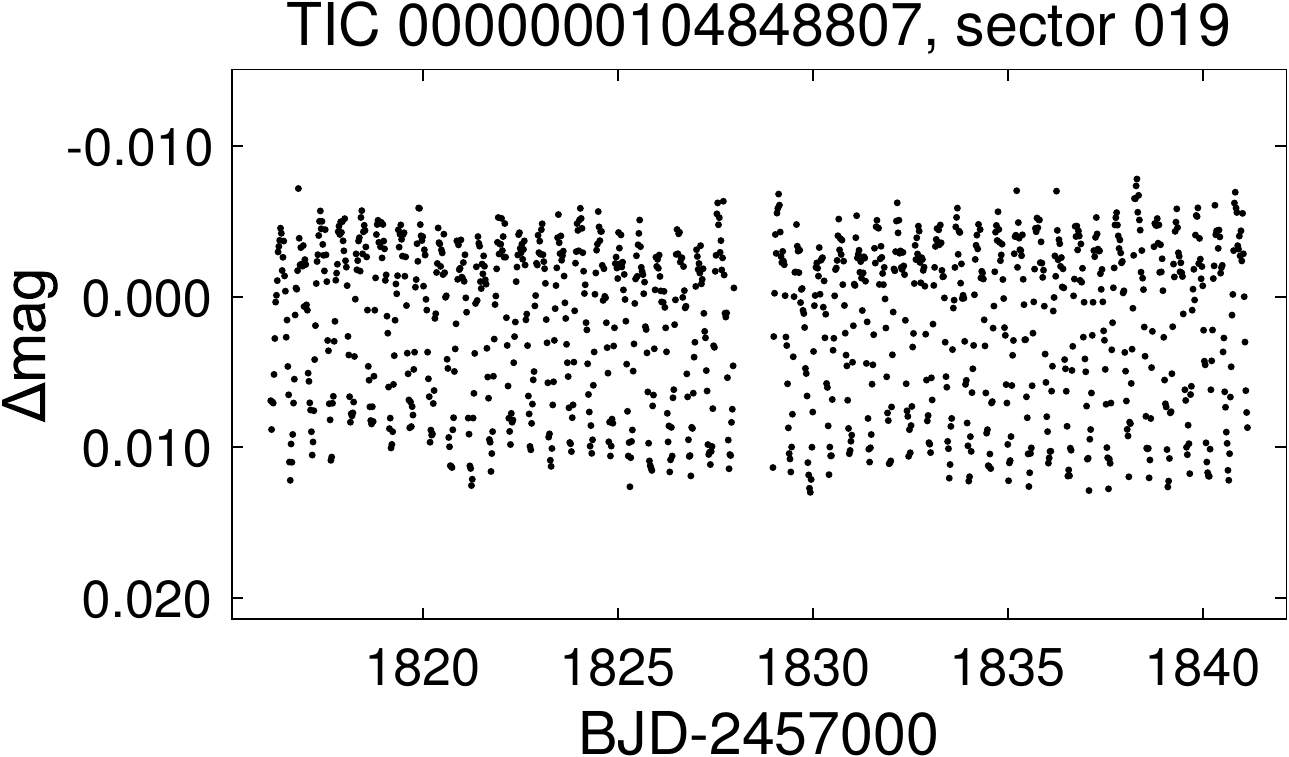}
\includegraphics[width=0.68\columnwidth]{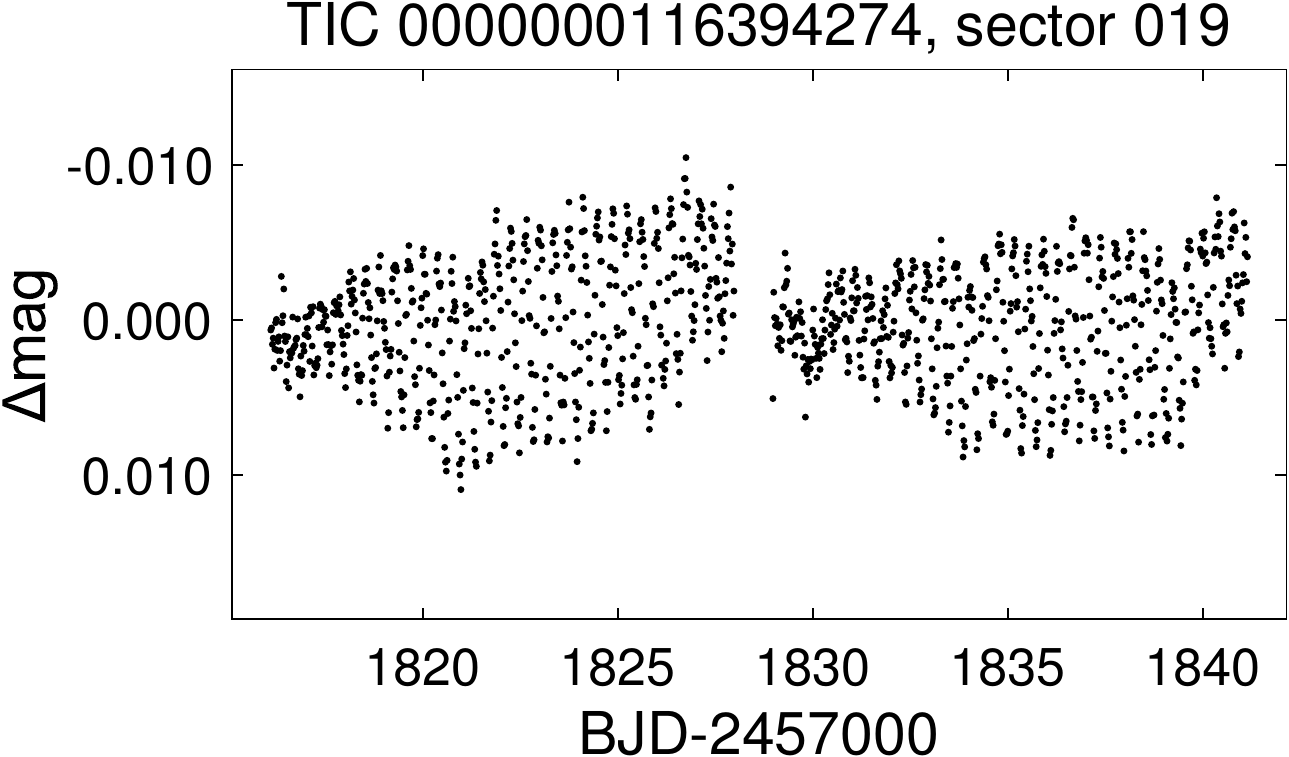}
\includegraphics[width=0.68\columnwidth]{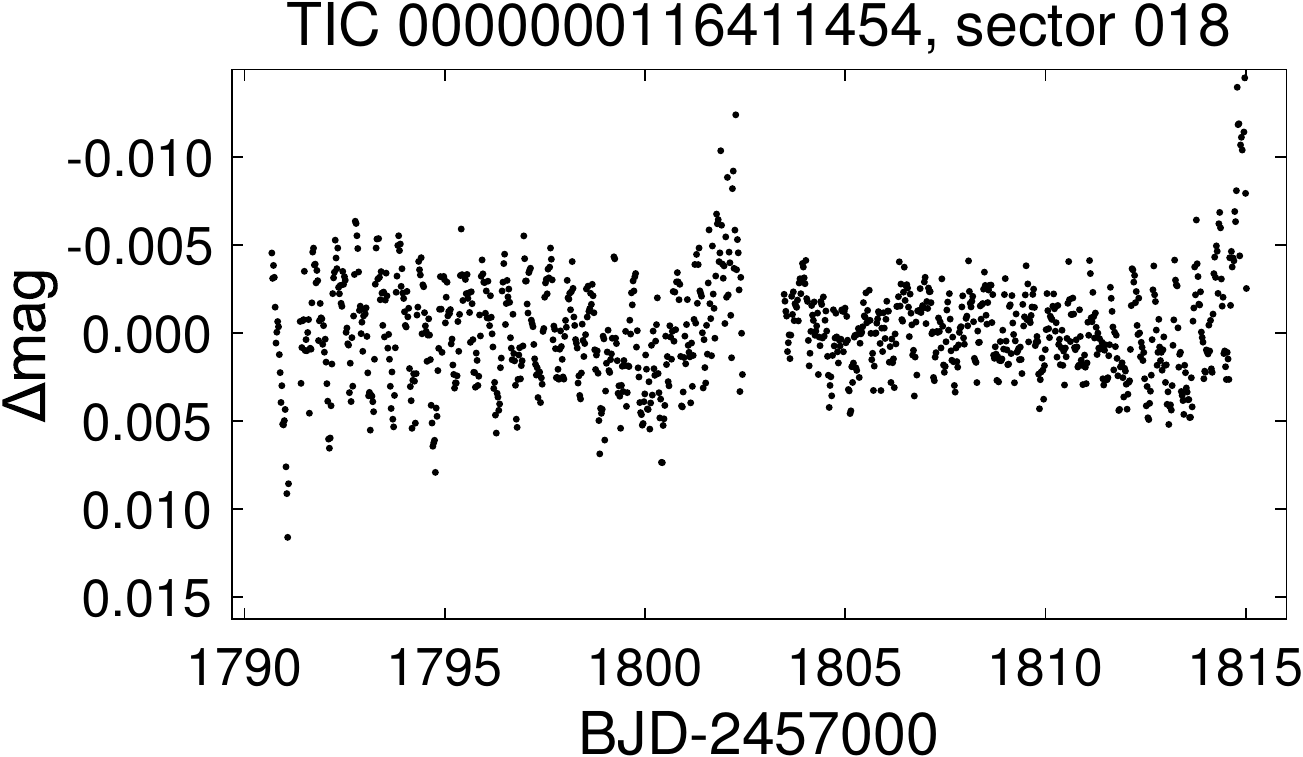}
\includegraphics[width=0.68\columnwidth]{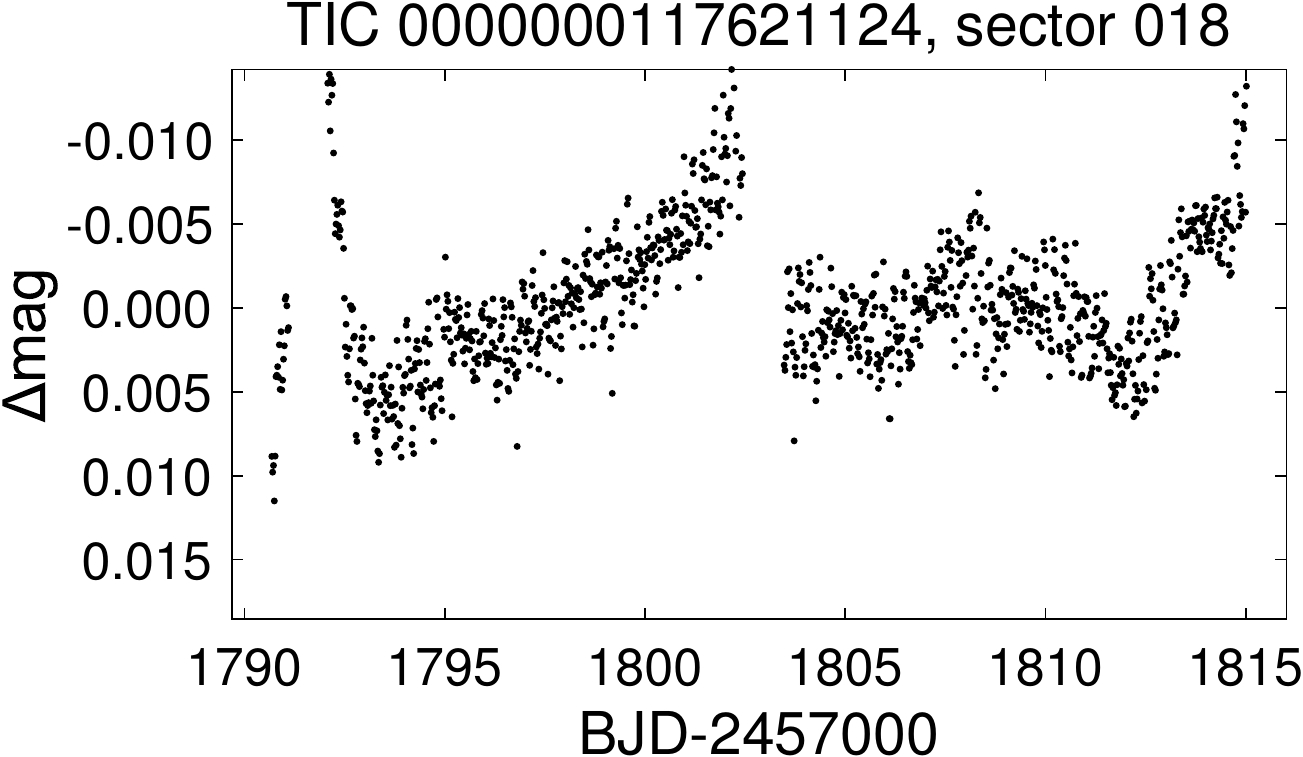}
\includegraphics[width=0.68\columnwidth]{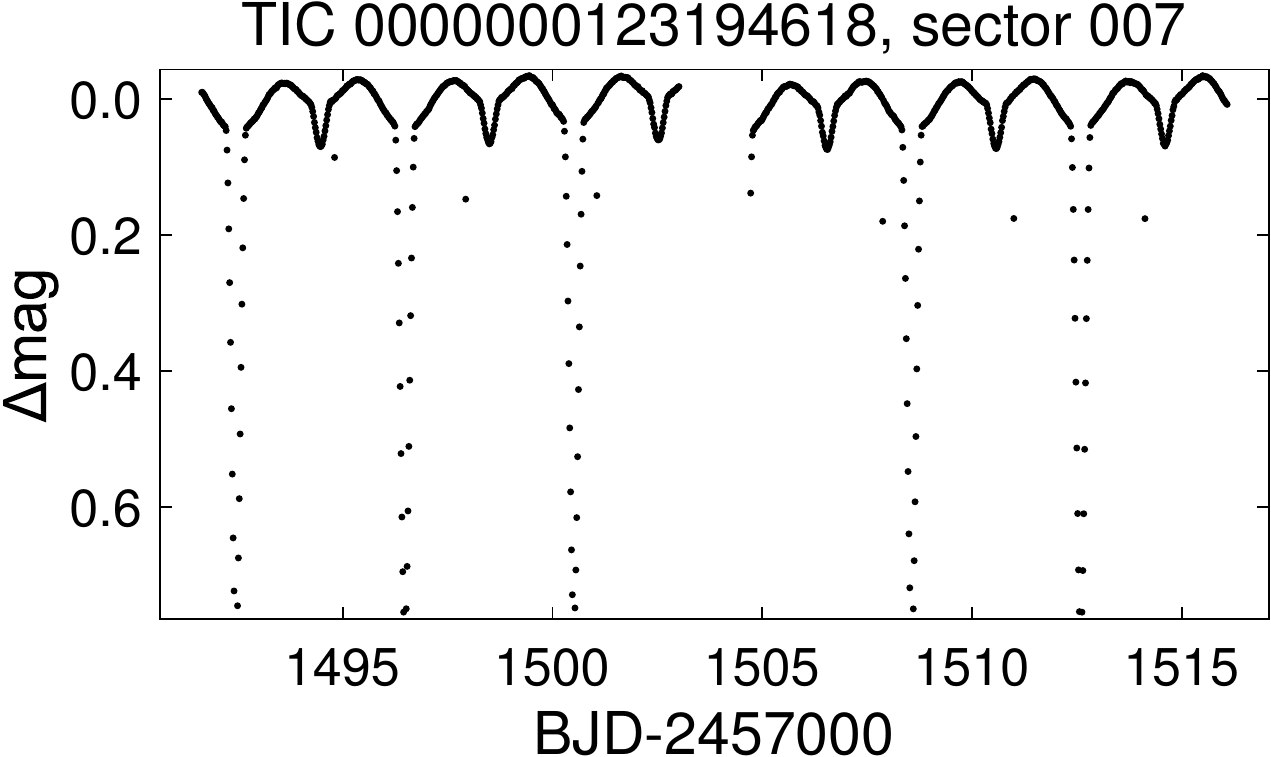}
\includegraphics[width=0.68\columnwidth]{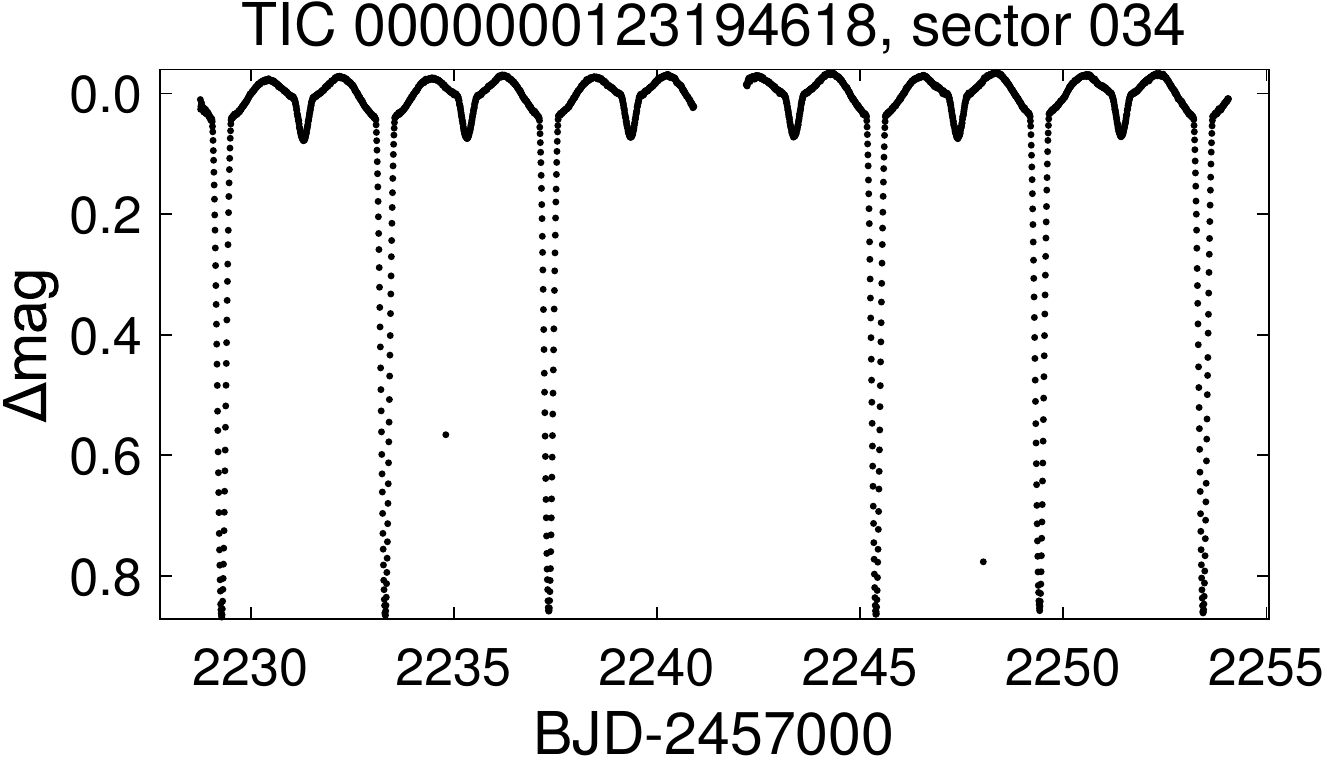}
\includegraphics[width=0.68\columnwidth]{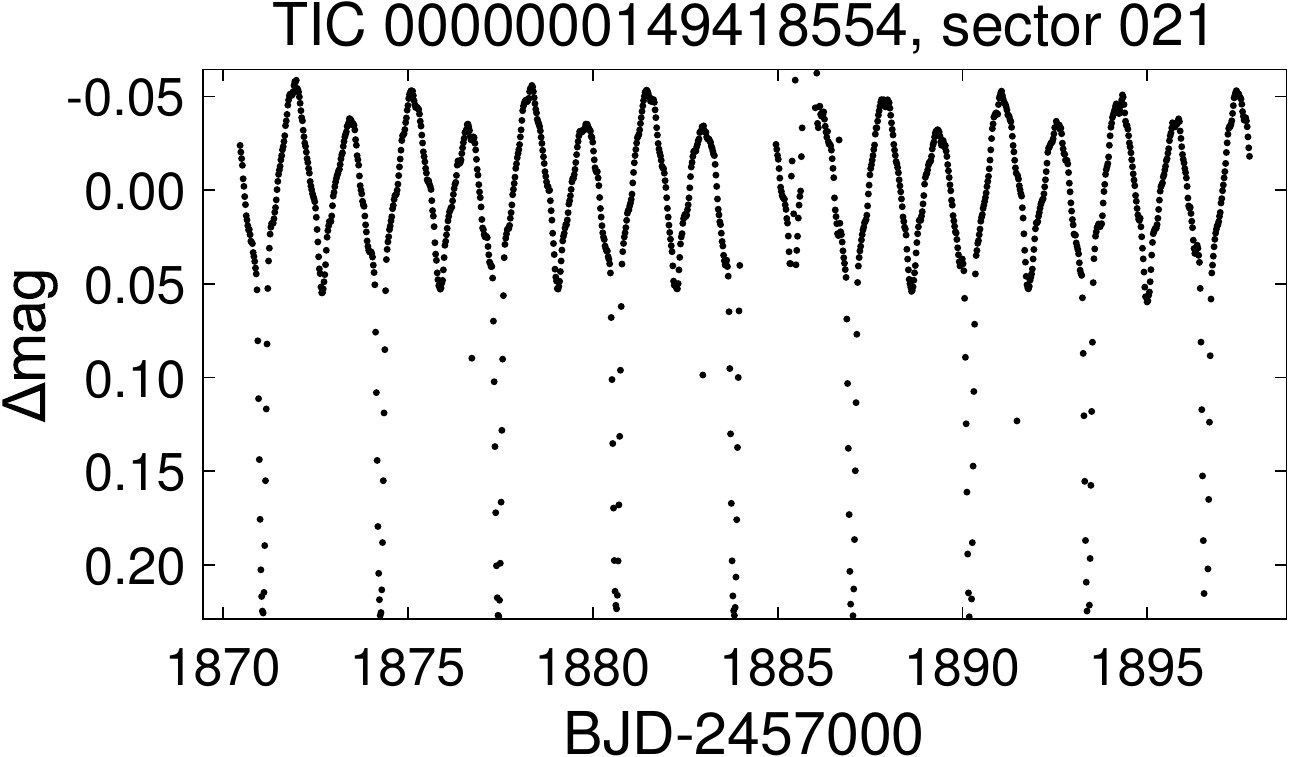}
\includegraphics[width=0.68\columnwidth]{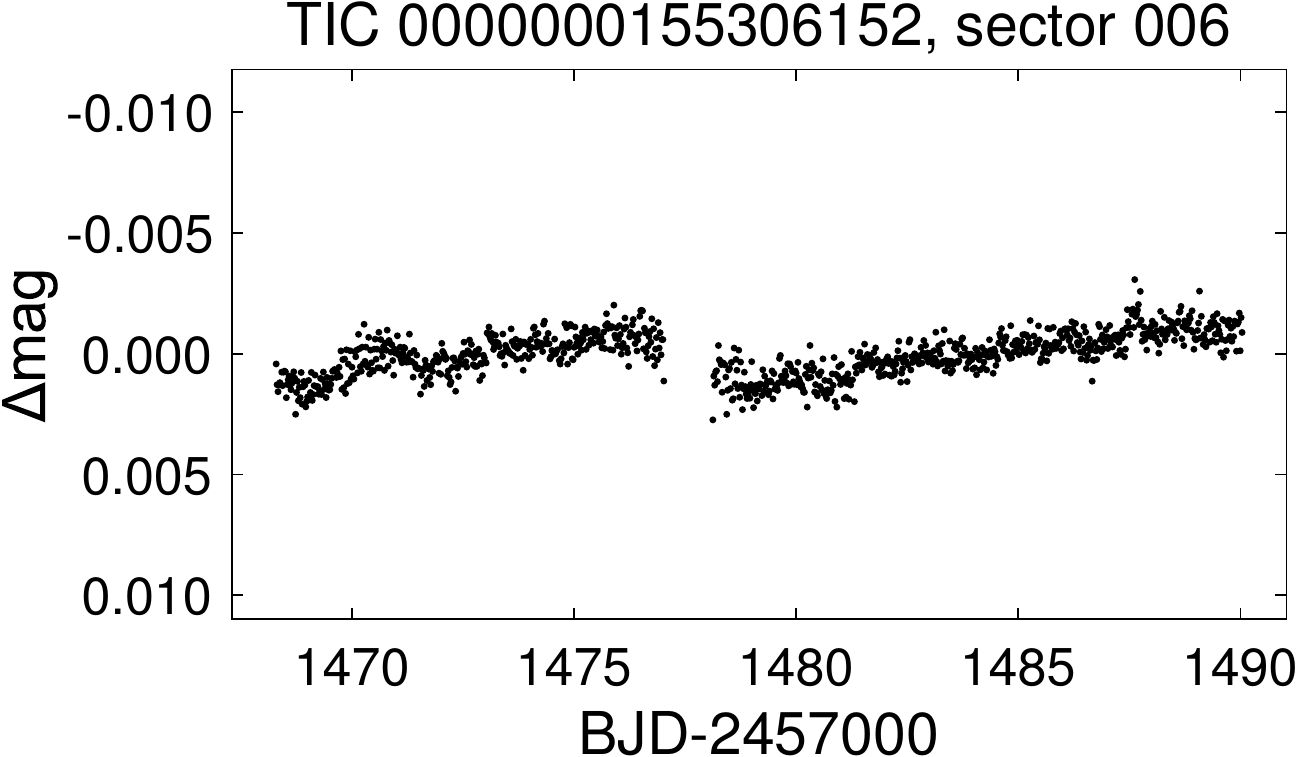}
    \caption{TESS light curves of all objects contained in the final shell star sample.}
		\label{lc1}
\end{figure*}

\begin{figure*}

\includegraphics[width=0.68\columnwidth]{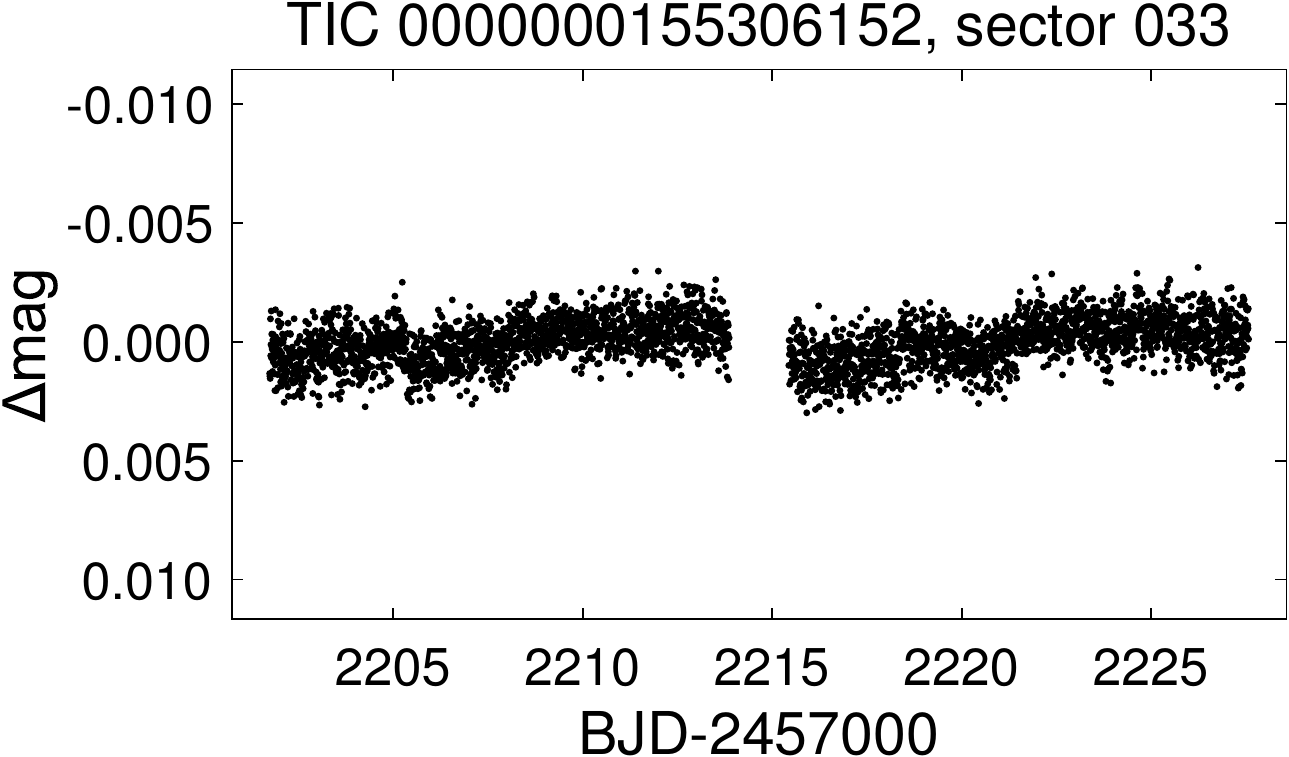}
\includegraphics[width=0.68\columnwidth]{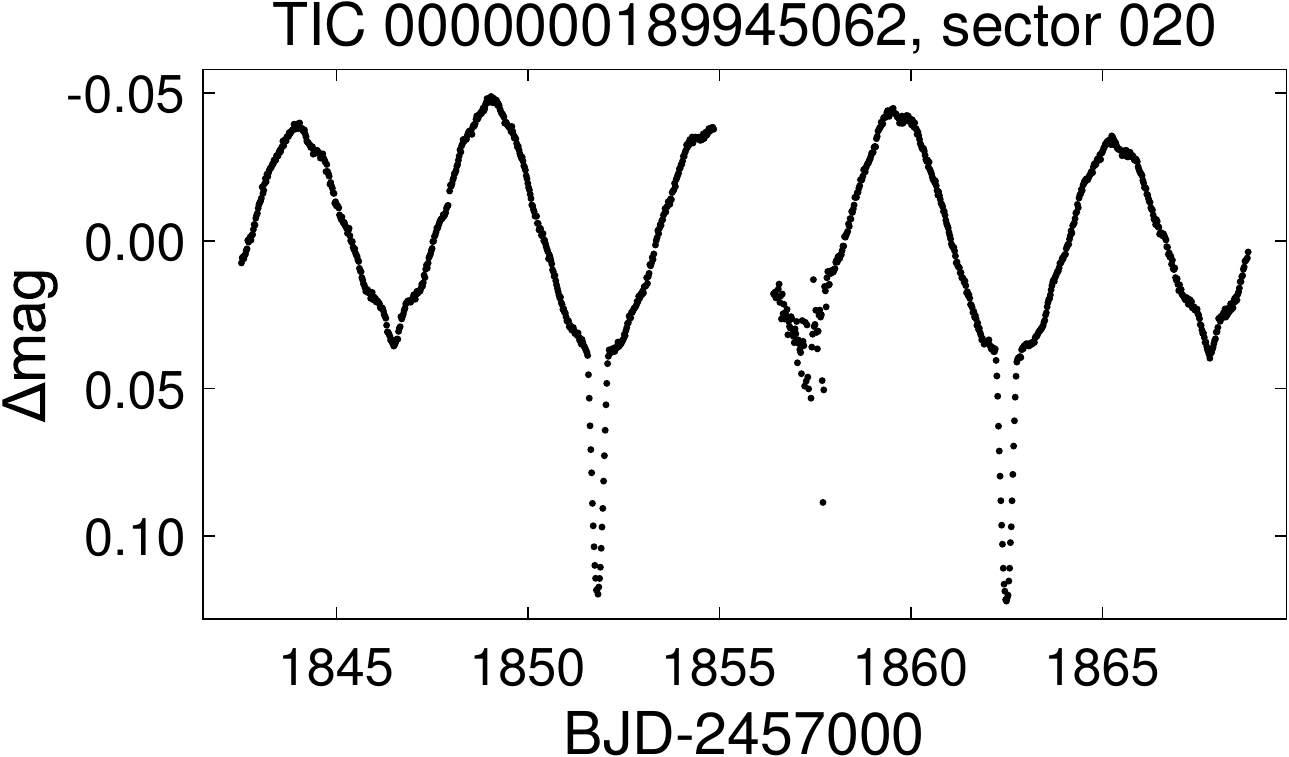}
\includegraphics[width=0.68\columnwidth]{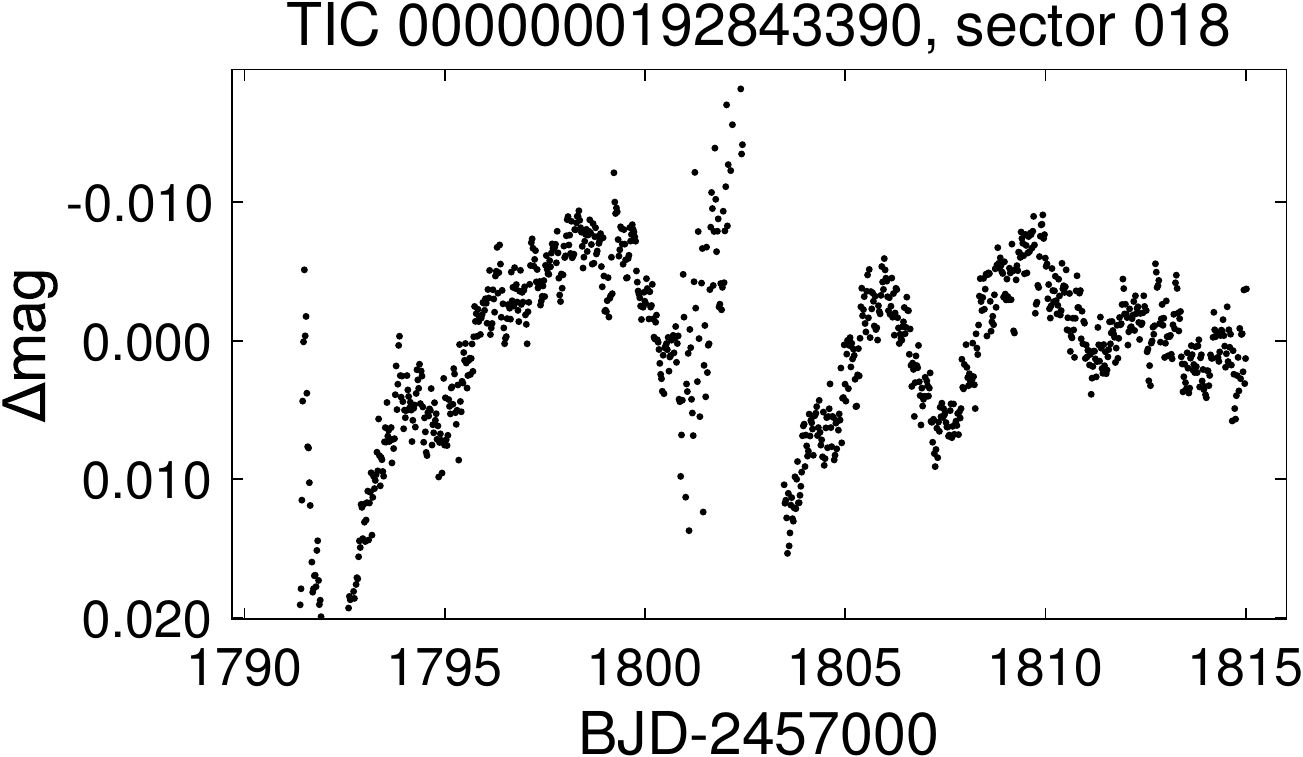}
\includegraphics[width=0.68\columnwidth]{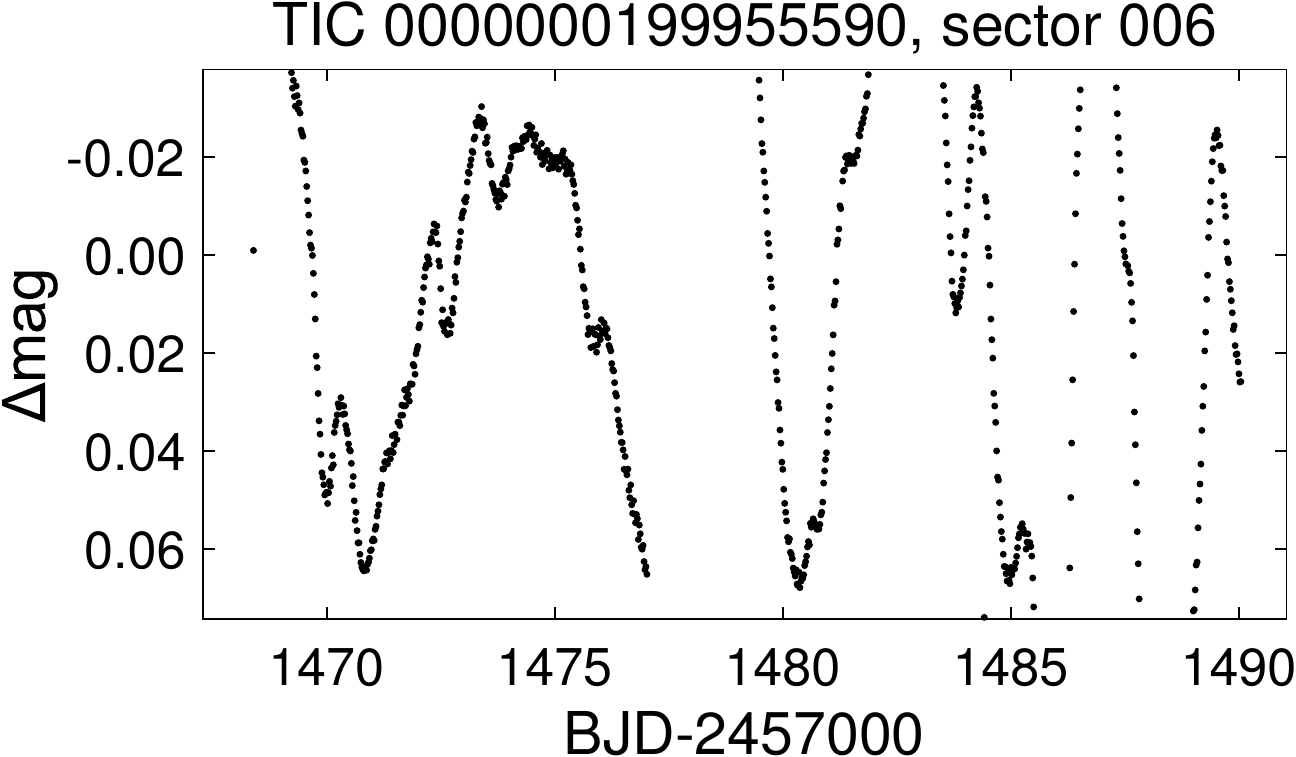}
\includegraphics[width=0.68\columnwidth]{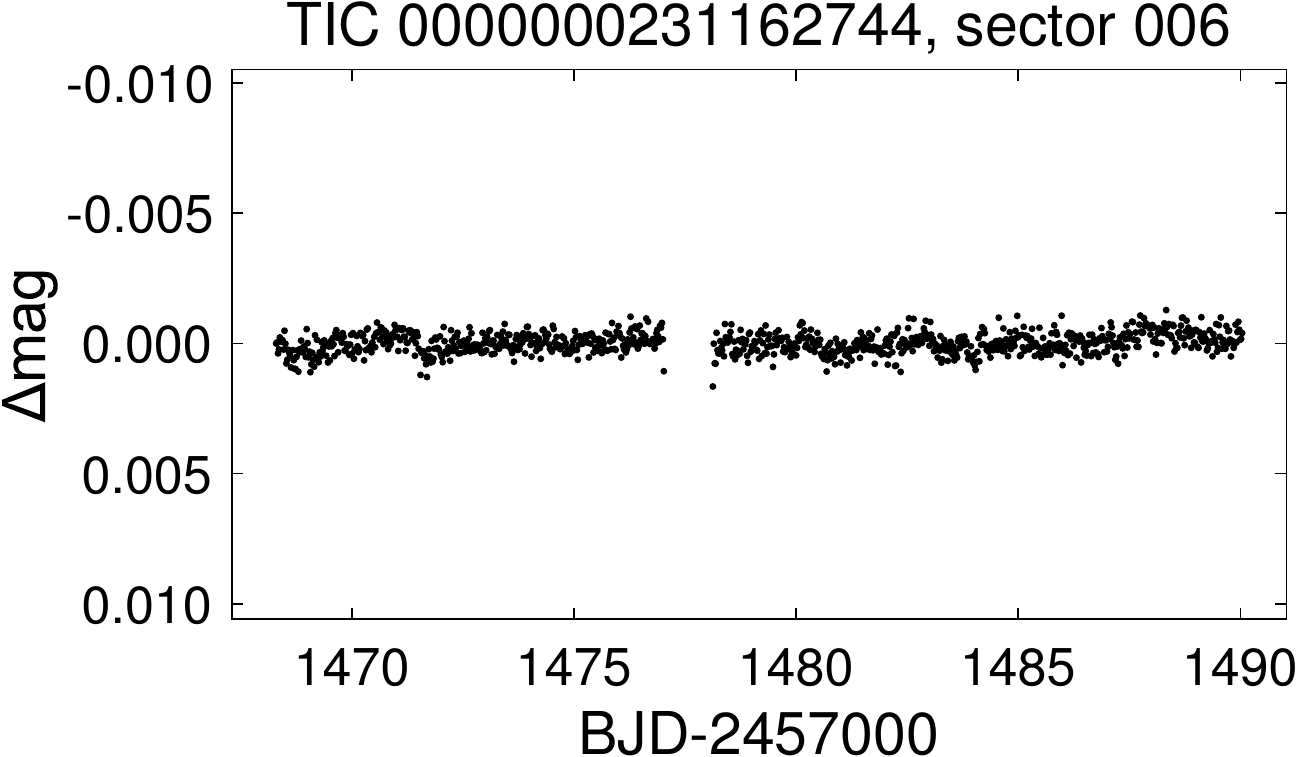}
\includegraphics[width=0.68\columnwidth]{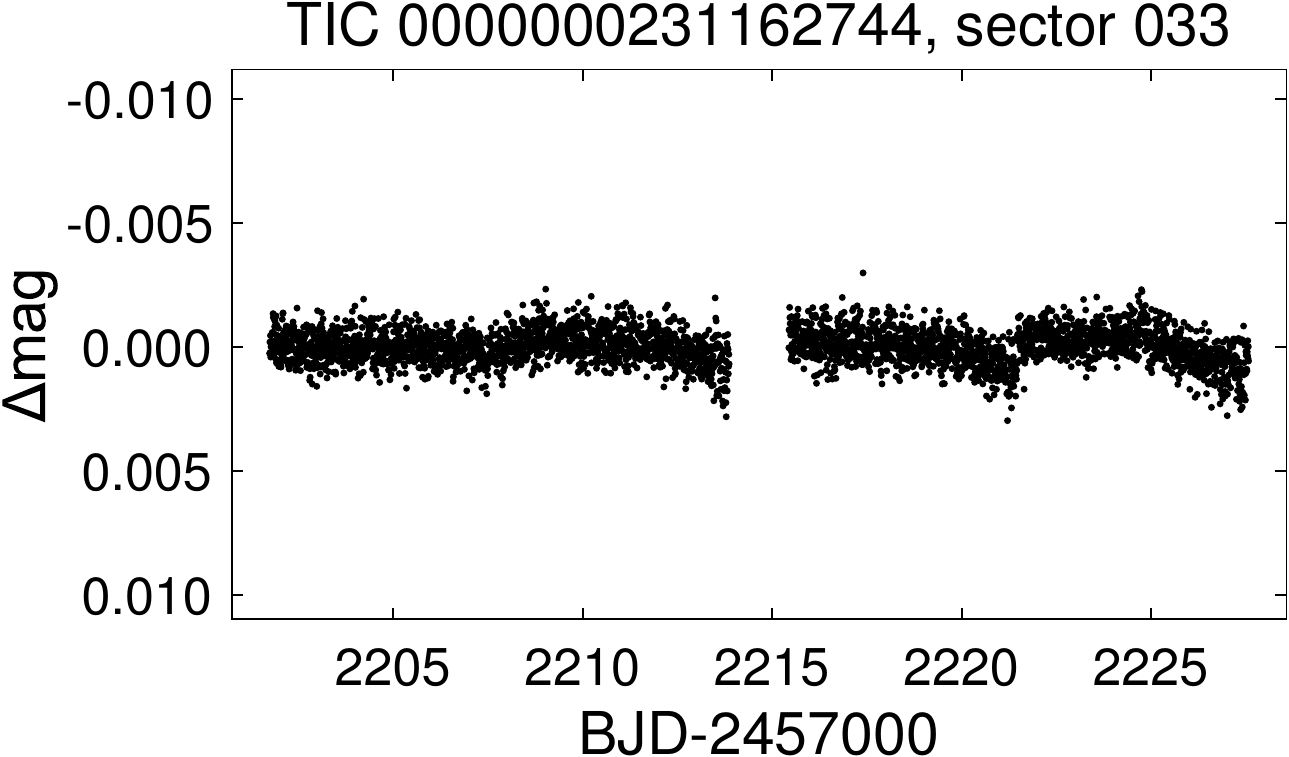}
\includegraphics[width=0.68\columnwidth]{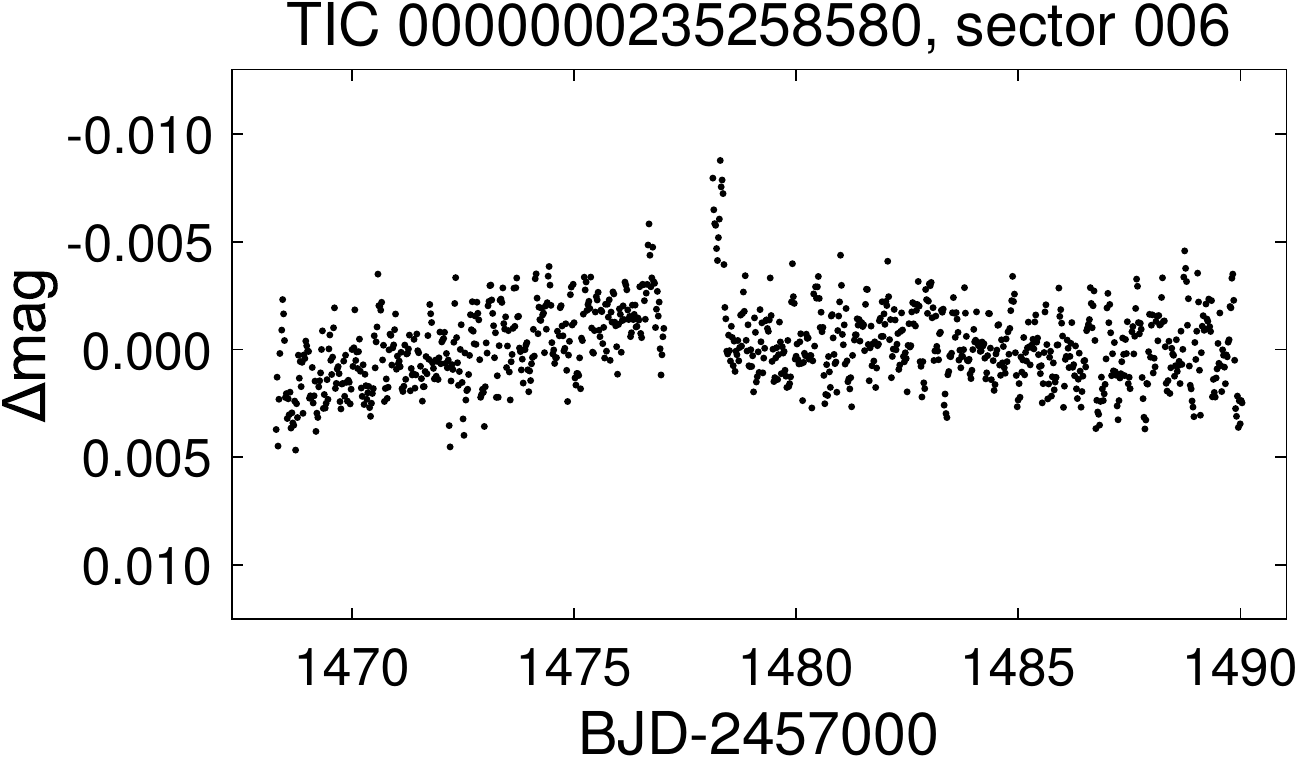}
\includegraphics[width=0.68\columnwidth]{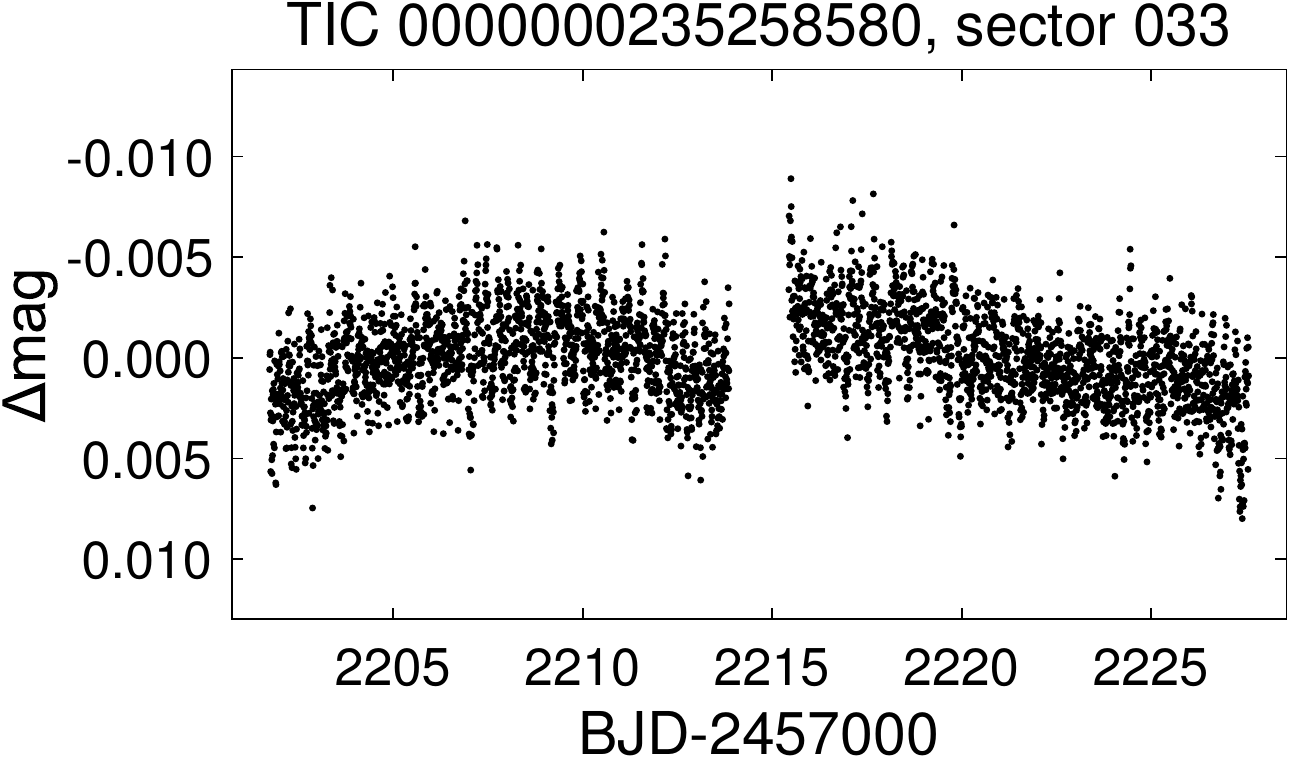}
\includegraphics[width=0.68\columnwidth]{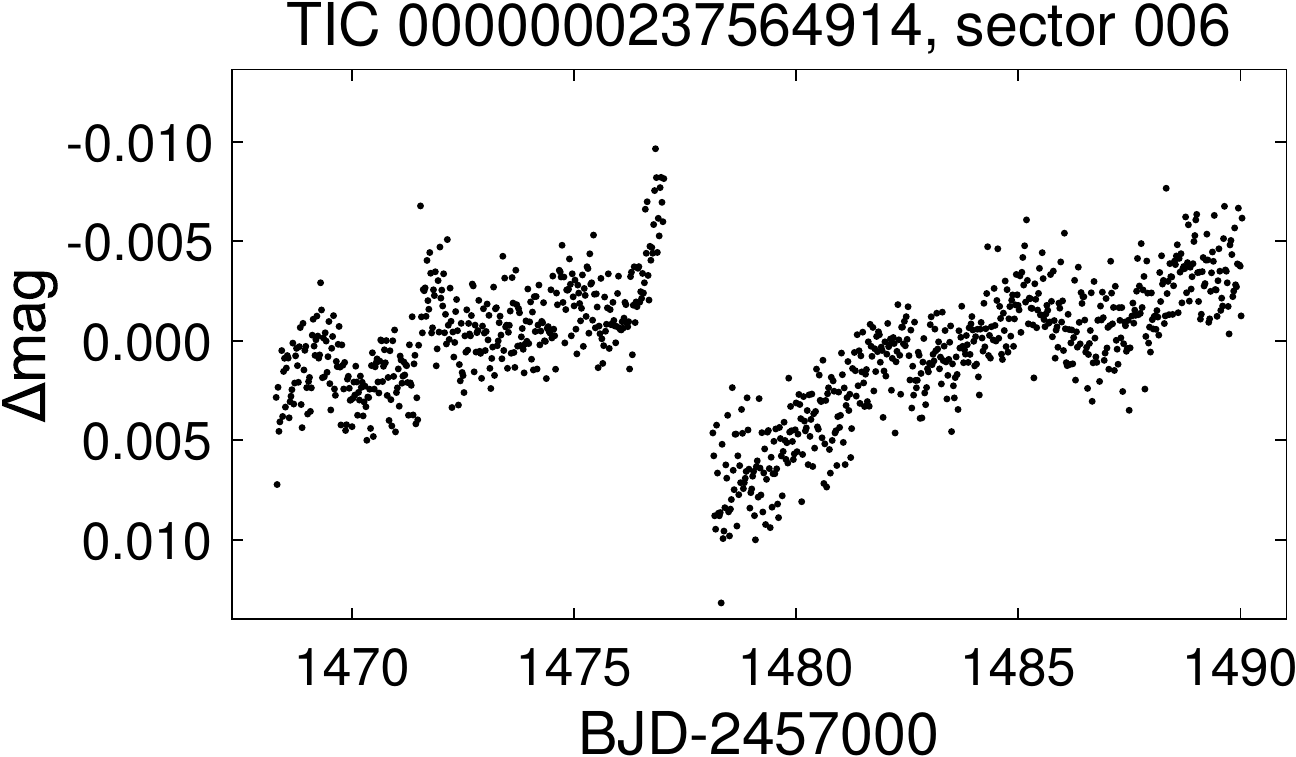}
\includegraphics[width=0.68\columnwidth]{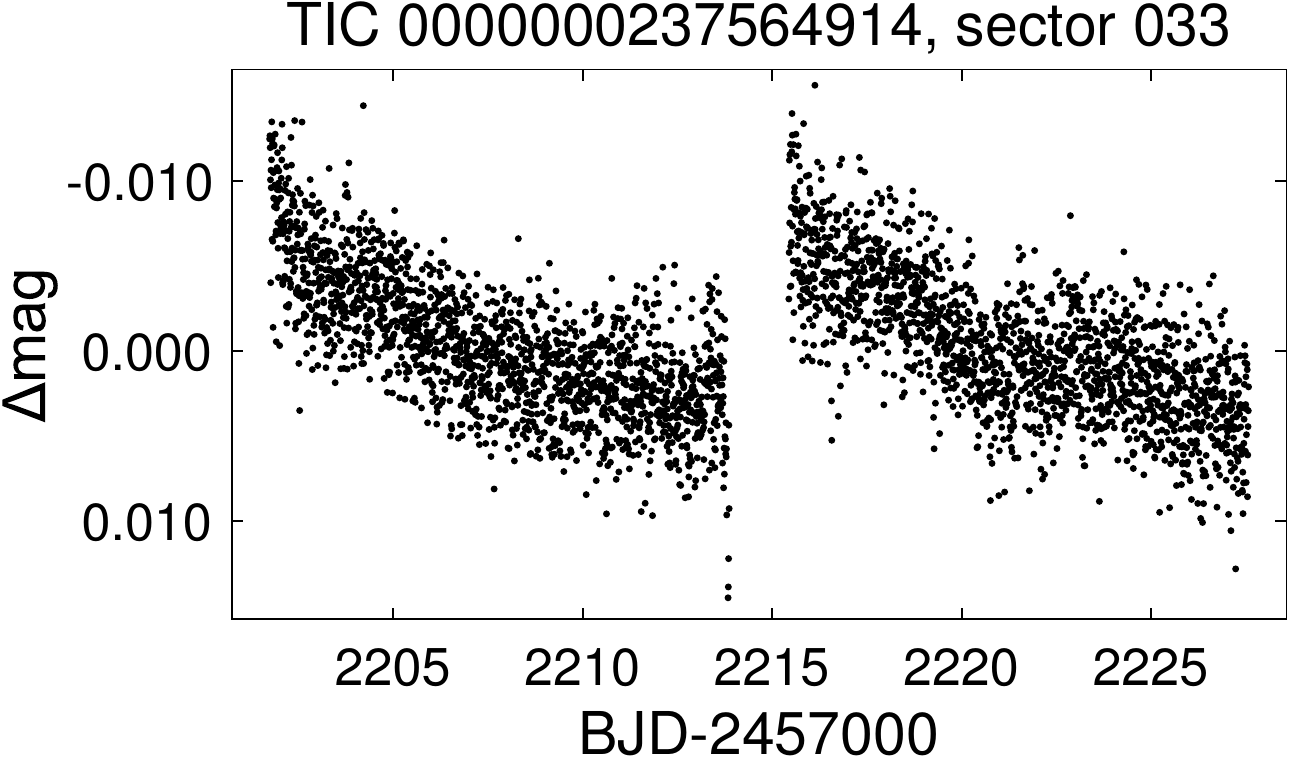}
\includegraphics[width=0.68\columnwidth]{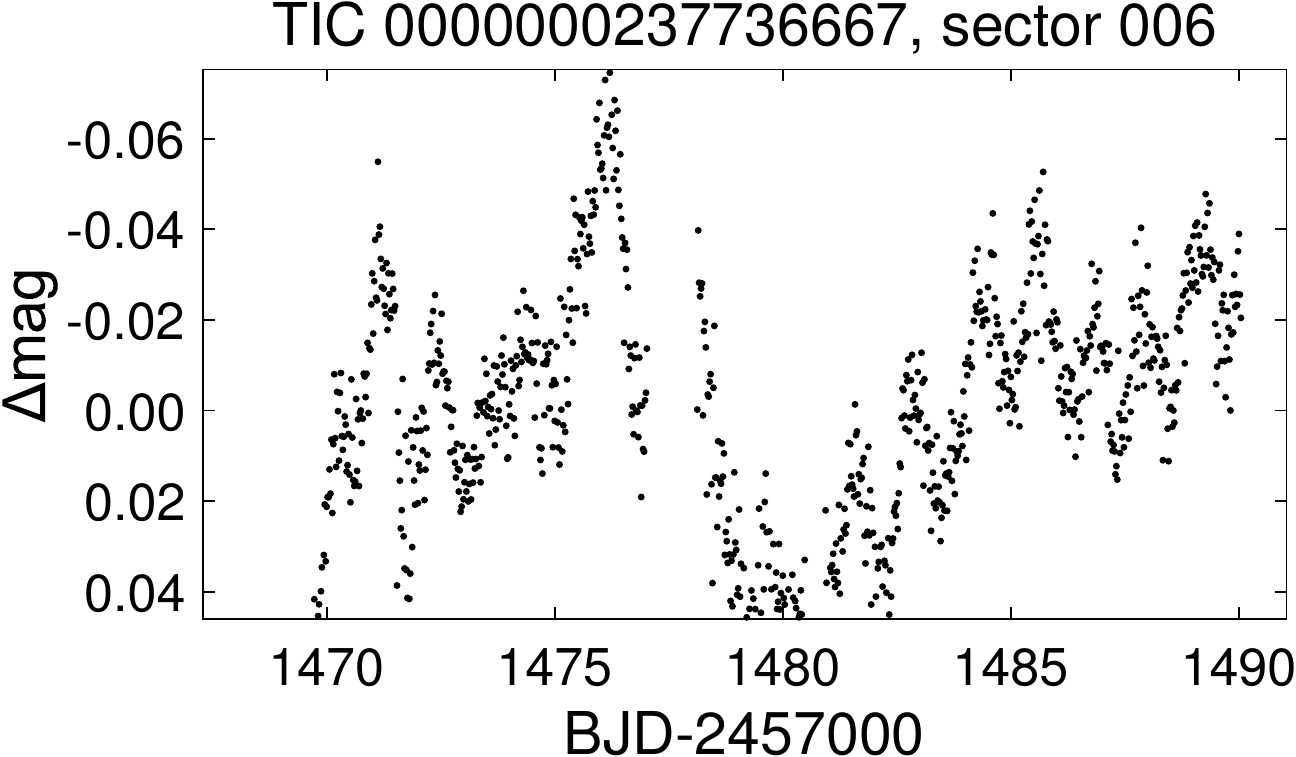}
\includegraphics[width=0.68\columnwidth]{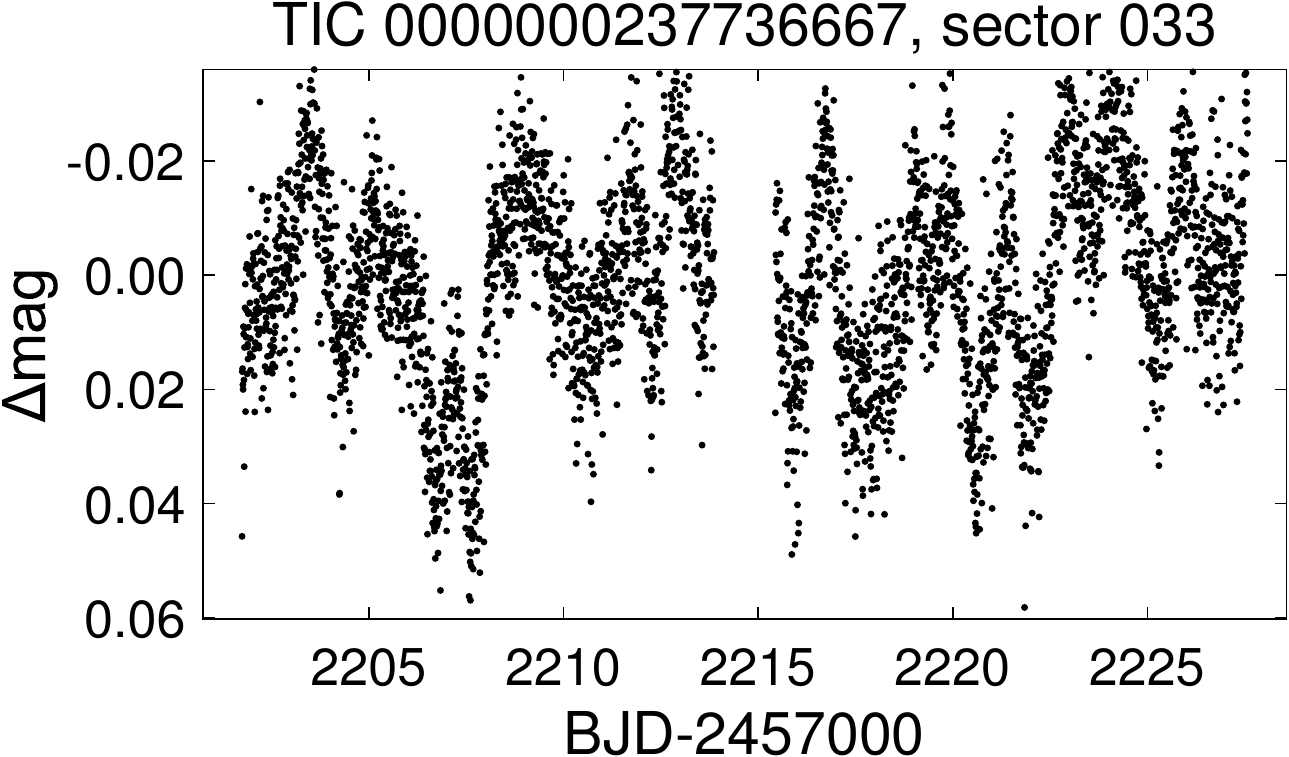}
\includegraphics[width=0.68\columnwidth]{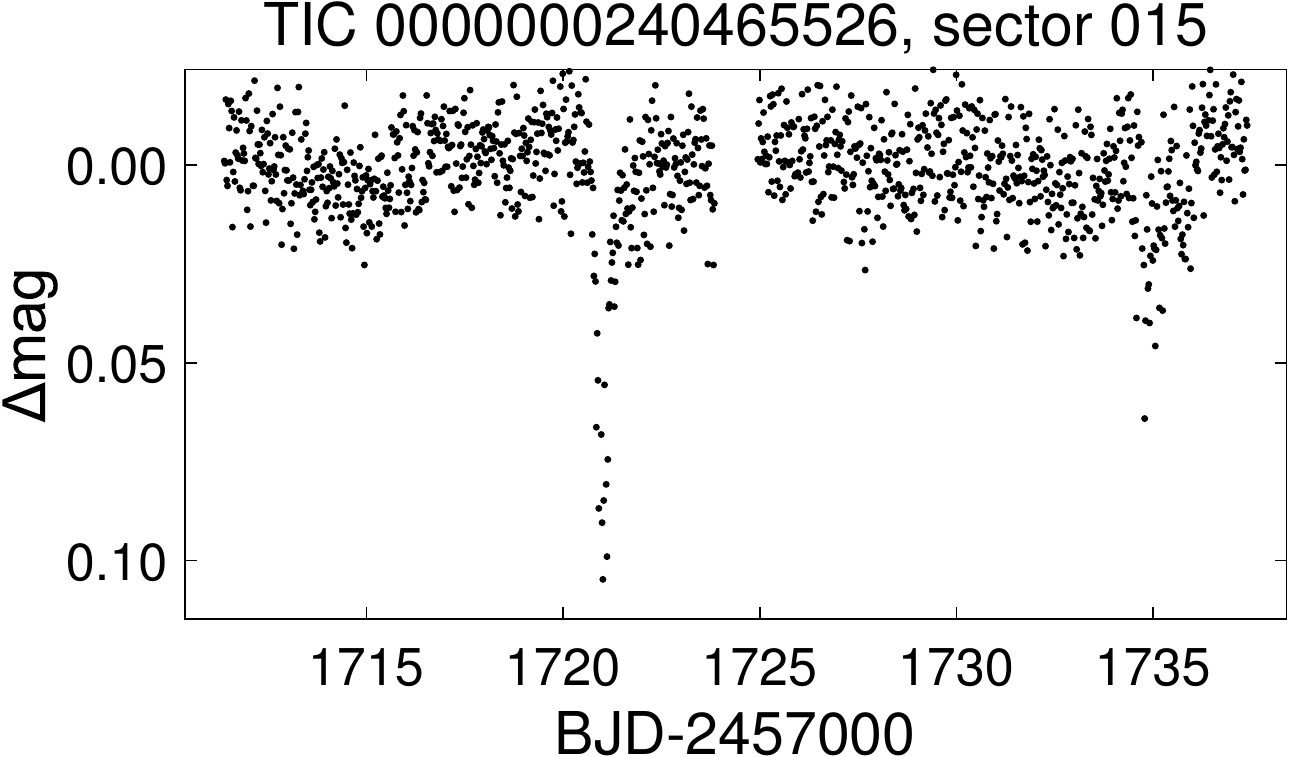}
\includegraphics[width=0.68\columnwidth]{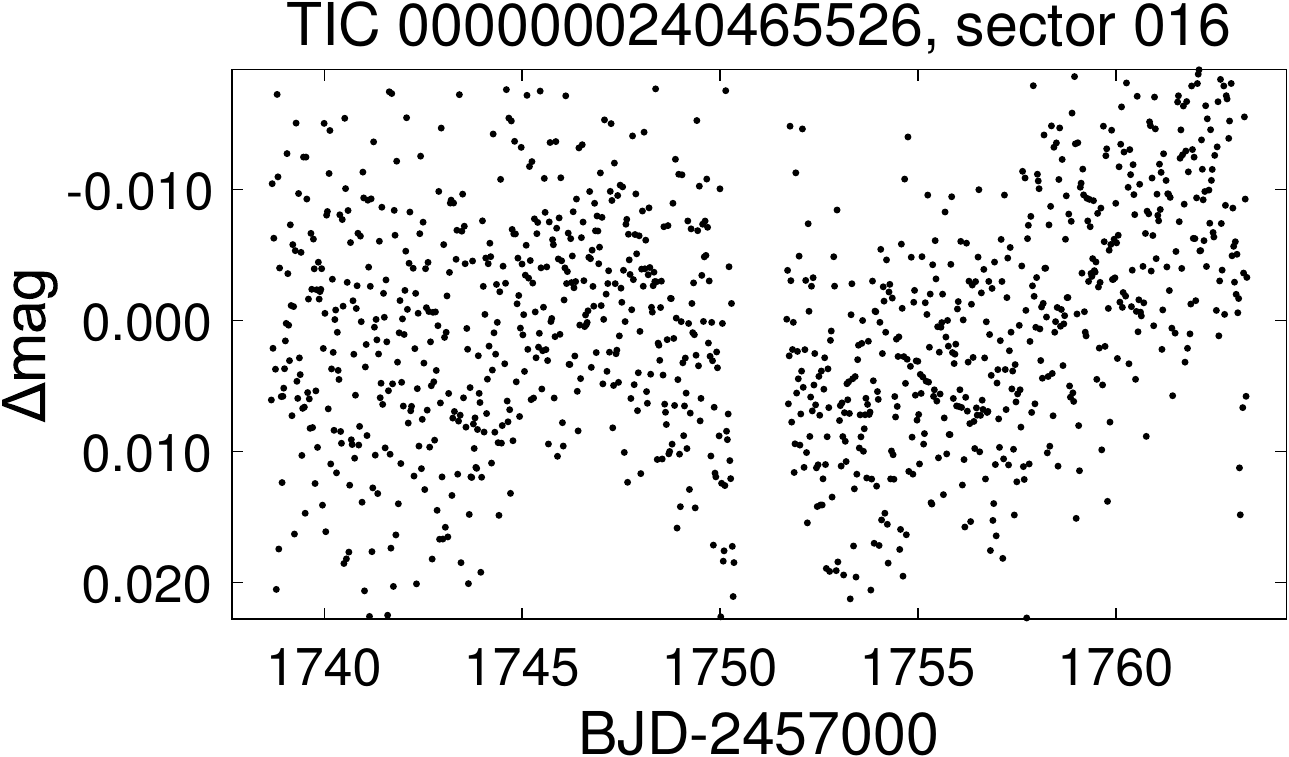}
\includegraphics[width=0.68\columnwidth]{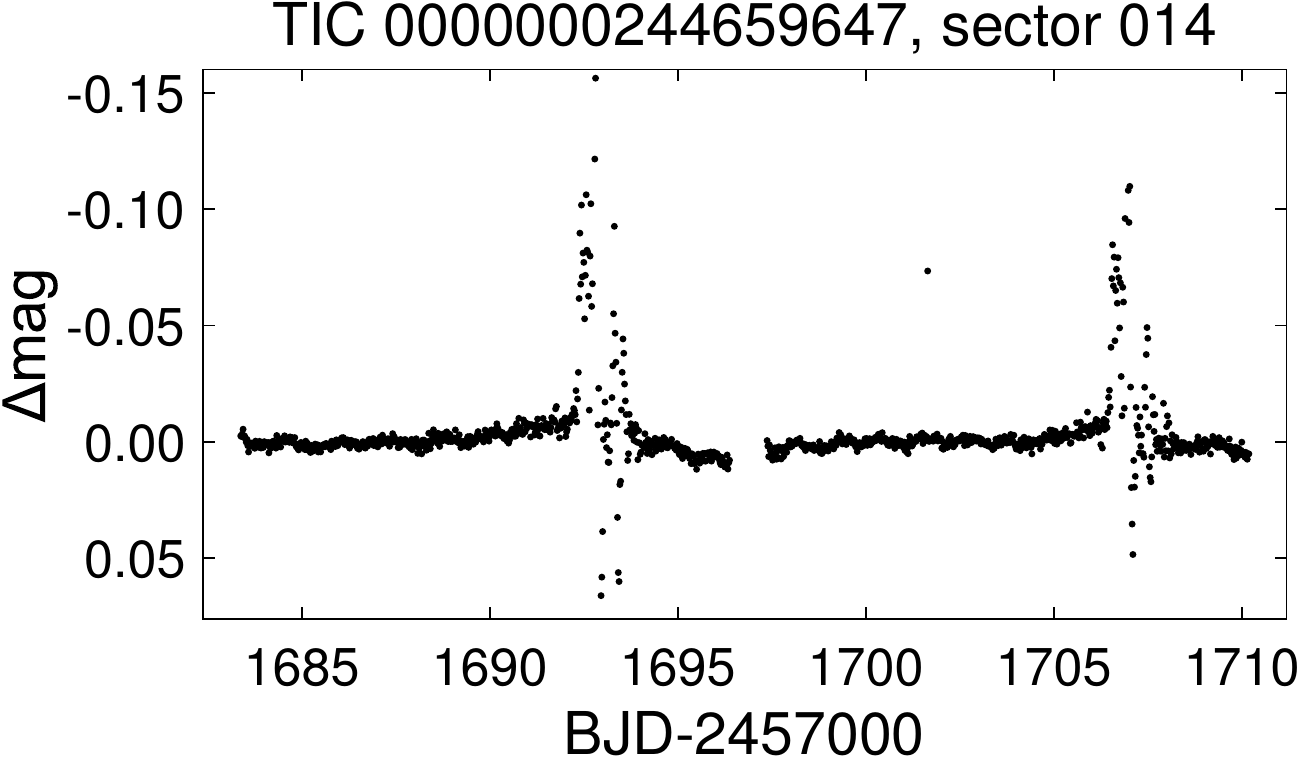}
\includegraphics[width=0.68\columnwidth]{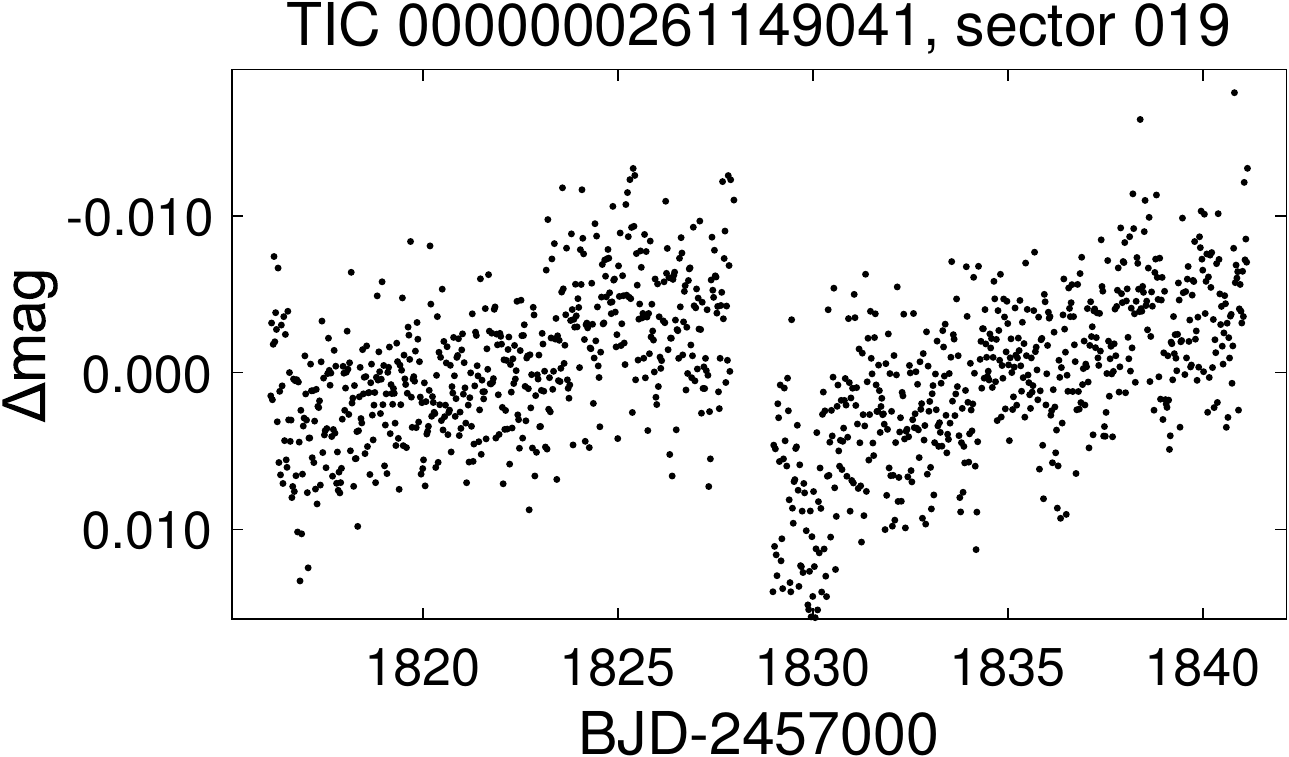}
\includegraphics[width=0.68\columnwidth]{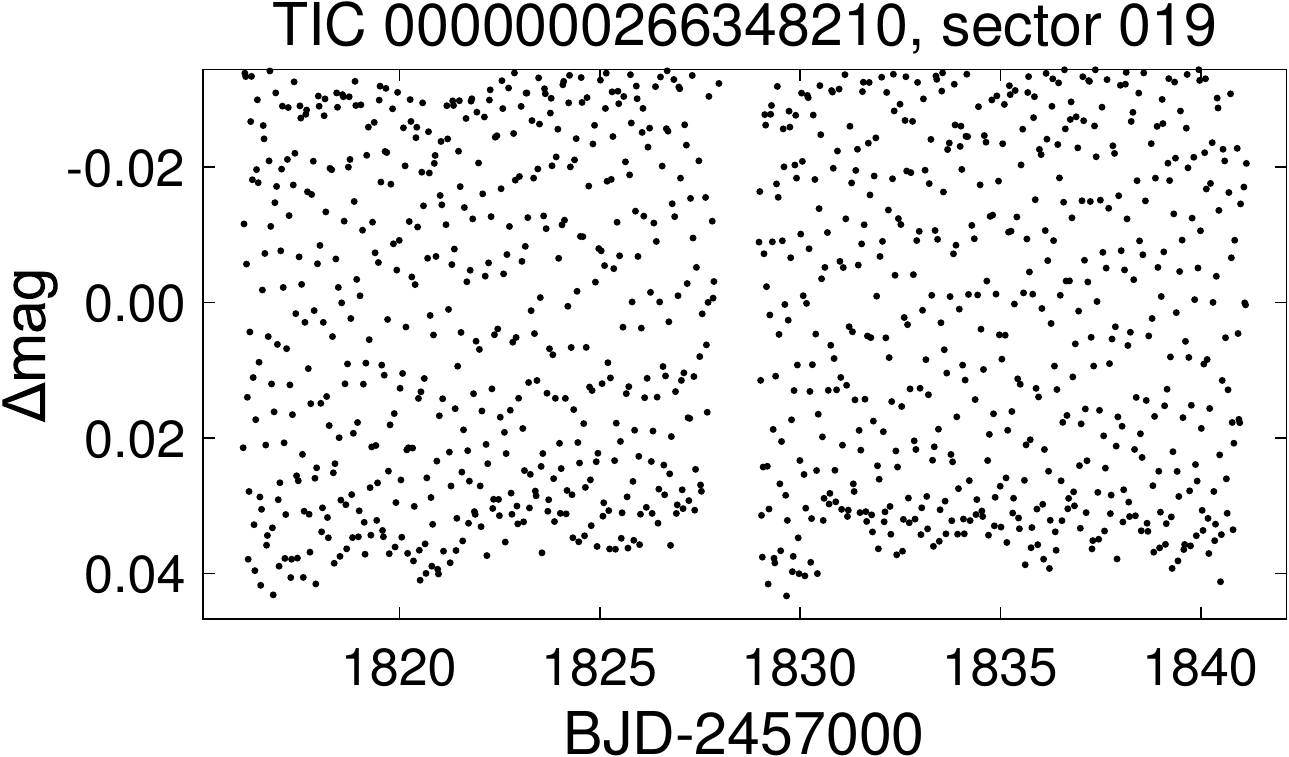}
\includegraphics[width=0.68\columnwidth]{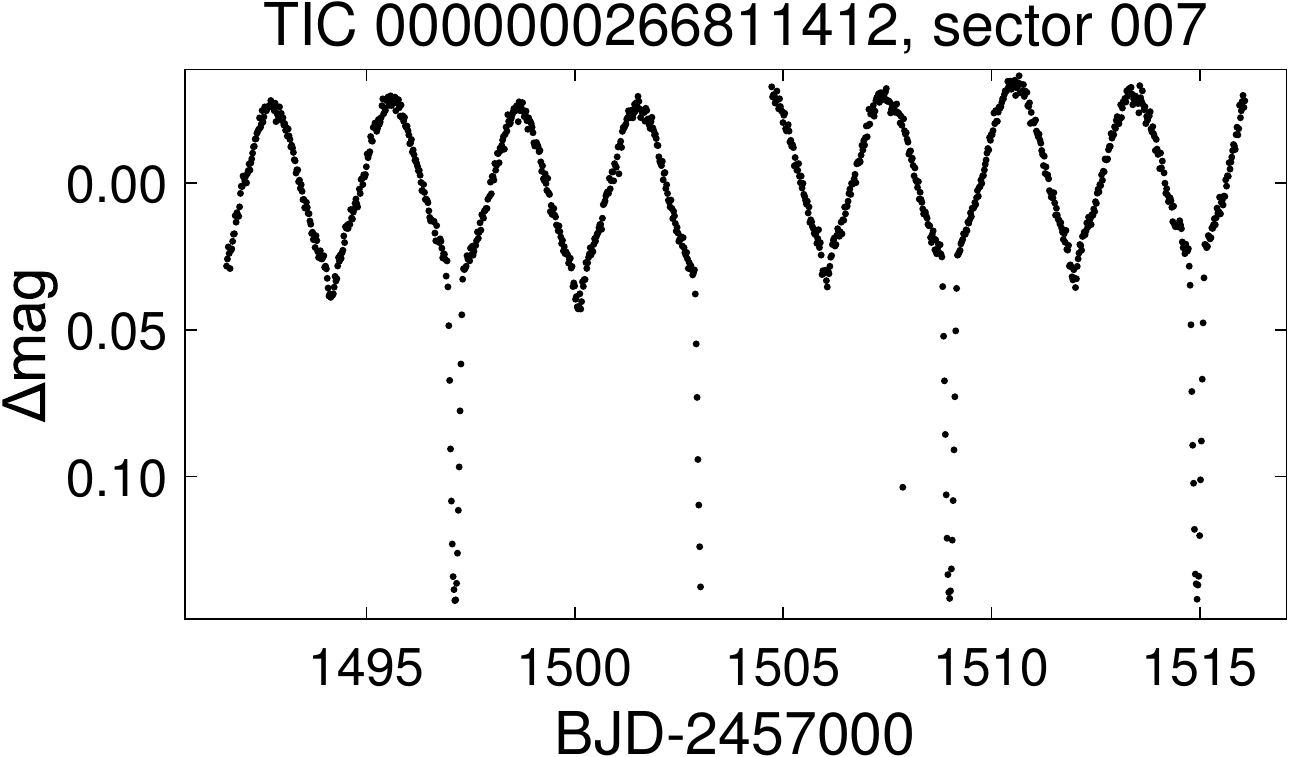}
    \caption{TESS light curves of all objects contained in the final shell star sample.}
		\label{lc2}
\end{figure*}

\begin{figure*}

\includegraphics[width=0.68\columnwidth]{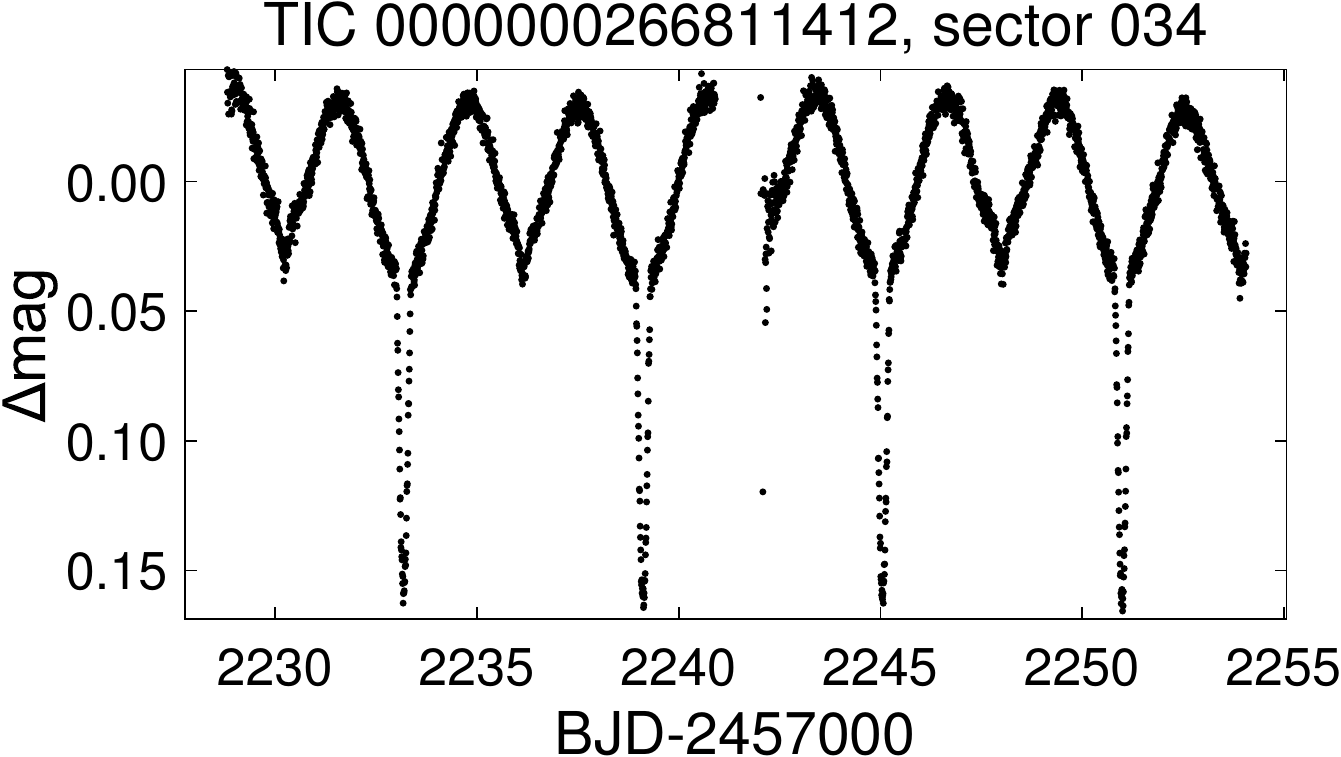}
\includegraphics[width=0.68\columnwidth]{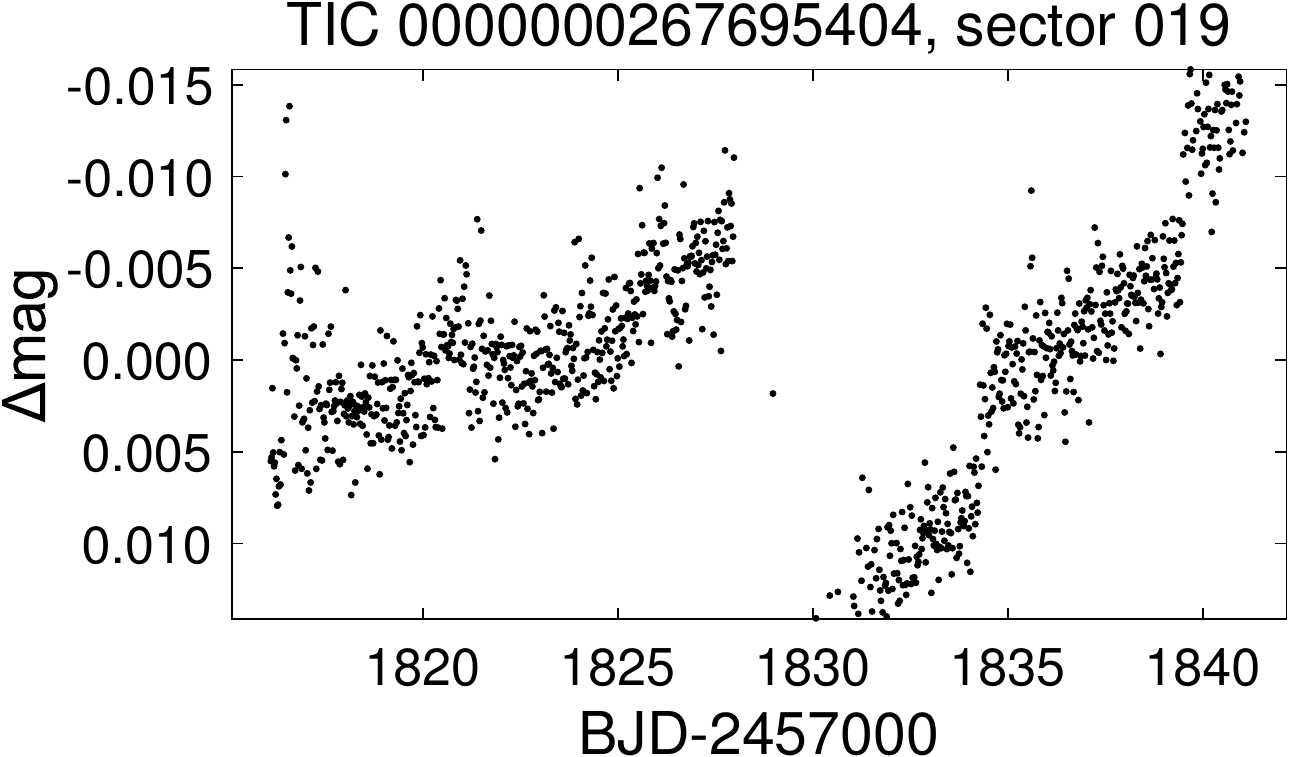}
\includegraphics[width=0.68\columnwidth]{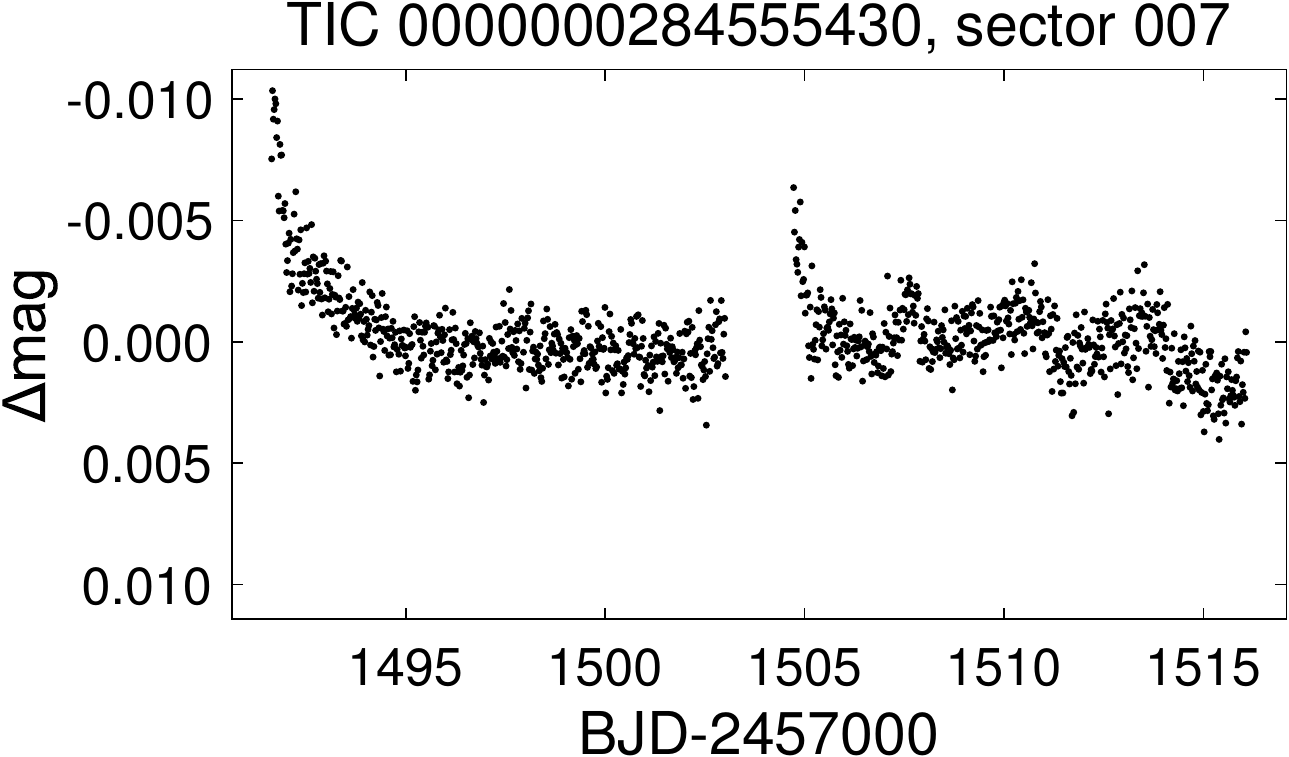}
\includegraphics[width=0.68\columnwidth]{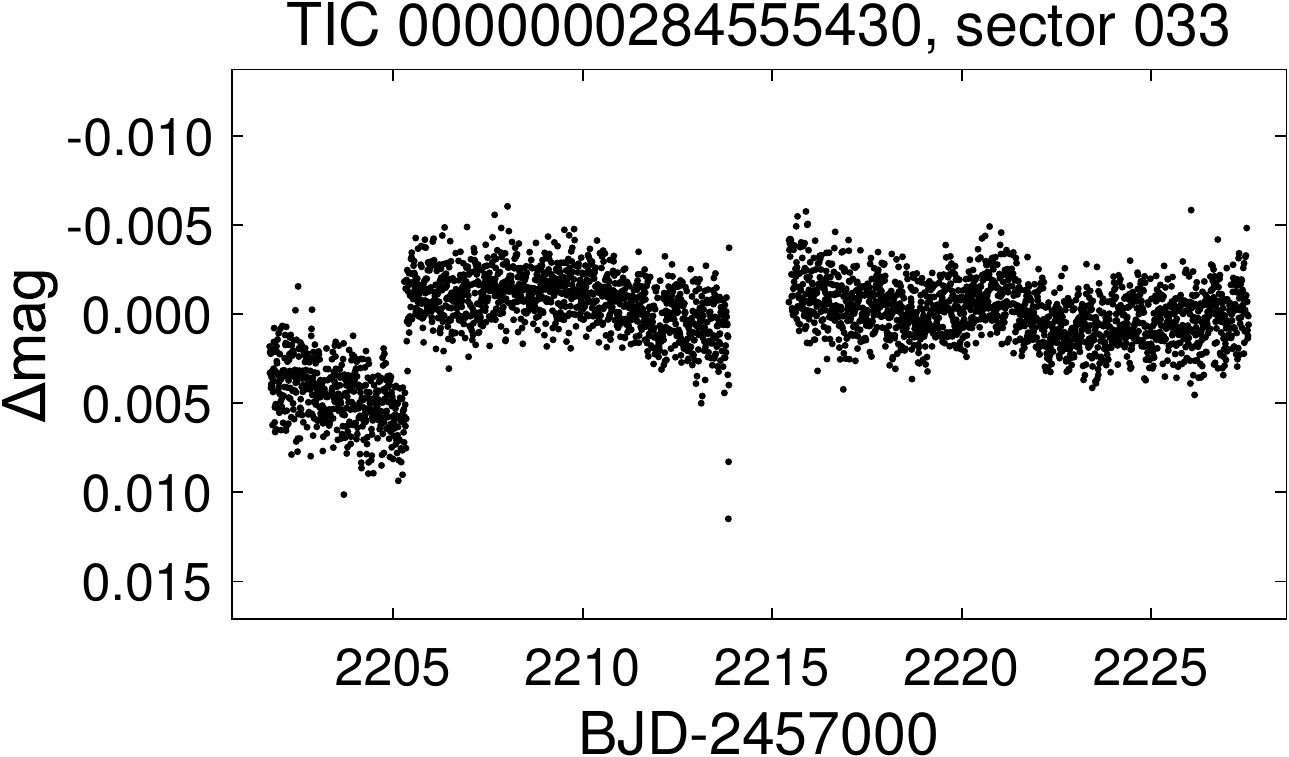}
\includegraphics[width=0.68\columnwidth]{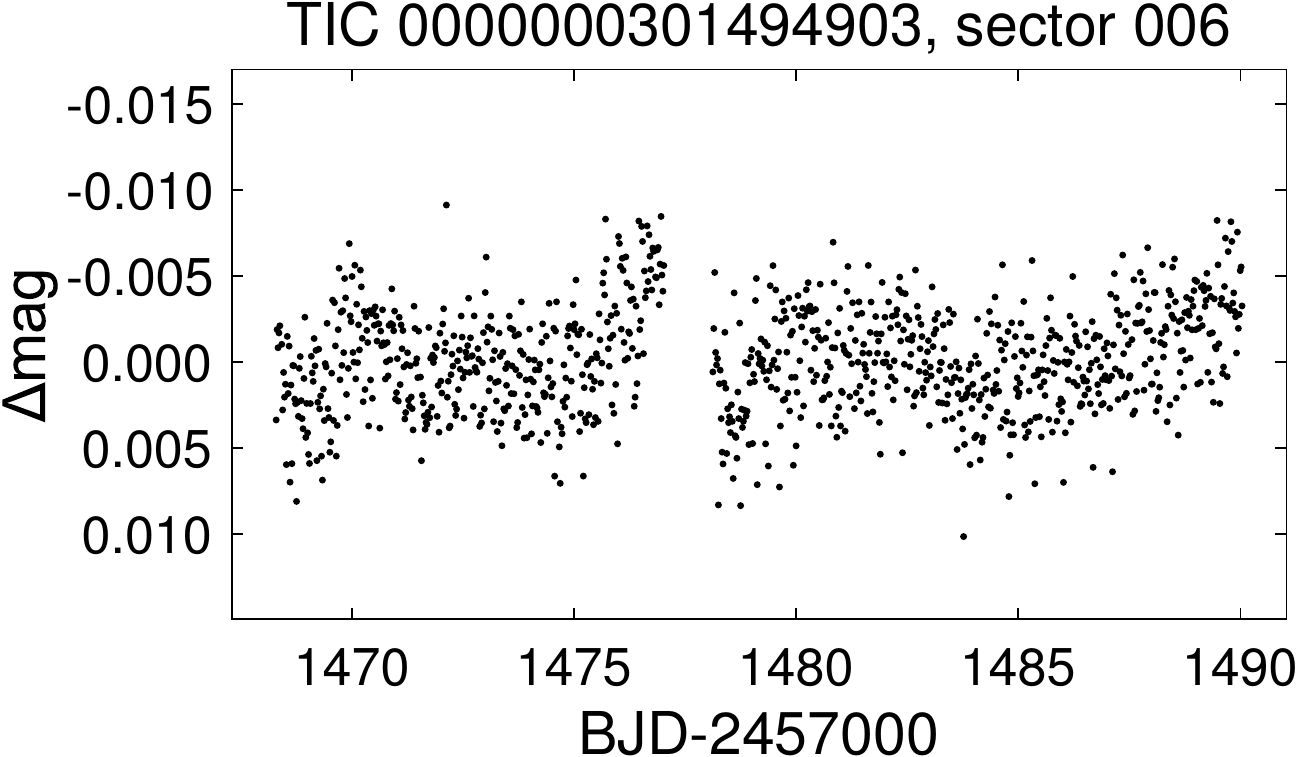}
\includegraphics[width=0.68\columnwidth]{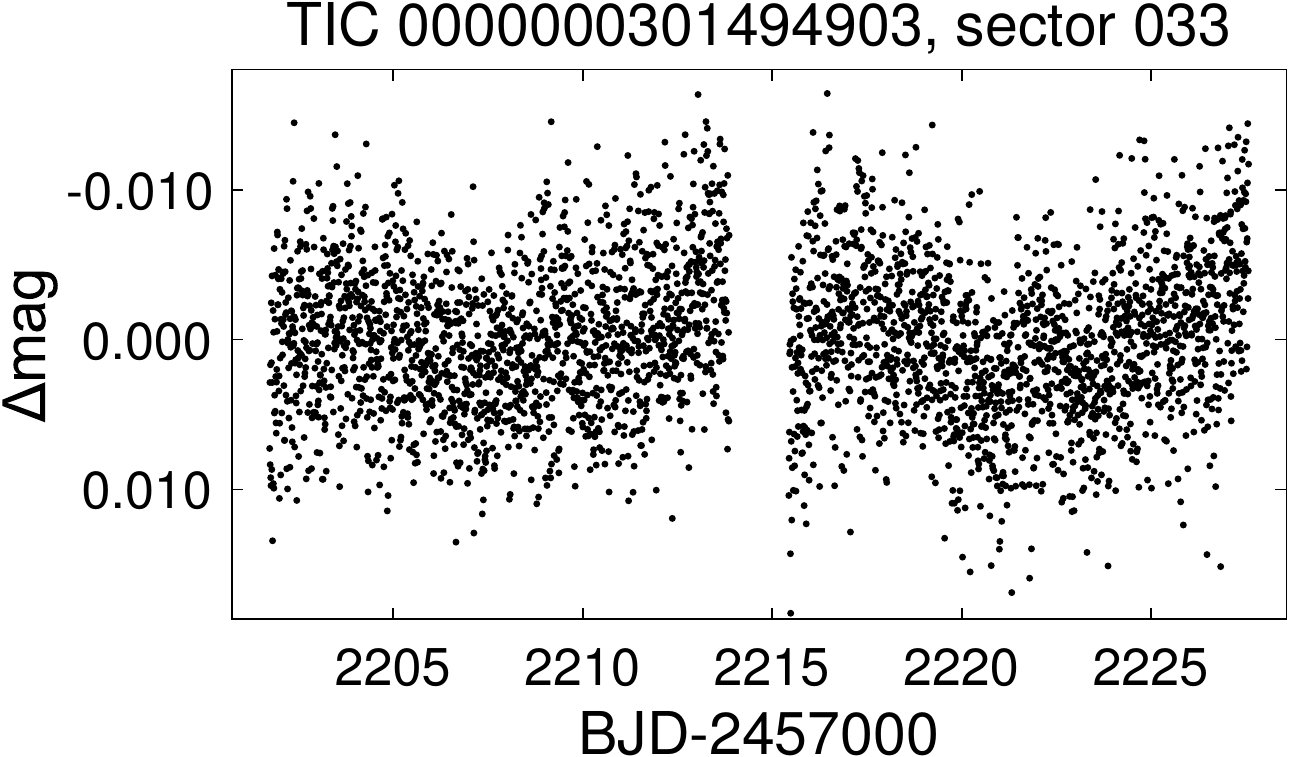}
\includegraphics[width=0.68\columnwidth]{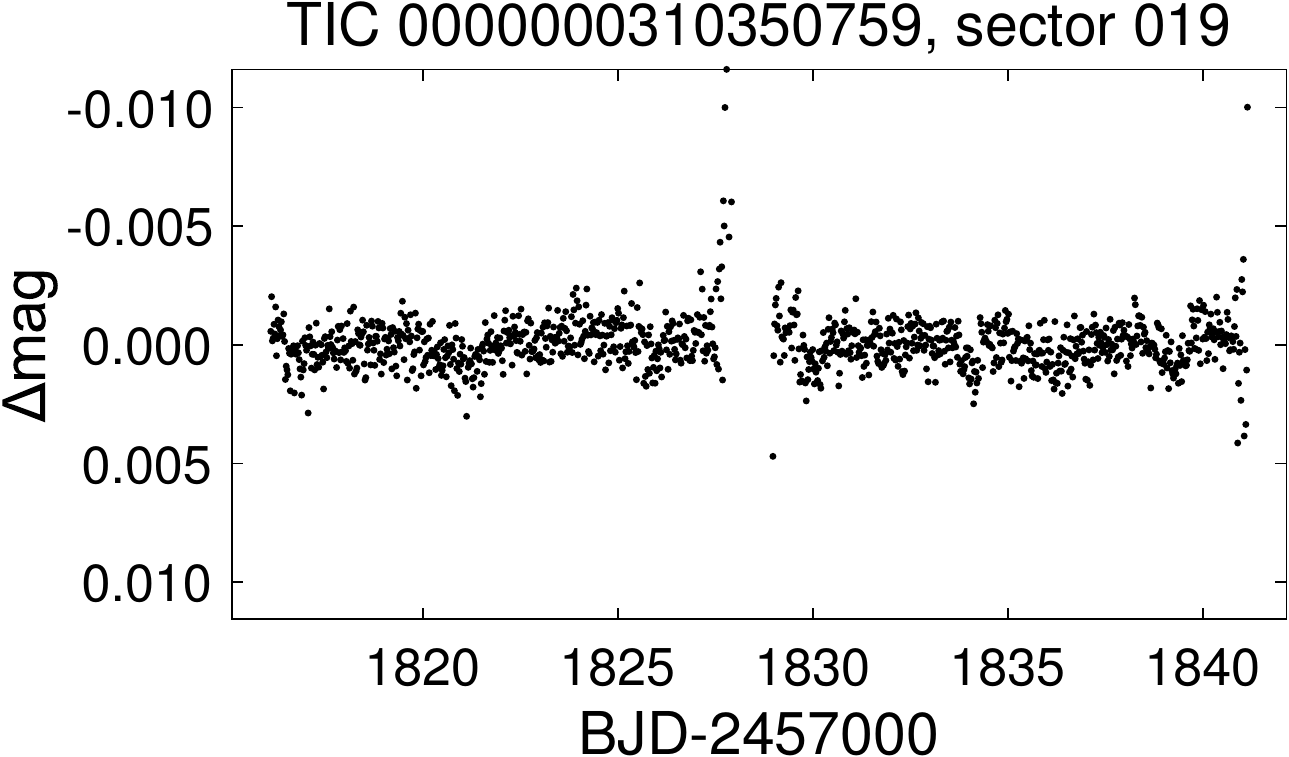}
\includegraphics[width=0.68\columnwidth]{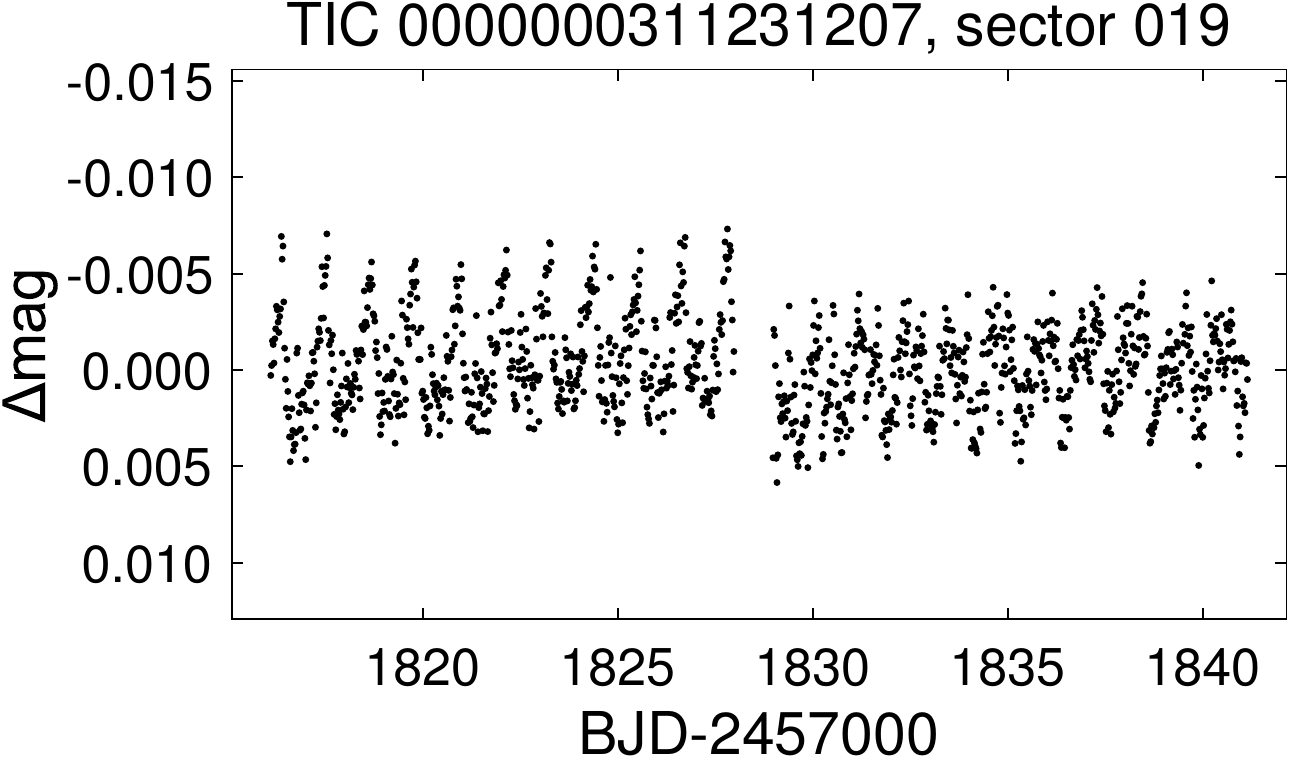}
\includegraphics[width=0.68\columnwidth]{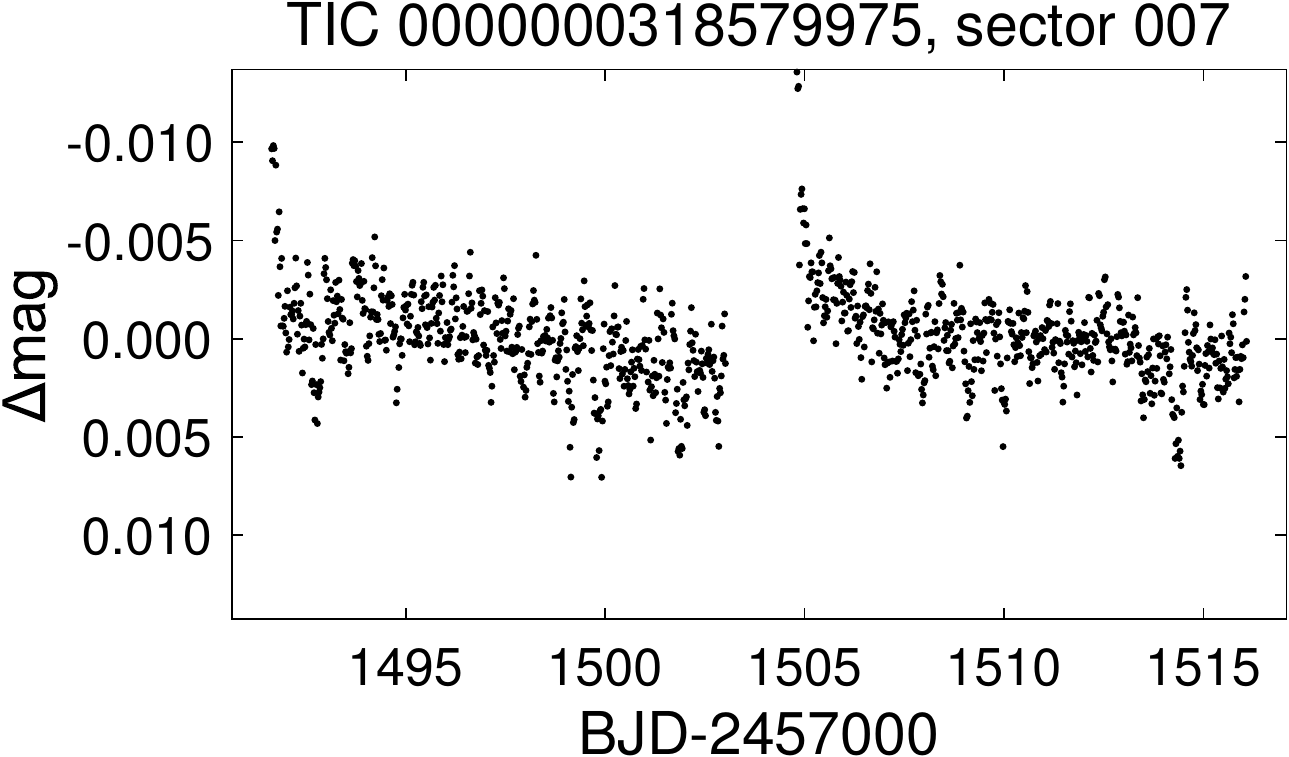}
\includegraphics[width=0.68\columnwidth]{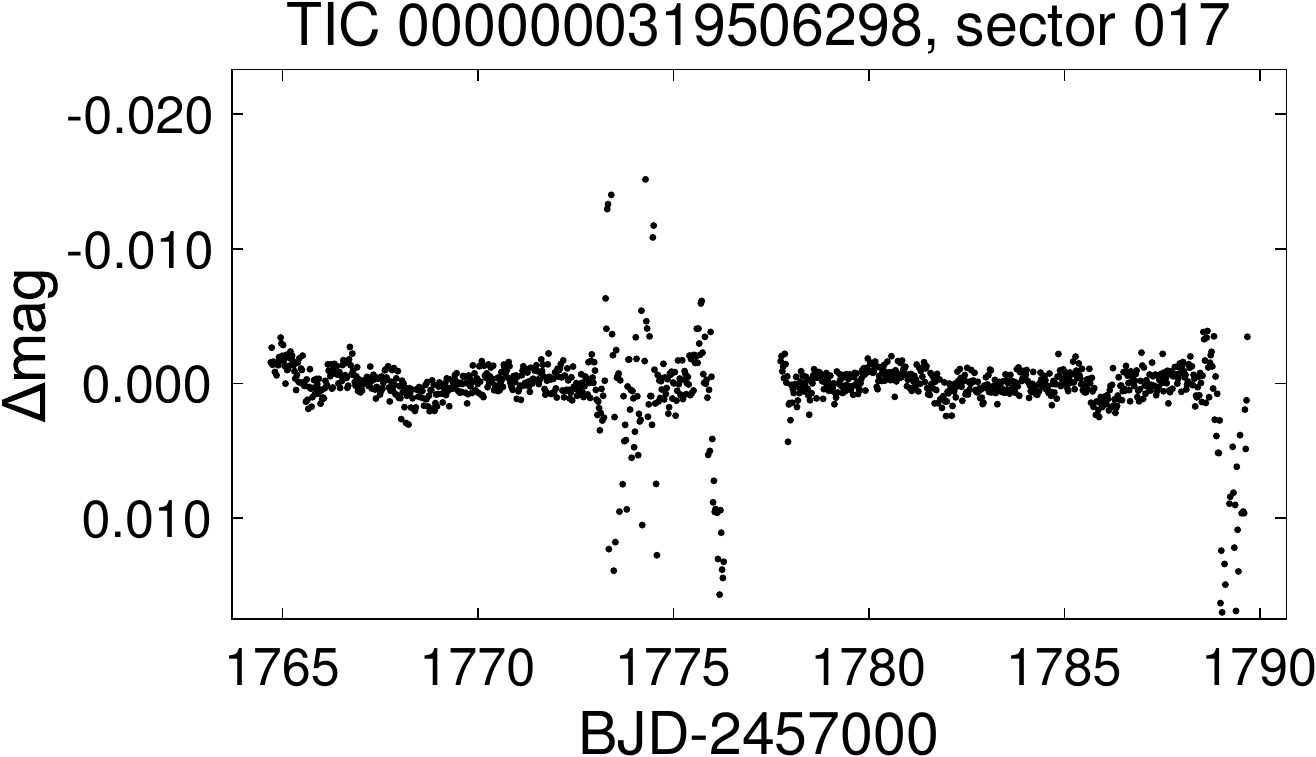}
\includegraphics[width=0.68\columnwidth]{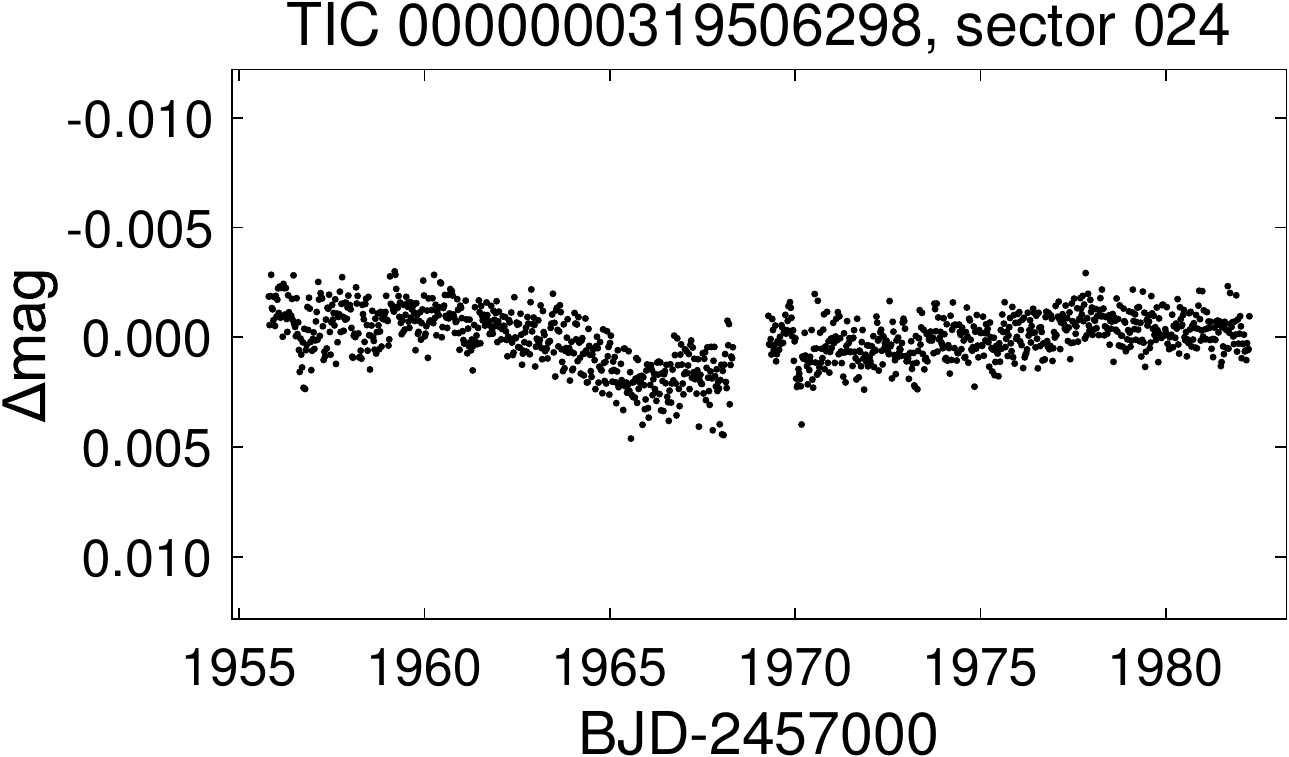}
\includegraphics[width=0.68\columnwidth]{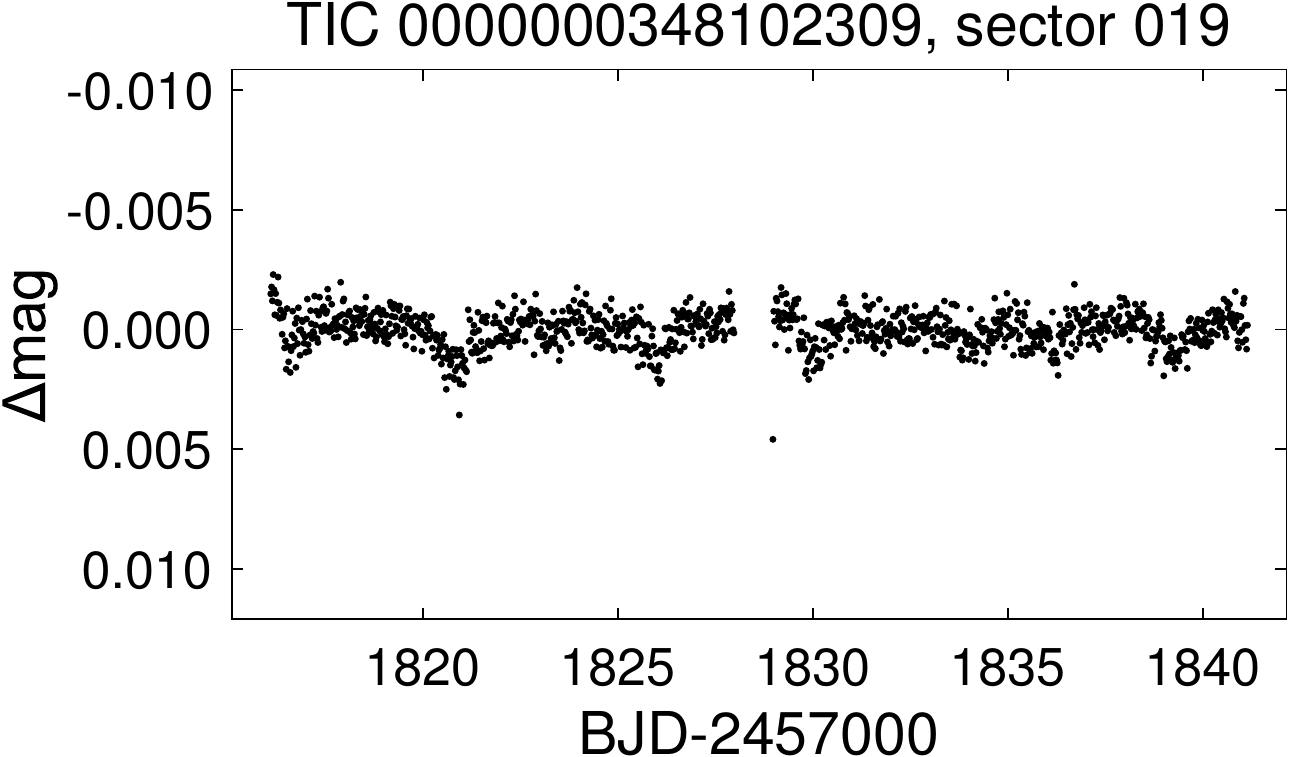}
\includegraphics[width=0.68\columnwidth]{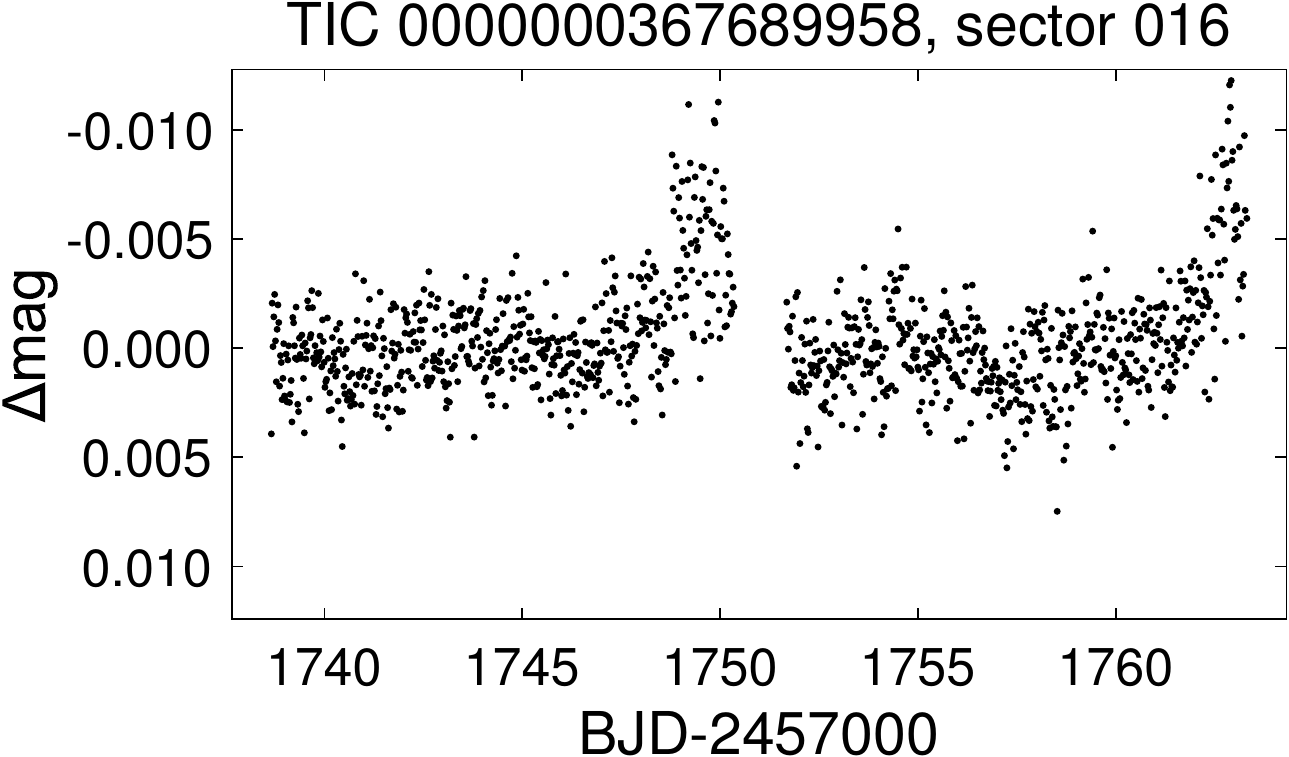}
\includegraphics[width=0.68\columnwidth]{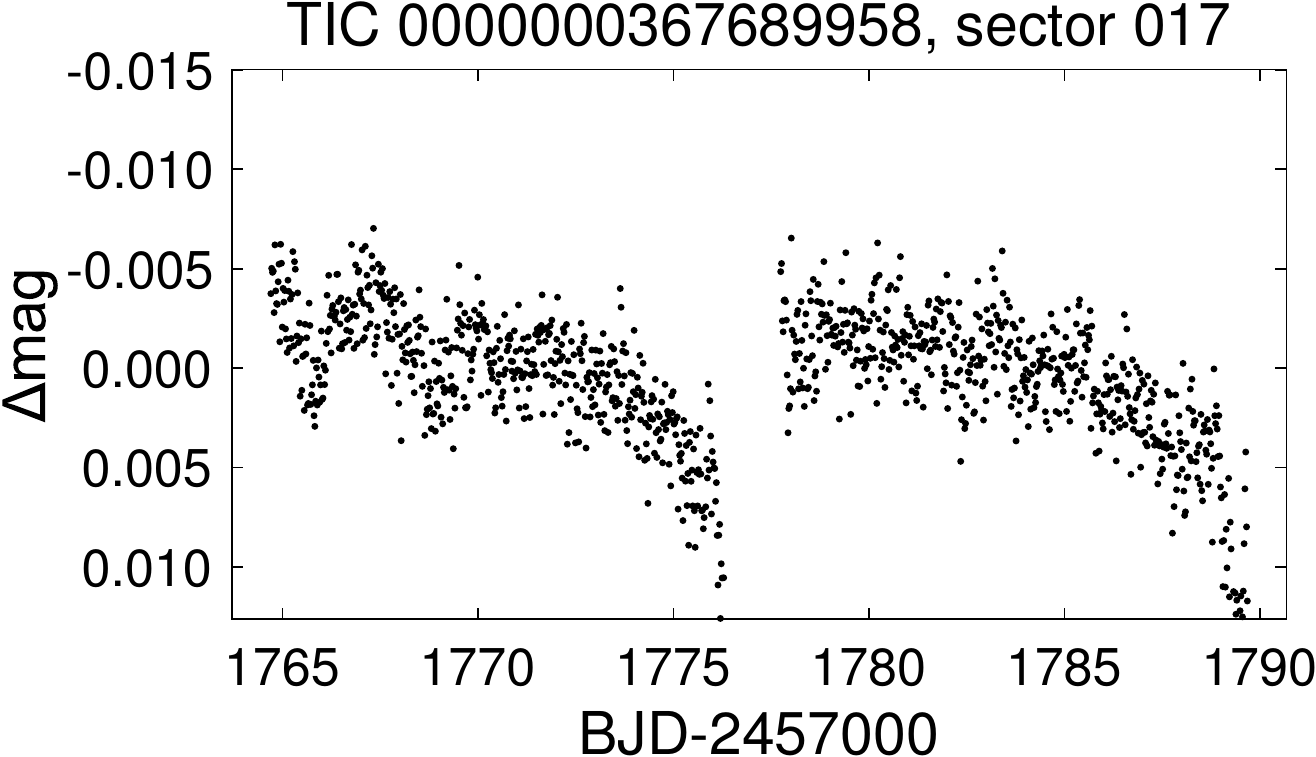}
\includegraphics[width=0.68\columnwidth]{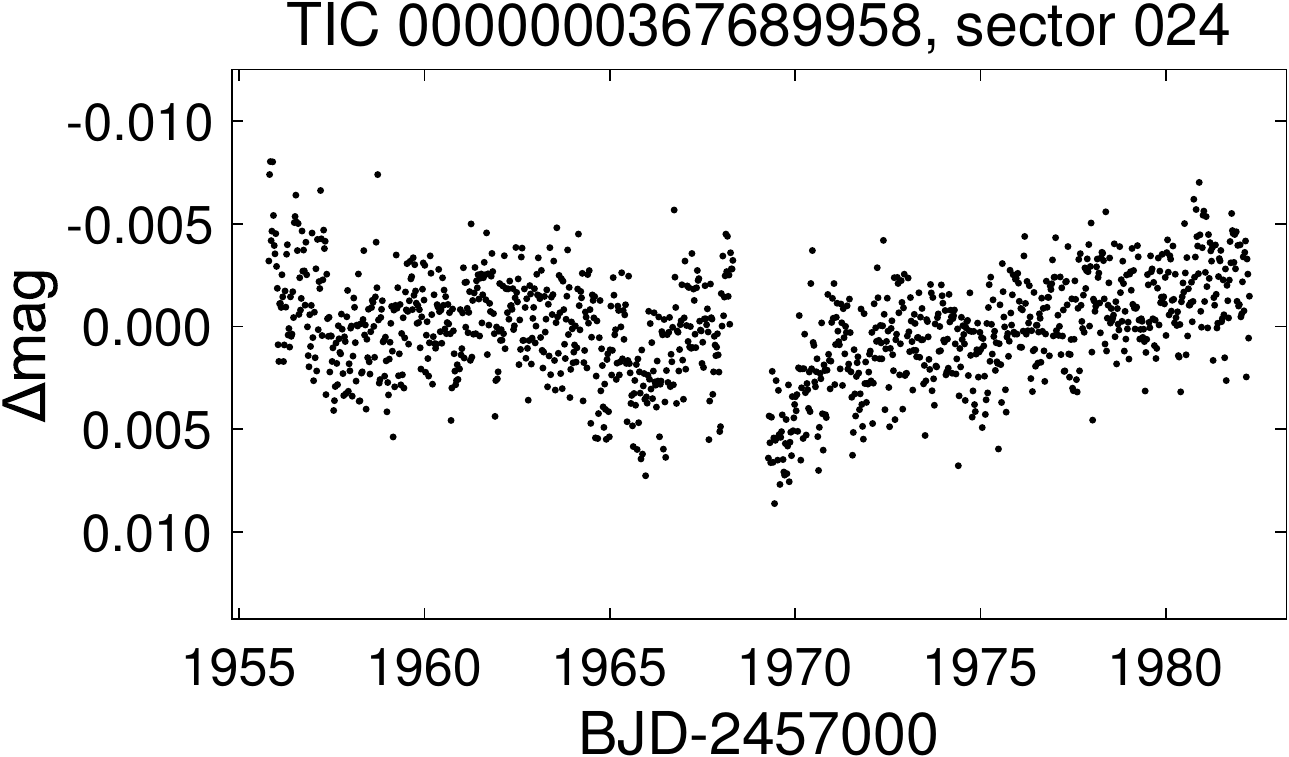}
\includegraphics[width=0.68\columnwidth]{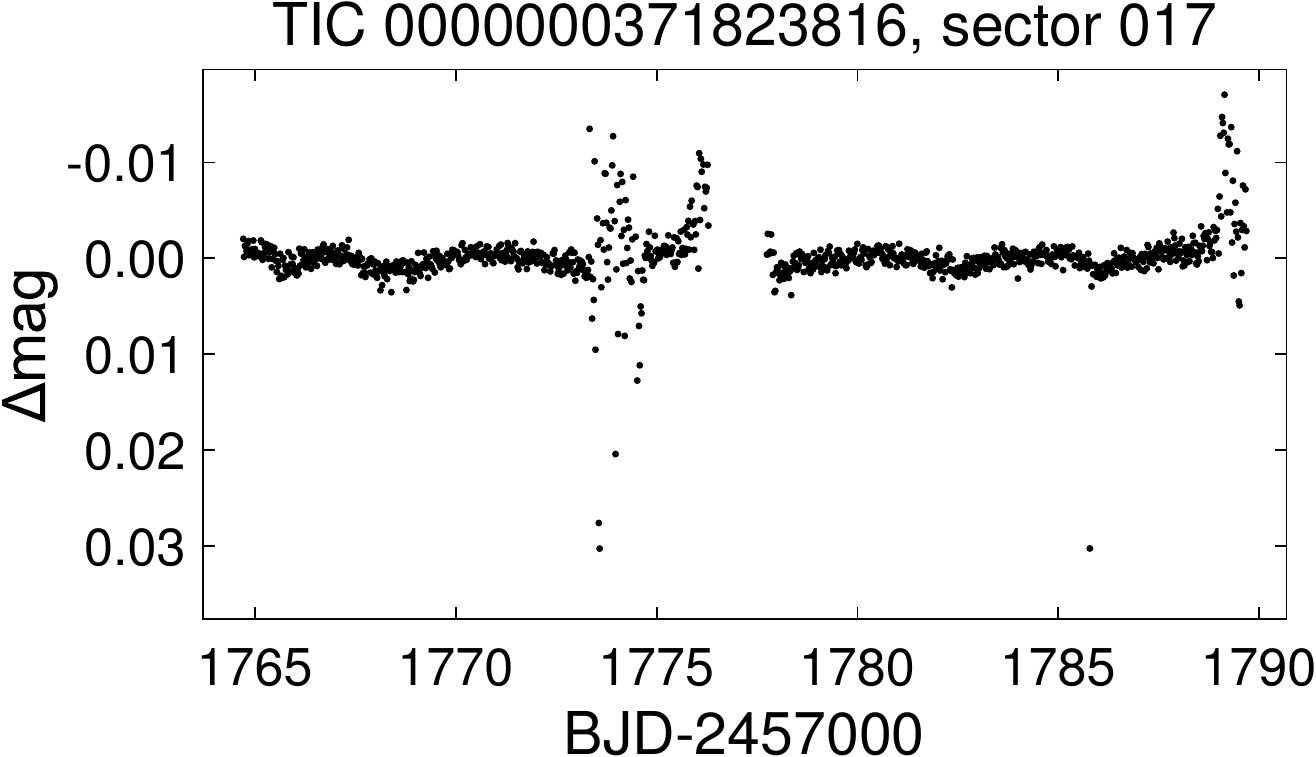}
\includegraphics[width=0.68\columnwidth]{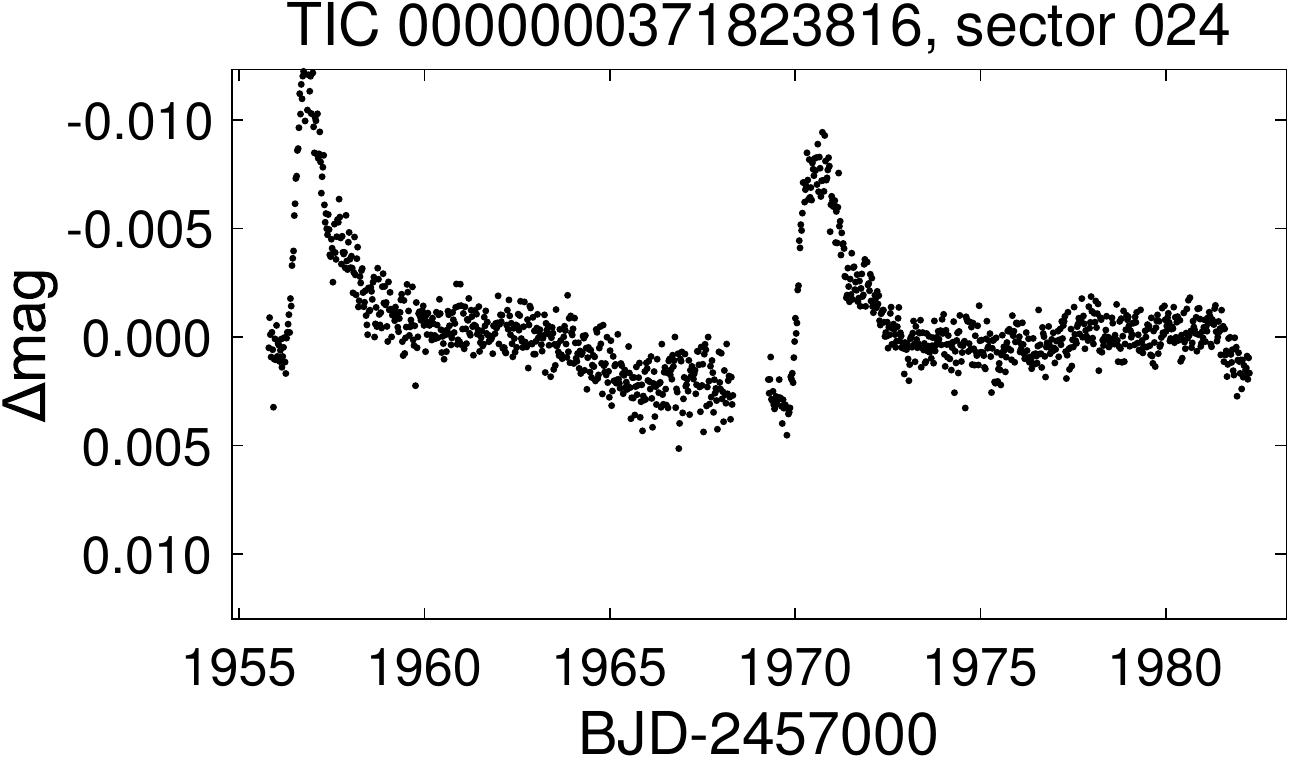}
\includegraphics[width=0.68\columnwidth]{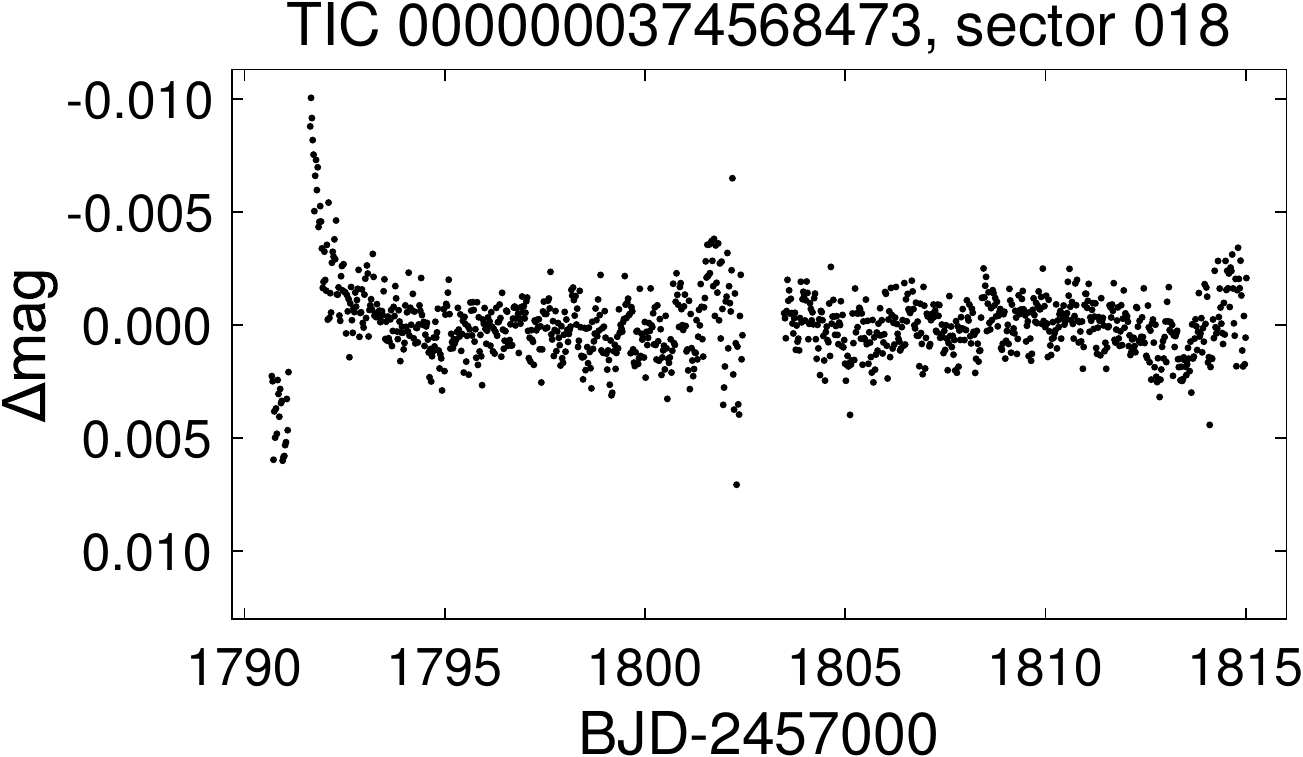}
    \caption{TESS light curves of all objects contained in the final shell star sample.}
		\label{lc3}
\end{figure*}

\begin{figure*}

\includegraphics[width=0.68\columnwidth]{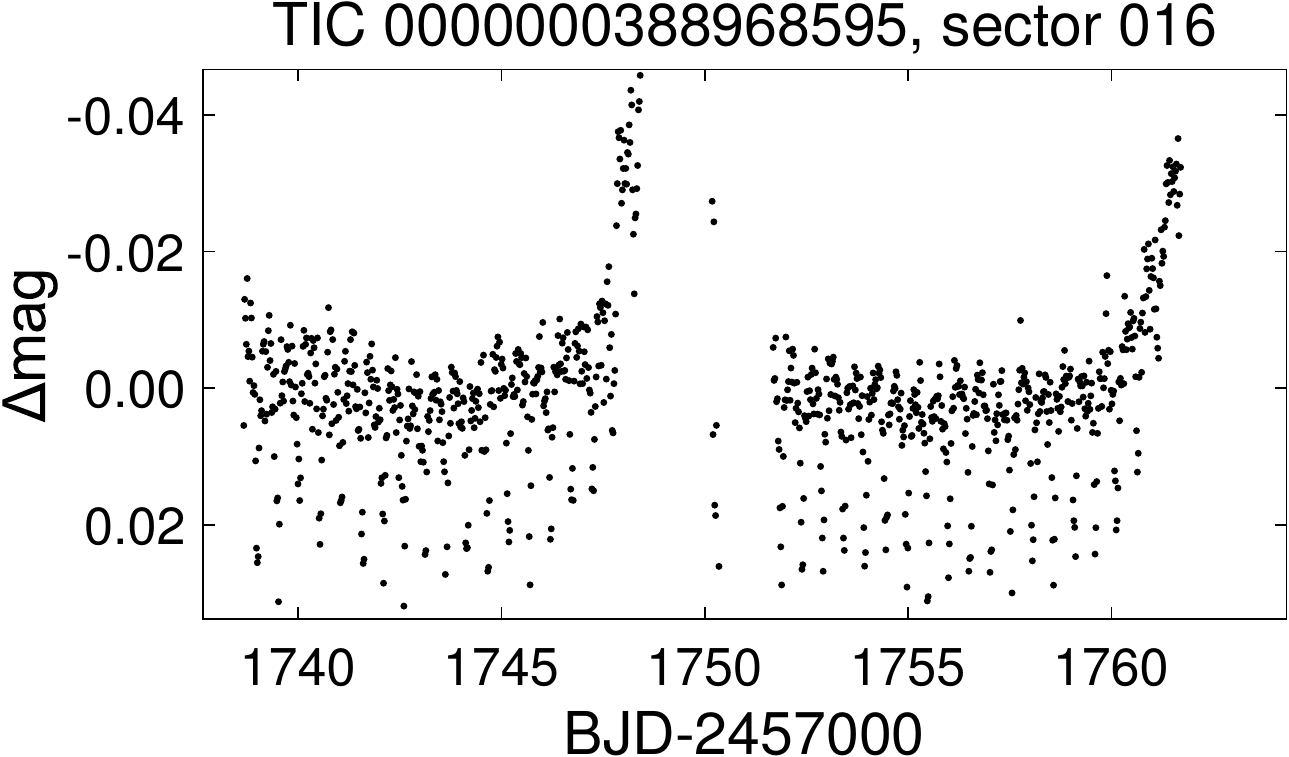}
\includegraphics[width=0.68\columnwidth]{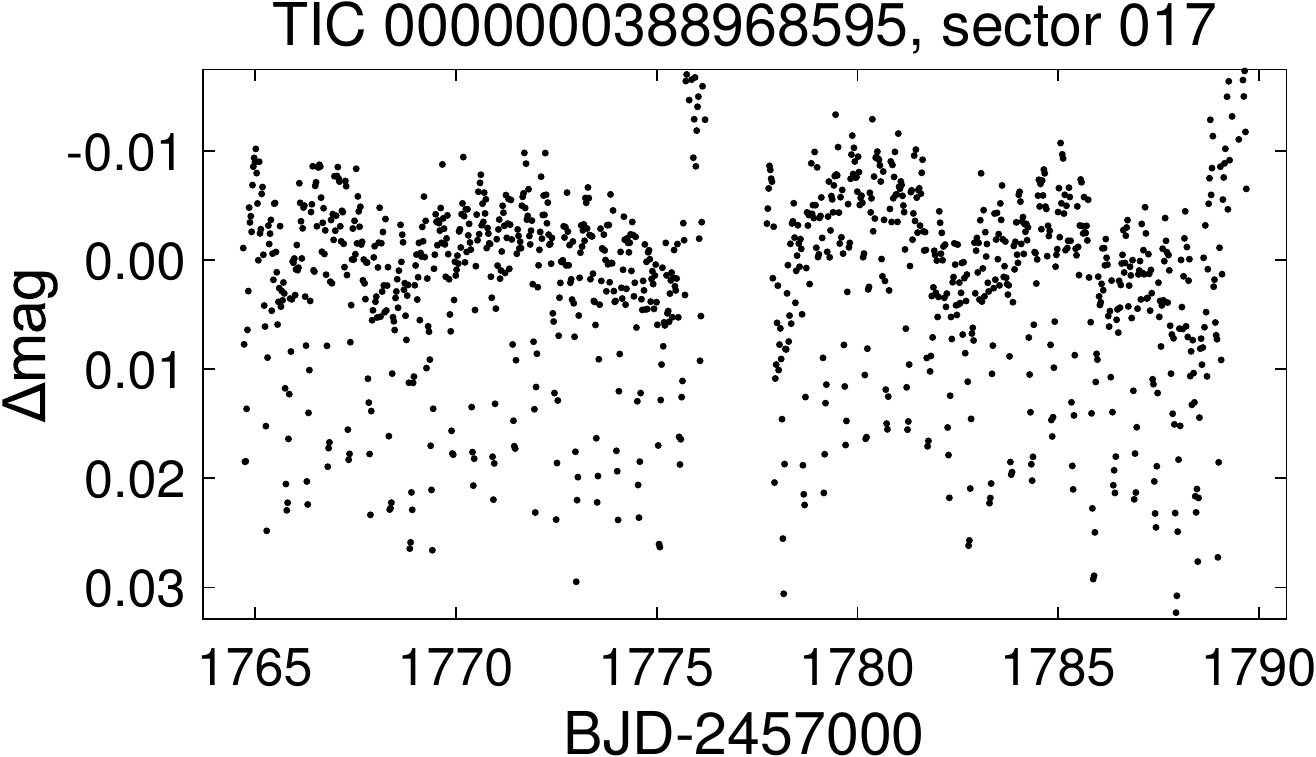}
\includegraphics[width=0.68\columnwidth]{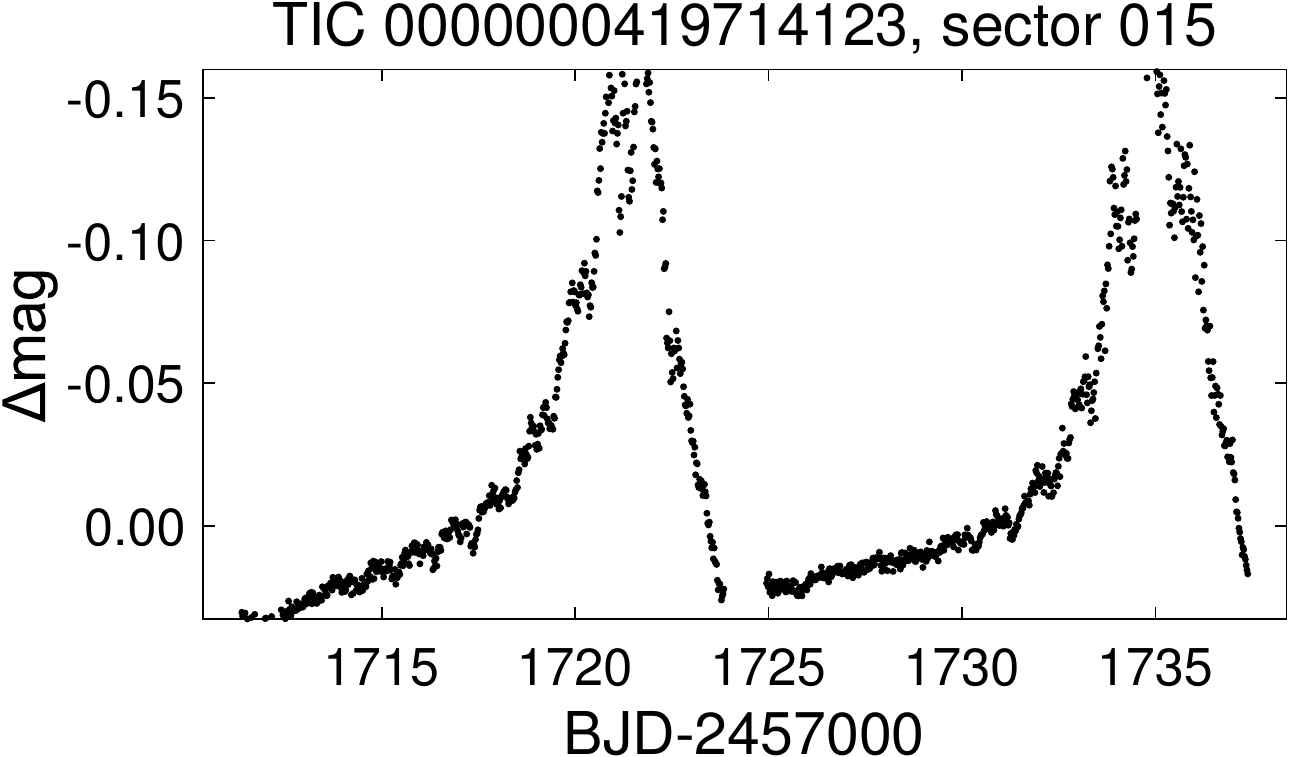}
\includegraphics[width=0.68\columnwidth]{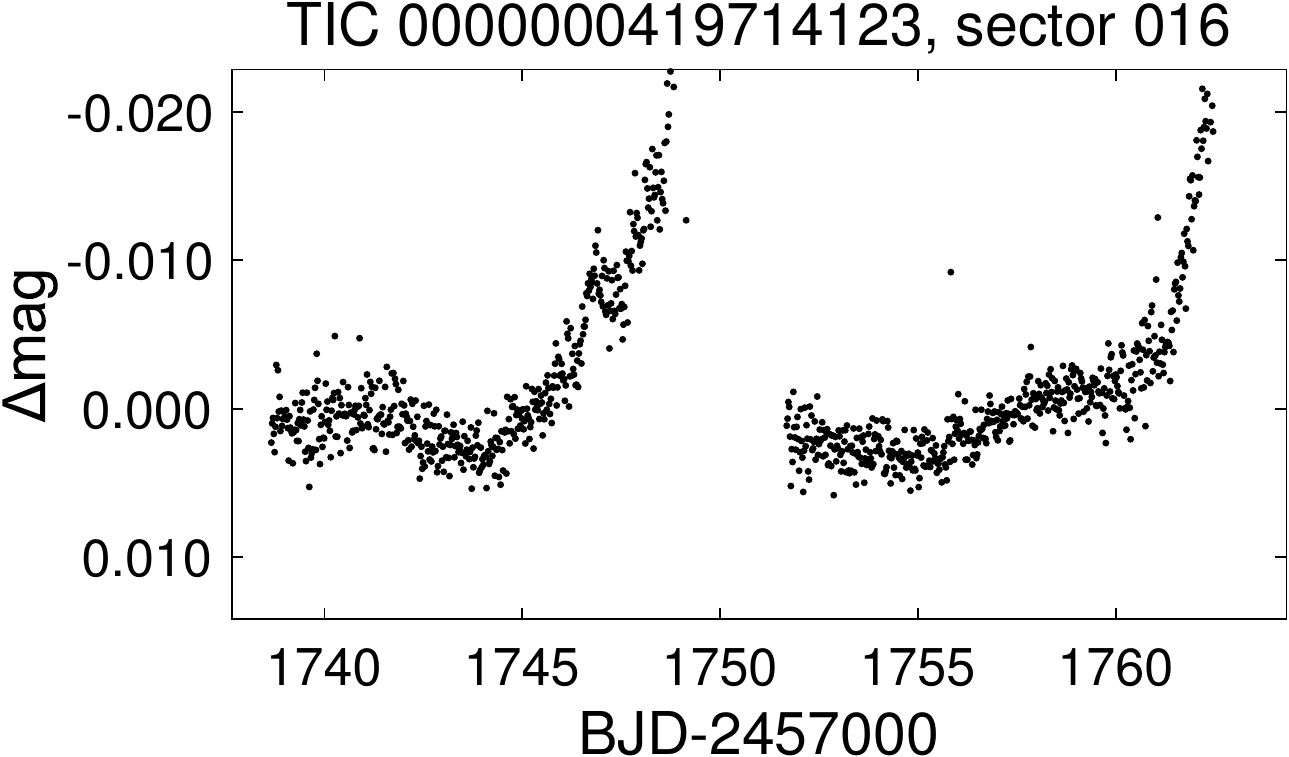}
\includegraphics[width=0.68\columnwidth]{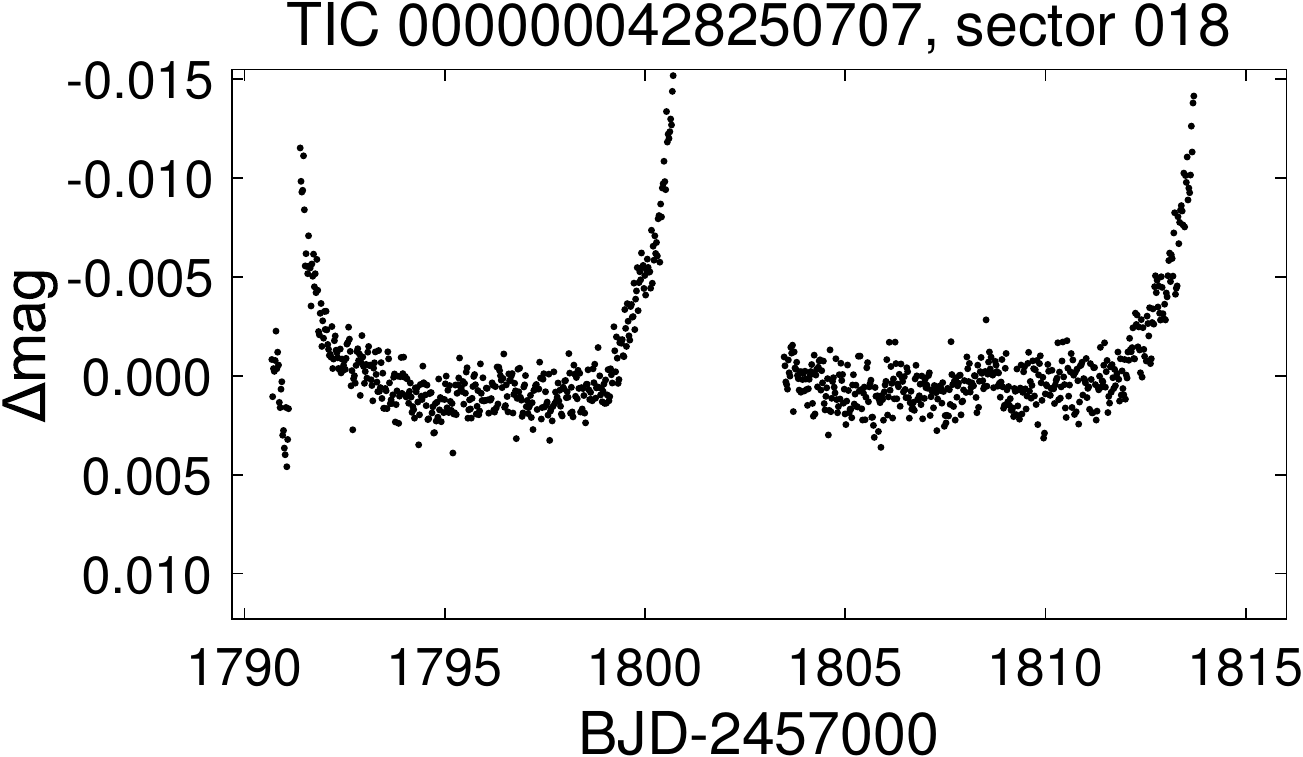}
\includegraphics[width=0.68\columnwidth]{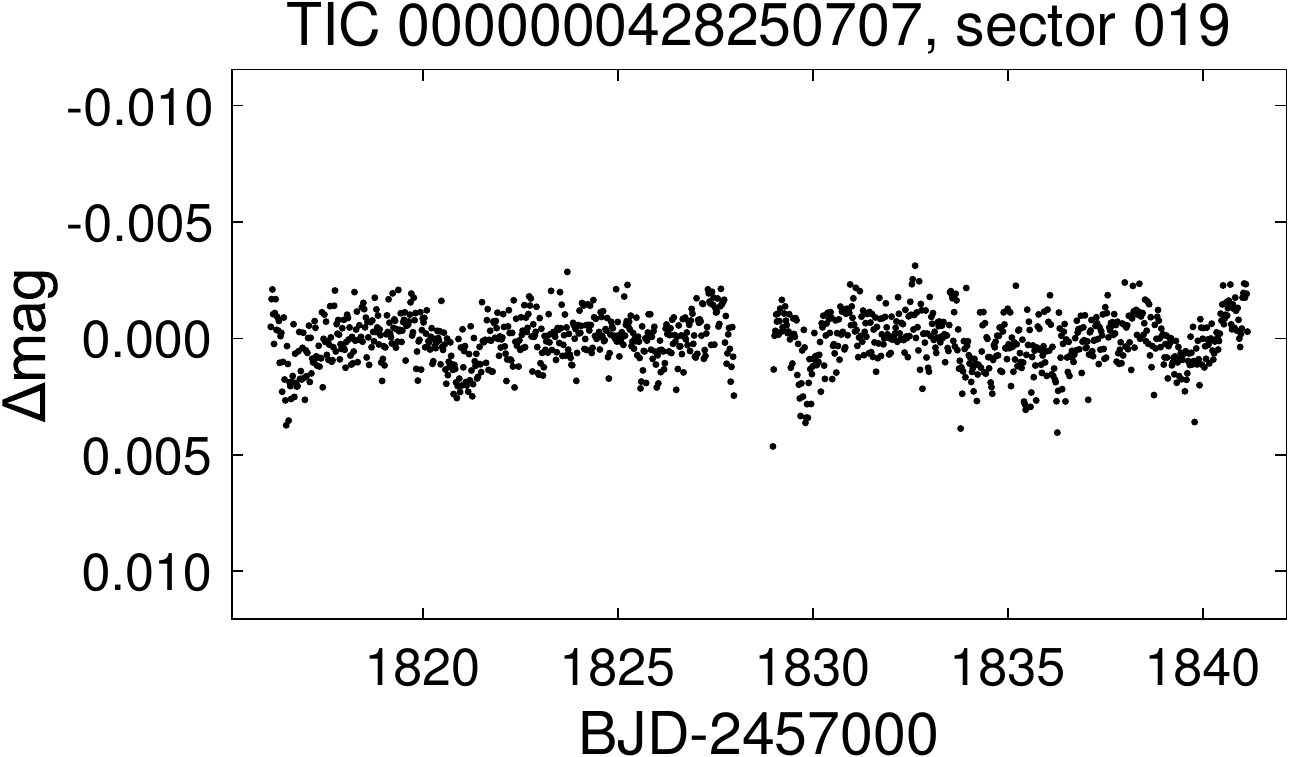}
\includegraphics[width=0.68\columnwidth]{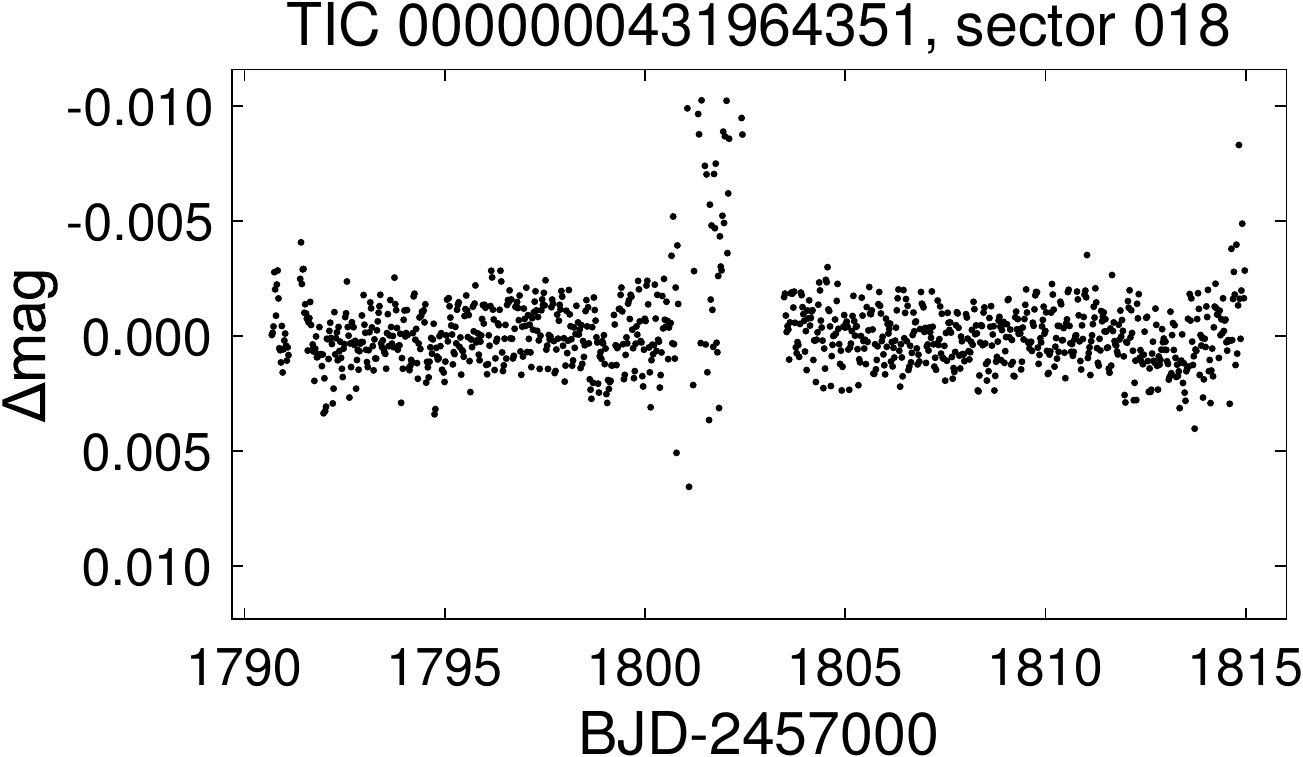}
\includegraphics[width=0.68\columnwidth]{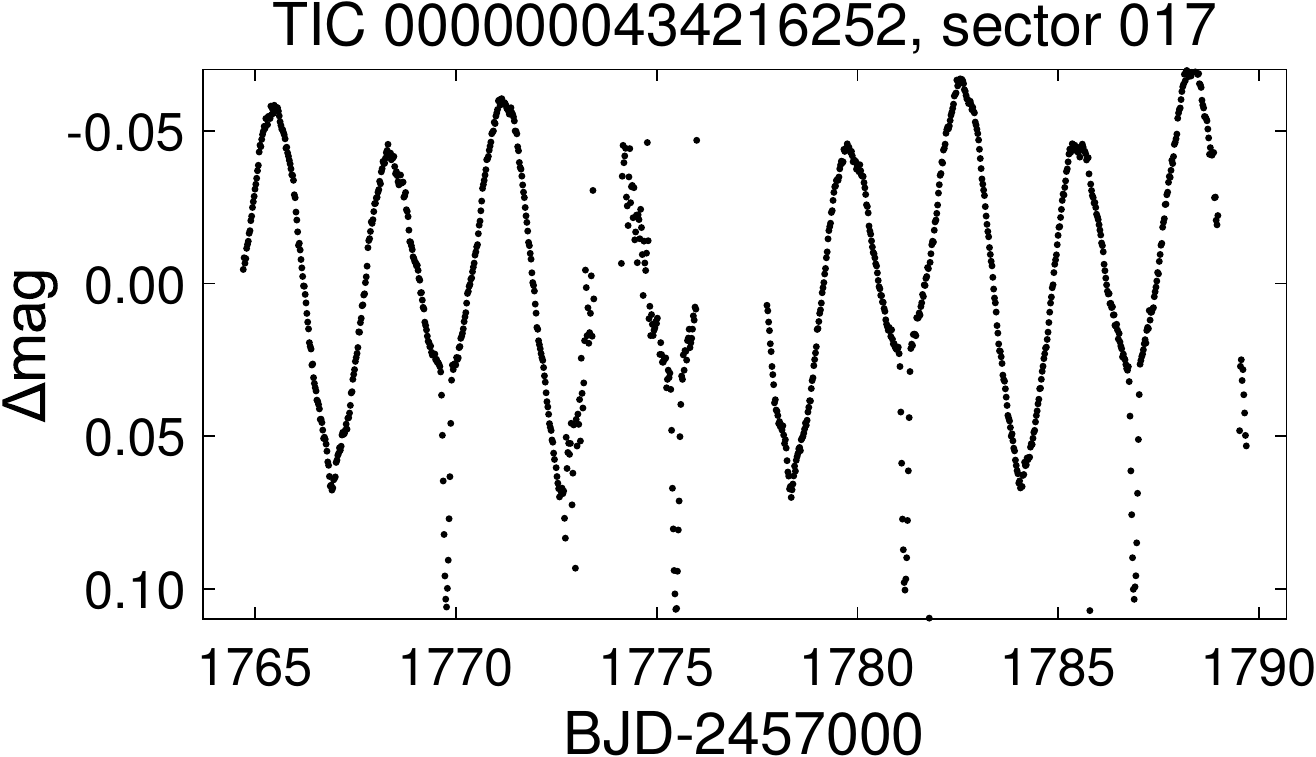}
\includegraphics[width=0.68\columnwidth]{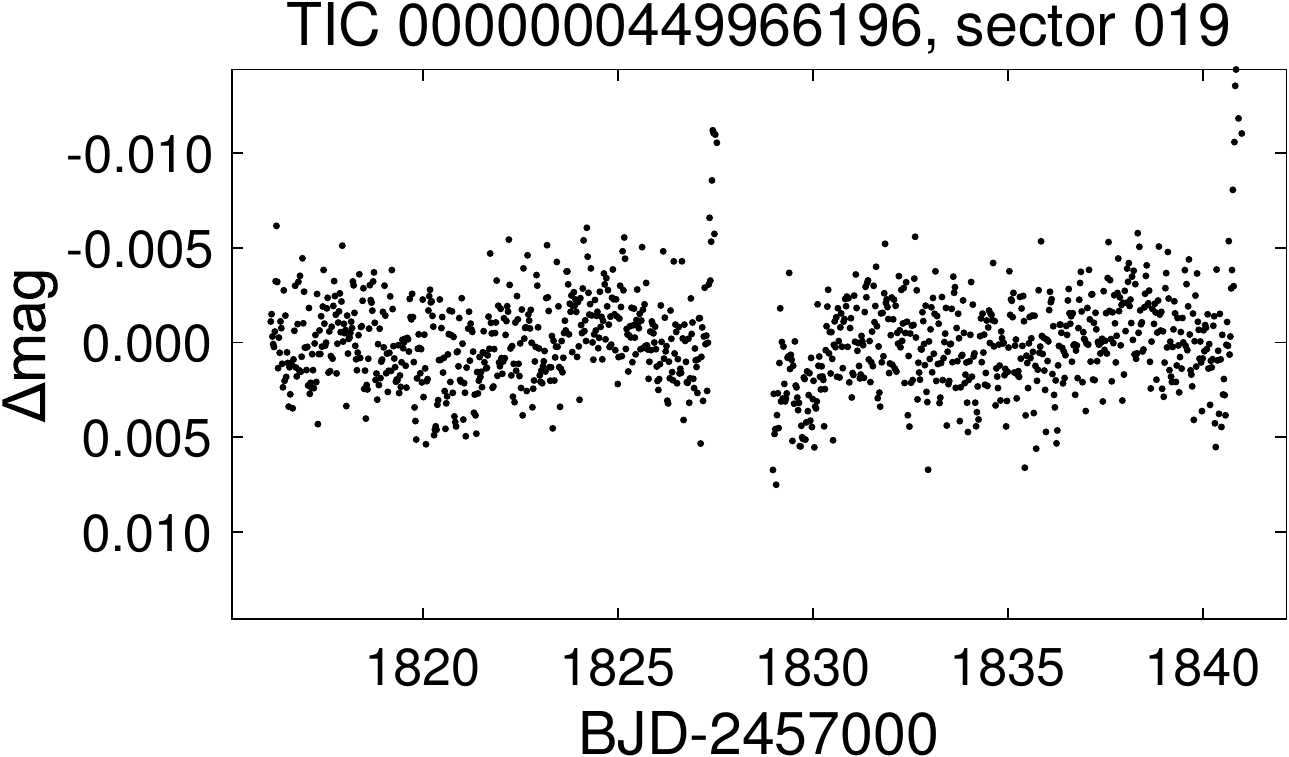}
\includegraphics[width=0.68\columnwidth]{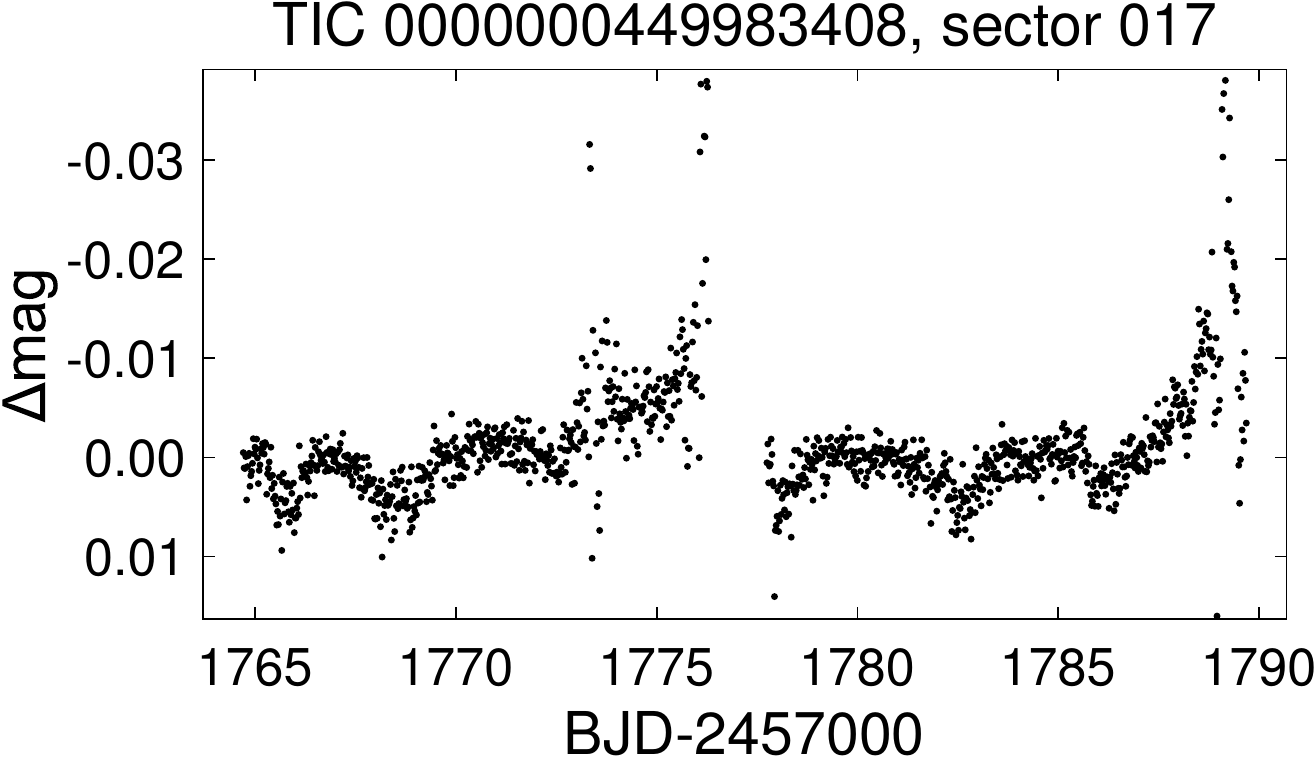}
\includegraphics[width=0.68\columnwidth]{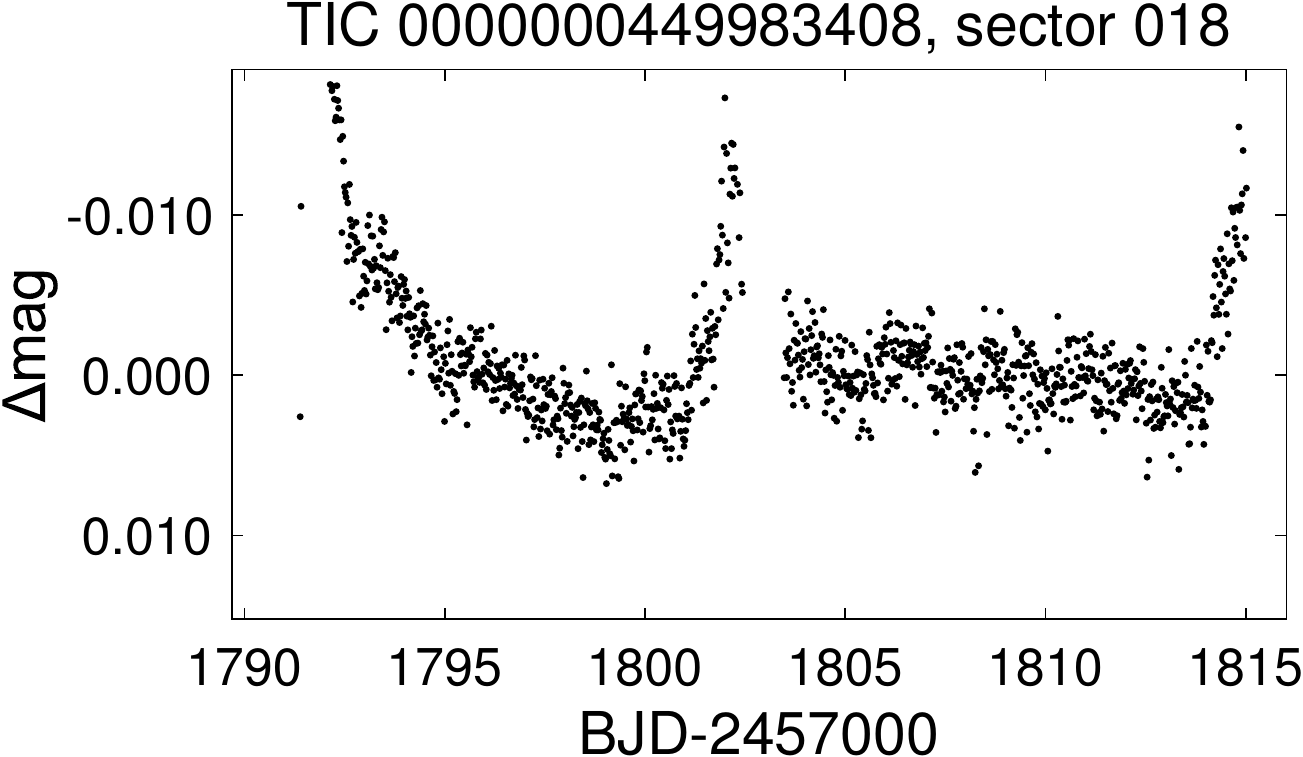}
\includegraphics[width=0.68\columnwidth]{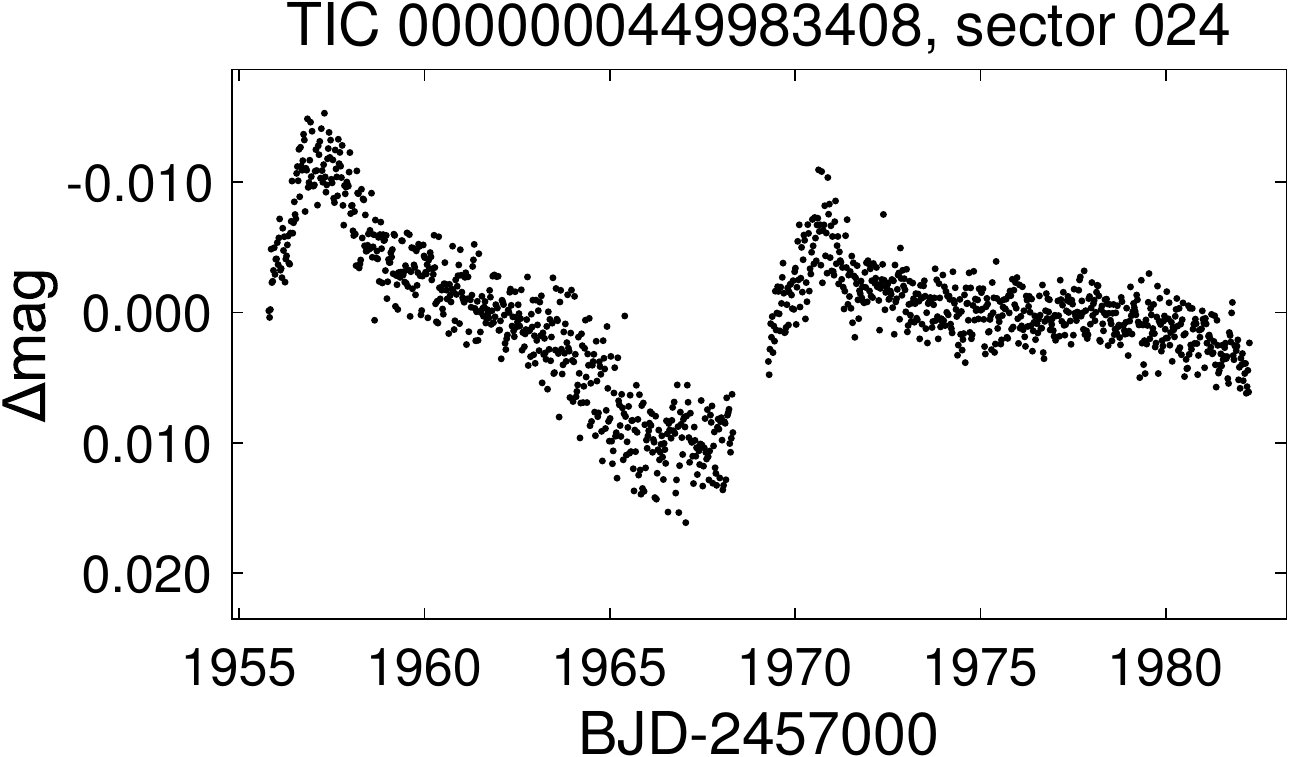}
\includegraphics[width=0.68\columnwidth]{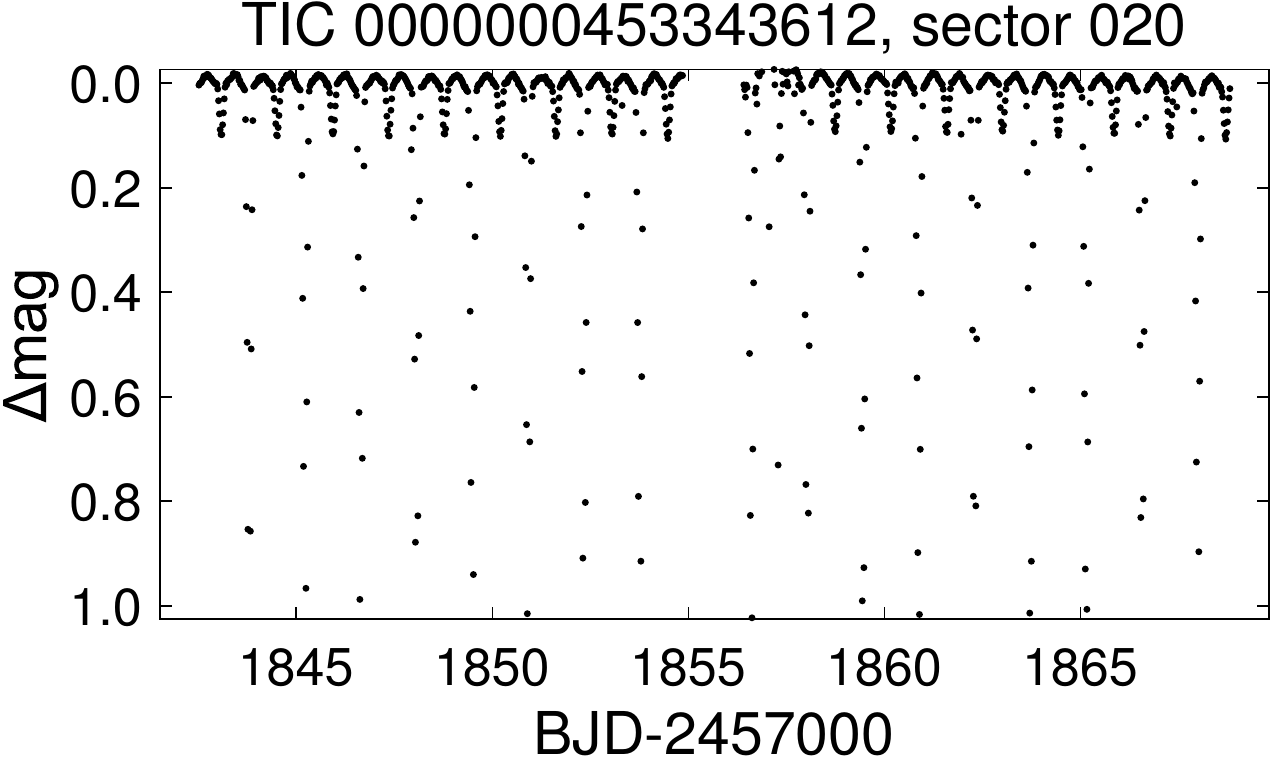}
\includegraphics[width=0.68\columnwidth]{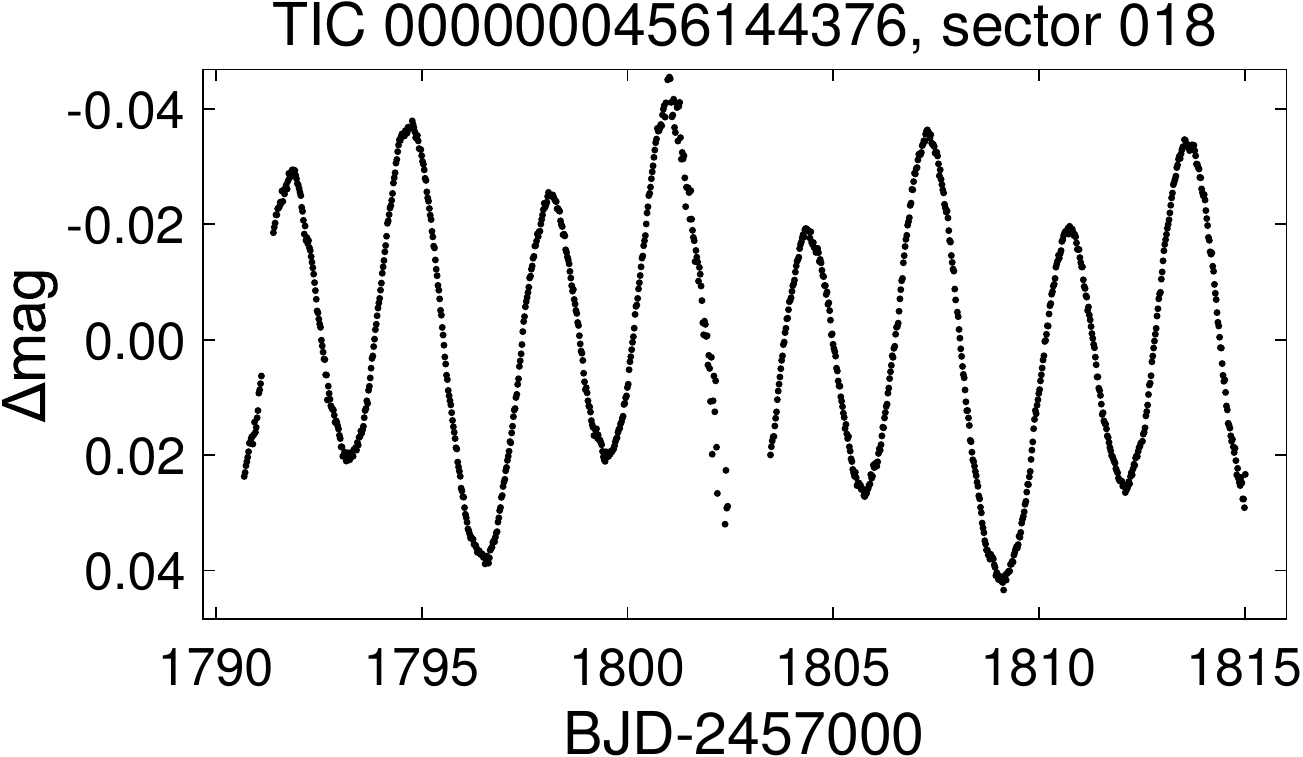}
\includegraphics[width=0.68\columnwidth]{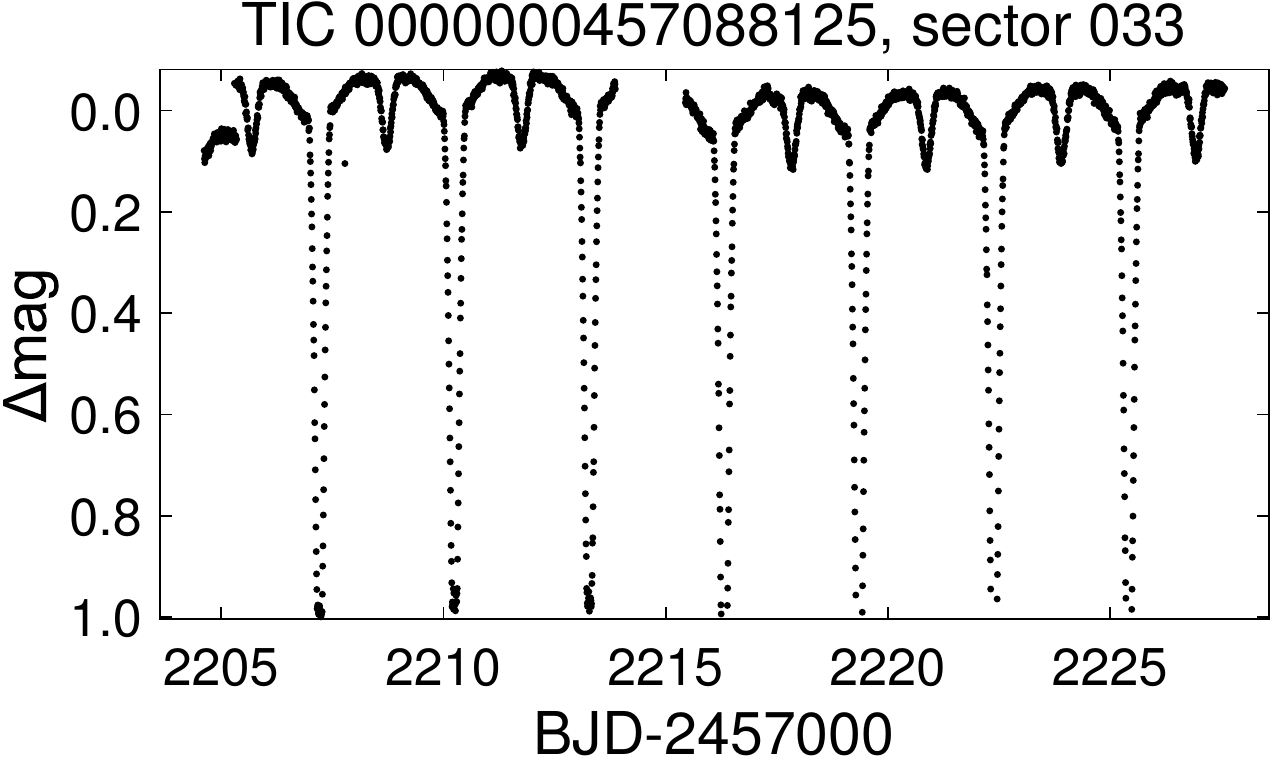}
\includegraphics[width=0.68\columnwidth]{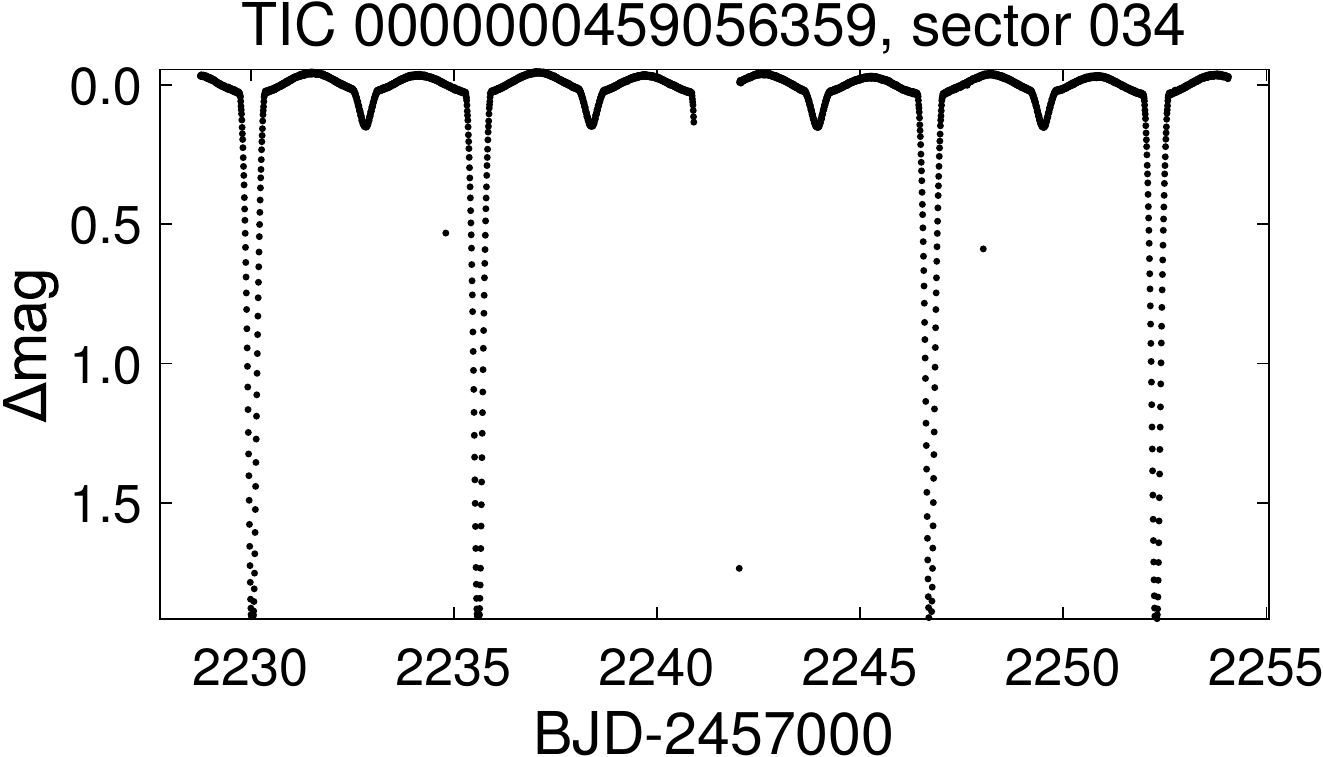}
\includegraphics[width=0.68\columnwidth]{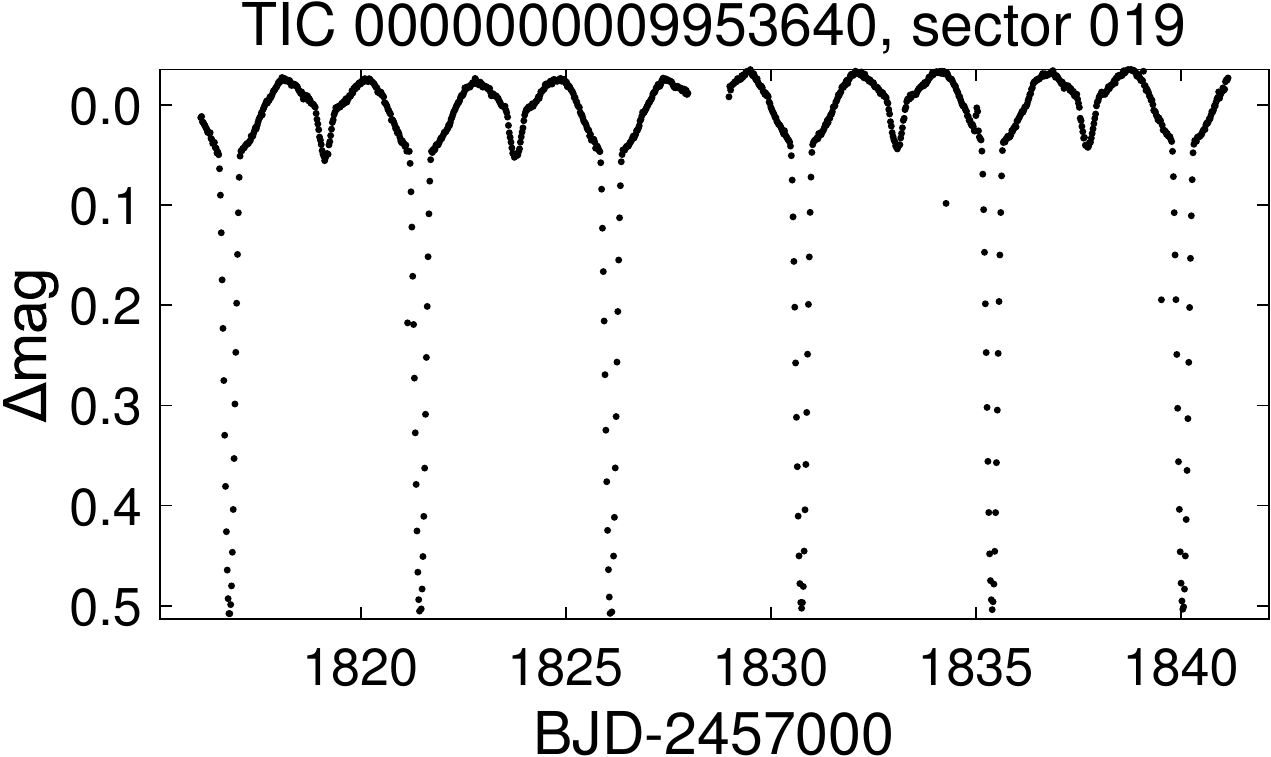}
\includegraphics[width=0.68\columnwidth]{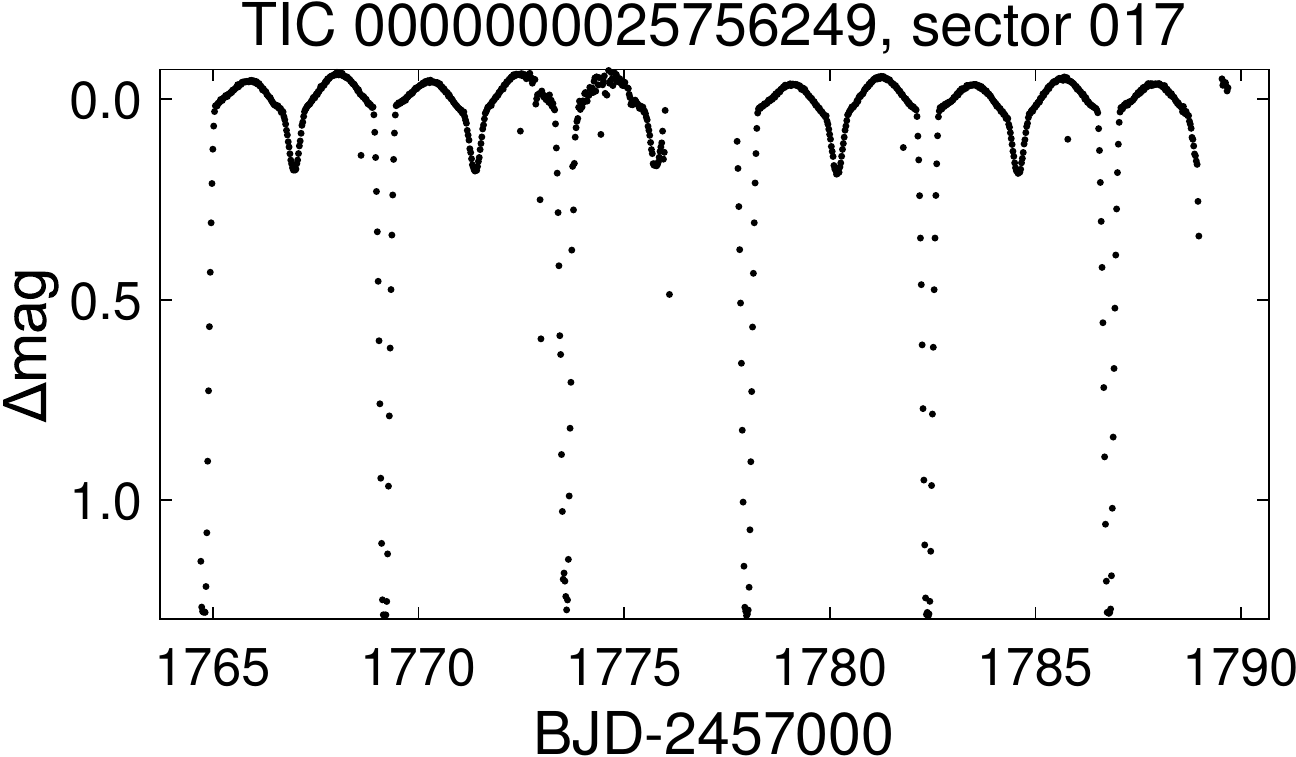}
    \caption{TESS light curves of all objects contained in the final shell star sample.}
		\label{lc4}
\end{figure*}




\bsp	
\label{lastpage}
\end{document}